\documentclass[aps,floatfix,showpacs,nofootinbib,twocolumn]{revtex4-1}
\usepackage{amsmath}
\usepackage{graphicx}
\usepackage{epstopdf}
\usepackage{longtable}

\begin{document}

\title{On the influence of the cosmological constant on trajectories of light and associated measurements in Schwarzschild de Sitter space}

\author{Dmitri Lebedev}
\email{dlebedev@astro.queensu.ca}

\author{Kayll Lake}
\email{lake@astro.queensu.ca}

\affiliation{Department of Physics, Queen's University, Kingston,
Ontario, Canada, K7L 3N6 }

\date{\today}

\begin{abstract}
In this paper we review and build on the common methods used to analyze null geodesics in Schwarzschild de Sitter space. We present a general technique which allows finding measurable intersection angles of null trajectories analytically, and as one of its applications we establish a general relativistic aberration relationship. The tools presented are used to analyze some standard setups of gravitational deflection of light and gain a clear understanding of the role that the cosmological constant, $\Lambda$, plays in gravitational lensing phenomena. Through reviewing some recent papers on the topic with the present results in mind, we attempt to explain the major sources of disagreement in the ongoing debate on the subject, which started with Rindler and Ishak's original paper, regarding the influence of $\Lambda$ on lensing phenomena. To avoid ambiguities and room for misunderstanding we present clear definitions of the quantities used in the present analysis as well as in other papers we discuss.
\end{abstract}

\maketitle

\section{Introduction}
In 1983 Islam, \cite{islam}, showed that the trajectory of light in Schwarzschild de Sitter, henceforth SdS, space is independent of the cosmological constant. As we shall see, this conclusion is, for the most part, correct but does not imply that physical measurements associated with trajectories of light do not depend on $\Lambda$ as well. Making this concept clear will enable us to see a source of confusion in some of the recent literature on the topic. It seems that merely based on Islam's work it was generally assumed that $\Lambda$ plays no role in gravitational lensing phenomena and has no place in the analysis; and that the appearance of $\Lambda$ in some equations could be transformed away, in one way or another, and therefore is artificial, revealing the true independence on $\Lambda$. This general belief turns out to be true only in situations where no measurements made by specific observers are considered. However, to study the phenomenon properly, it is important to consider measurements made by observers and the dependence of measurable quantities on the system parameters. In 2007 Rindler and Ishak, \cite{ri1}, showed that if measurable intersection angles are considered, in a standard simple setup of gravitational deflection of light, then results of interest do depend on $\Lambda$. Rindler and Ishak's conclusions immediately led to both enthusiasm and scepticism; perhaps they were mistakenly seen to be in direct contradiction to the common belief that followed after Islam's work. Since their original paper, there was much activity surrounding this topic. Some authors searched for other setups and methods of analysis in which results of interest depend on $\Lambda$ in support of Rindler and Ishak's conclusions, see for example \cite{sereno1}, \cite{bbs}, \cite{schucker1}, \cite{kcd}. Others tried to find errors in Rindler and Ishak's work and explain the invalidity of their conclusions, and ultimately show that the traditional approach to the topic needs no modification, see for example \cite{park}, \cite{ak}, \cite{kp}, \cite{simpson}. All together, the papers that followed \cite{ri1} amount to a very interesting discussion of the subject, in which, unfortunately, there are no definitively agreed upon answers to many important questions. In what follows we attempt to make the theory abundantly clear and explain the exact role of $\Lambda$ in gravitational lensing phenomena. We discuss and clarify key issues and illuminate sources of disagreement in the recent literature. In turn we hope to settle the ongoing debate on the influence of $\Lambda$ and present a clear description of light deflection phenomenon in SdS space together with all the necessary tools for analyzing any setup.

Along the course of our investigation, we derive and introduce an invariant general formula, which allows the determination of a measurable intersection angle from fundamental parameters. This formula seems to be essential in the study of the present topic, but quite surprisingly is missing from the current literature. We also address the role of relativistic aberration of light in the analysis and demonstrate how our general formula encompasses this effect and allows for a simple way to account for it. In fact, the general formula can be used to derive an invariant aberration equation, applicable to any background geometry and orientation, and which reduces to the known aberration equation as a special case. The general angle formula and the general aberration equation we present may be considered as some of the most significant results of this paper; their applicability may extend to multiple areas well beyond the current topic.

Our presentation is organized as follows. In section \ref{sec2} we discuss the influence of $\Lambda$ on the geometry and build an intuitive understanding of how this may lead to the appearance of $\Lambda$ in results of interest. In section \ref{sec3} we turn our attention to null geodesics and address the fundamental issue regarding the appearance of $\Lambda$ in the orbital equation of light and its solution. In section \ref{sec4} we continue the discussion of the above issue and present the necessary tools needed to pose and answer some important questions. In section \ref{sec5} we derive the general formula for measurable intersection angles and demonstrate its use in a few applications. Finally, in section \ref{sec6} we discuss some of the recent papers on the topic and respond to their results and conclusions.

\section{Underlying geometry and diagrams} \label{sec2}
Consider the Kottler metric \cite{kot}, describing SdS spacetime,
\begin{equation}
\mathrm{d}s^2 = -f(r)\mathrm{d}t^2+\frac{\mathrm{d}r^2}{f(r)}+r^2\sin^2(\theta)\mathrm{d}\phi^2+r^2\mathrm{d}\theta^2, \label{tp1e1}
\end{equation}
where
\begin{equation*}
f(r)=1-\frac{2m}{r}-\frac{\Lambda}{3}r^2.
\end{equation*}
Here we have an object of mass $m$ at the centre of the coordinates, in a universe with a cosmological constant $\Lambda > 0$. The range that we are interested in is $f(r)>0$; for the case where both $m$ and $\Lambda$ are sufficiently small, this implies that $r_{Sch.}<r<r_{dS}$, where $r_{Sch.} \approx 2m$ and $r_{dS} \approx \sqrt{\frac{3}{\Lambda}}$. In this range, $t$ is a time-like coordinate while $r$, $\phi$ and $\theta$ are space-like coordinates. $r_{Sch.}$ and $r_{dS}$ are known as the Schwarzschild and the de Sitter horizons, respectively. Sometimes also called the inner and outer horizons, respectively, in the context of SdS space. It is easily verified that any orbit in this geometry can be confined to a single azimuthally symmetric spatial slice containing the origin. Therefore, without loss of generality we can take $\theta=\frac{\pi}{2}$, and consider motion in the sub spacetime, with the metric
\begin{equation}
\mathrm{d}s^2 = -f(r)\mathrm{d}t^2+\frac{\mathrm{d}r^2}{f(r)}+r^2\mathrm{d}\phi^2. \label{tp1e2}
\end{equation}
It is useful to take slices of constant $t$ in this spacetime and study orbits in the two dimensional subspace, parametrized by $r$ and $\phi$. The metric on such a slice of space is
\begin{equation}
\mathrm{d}s^2 = \frac{\mathrm{d}r^2}{f(r)}+r^2\mathrm{d}\phi^2. \label{tp1e3}
\end{equation}
It is immediately evident that this space is not flat, however since it is parametrized by polar coordinates, ($r$,$\phi$), we can construct flat diagrams depicting the orbits taking place in the slice. We must however keep in mind the difference between our flat diagrams and the curved physical space in which measurements may take place. That is, diagrams will be drawn on a flat ($r$,$\phi$) plane, real events will be taking place in the curved spacetime, a slice of which is represented by metric \eqref{tp1e3}. Making this distinction is particularly important when considering angles. Intersection angles between curves appearing on the flat diagram may be considerably different when projected onto the curved physical space. For a visual demonstration of the issue let us consider the portion of SdS space in between the two horizons and isometrically embed the two dimensional slice with metric \eqref{tp1e3} in flat three dimensional space. Through an isometric embedding, which preserves distances, we can picture the structure of the underlying geometry in which the physical events take place. To this end, let us take a flat 3-space with cylindrical coordinates ($\rho$,$\varphi$,$z$) and metric
\begin{equation}
\mathrm{d}s^2 = \mathrm{d}\rho^2+r^2\mathrm{d}\varphi^2+\mathrm{d}z^2. \label{tp1e4}
\end{equation}
The complete description of the embedding is complicated for $m \neq 0$ and $\Lambda \neq 0$, but when both parameters are small enough, specifically when the product $m\sqrt{\Lambda}$ is negligibly small, then a convenient approximation can be used to get the shape of the surface. We consider this, realistic, case of such small parameters and approximate the embedded surface for small and large $r$ in turn.

\begin{figure}[!ht]
\includegraphics[width=85mm]{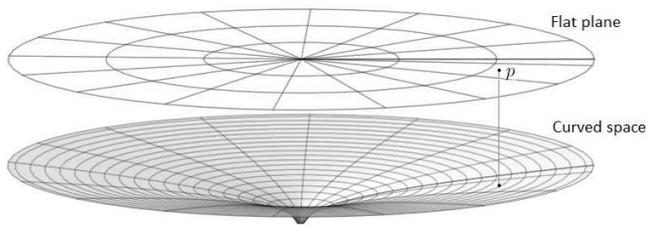}
\caption{Flamm's paraboloid and a flat plane. The curved surface represents the physical space close to the mass while the plane above is useful in making flat diagrams. The correspondence between the points of the plane and the points of the curved surface is by direct projection, as depicted in the diagram.} \label{fig1}
\end{figure}

For small $r$, $\Lambda r^2 \approx 0$, and the metric of \eqref{tp1e3} is approximately
\begin{equation}
\mathrm{d}s^2 = \frac{\mathrm{d}r^2}{f_{\Lambda=0}(r)}+r^2\mathrm{d}\phi^2, \label{tp1e5}
\end{equation}
where
\begin{equation*}
f_{\Lambda=0}(r)=1-\frac{2m}{r}.
\end{equation*}
Embedding this 2-surface in the flat 3-space of metric \eqref{tp1e4} yields the following relationships:
\begin{equation*}
\rho=r,\qquad \varphi=\phi,\qquad z=2\sqrt{2m(r-2m)}.
\end{equation*}
The embedded surface is therefore the set in flat 3-space satisfying
\begin{equation*}
z=2\sqrt{2m(\rho-2m)}.
\end{equation*}
It is known as Flamm's paraboloid. To ensure a one to one correspondence of points we only consider one half of the paraboloid, allowing only positive $z$ on the embedded surface. Hence, at small $r$ the intrinsic geometry of the surface described by metric \eqref{tp1e3} can be approximated by Flamm's paraboloid, shown in Figure \ref{fig1}.

\begin{figure}[!ht]
\includegraphics[width=85mm]{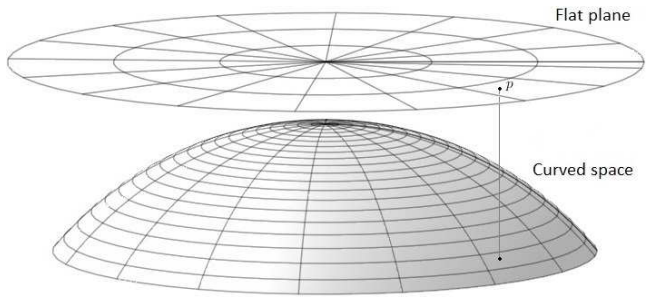}
\caption{Half spherical shell and a flat plane. The curved surface represents the physical space far away from the mass while the plane above is useful in making flat diagrams. The correspondence between the points of the plane and the points of the curved surface is by direct projection, as depicted in the diagram.} \label{fig2}
\end{figure}

For large $r$, $\frac{m}{r} \approx 0$, and the metric of \eqref{tp1e3} is approximately
\begin{equation}
\mathrm{d}s^2 = \frac{\mathrm{d}r^2}{f_{m=0}(r)}+r^2\mathrm{d}\phi^2, \label{tp1e6}
\end{equation}
where
\begin{equation*}
f_{m=0}(r)=1-\frac{\Lambda}{3}r^2.
\end{equation*}
Embedding this 2-surface in the flat 3-space of metric \eqref{tp1e4} yields the following relationships.
\begin{equation*}
\rho=r,\qquad \varphi=\phi,\qquad z=\sqrt{\frac{3}{\Lambda}-r^2}.
\end{equation*}
The embedded surface is therefore the set in flat 3-space satisfying
\begin{equation*}
z=\sqrt{\frac{3}{\Lambda}-\rho^2}.
\end{equation*}
It describes half of a spherical shell. To ensure a one to one correspondence of points we only consider positive values of $z$ on the embedded surface. Hence, at large values of $r$, the intrinsic geometry of the surface described by metric \eqref{tp1e3} can be approximated by half of a spherical shell, shown in Figure \ref{fig2}.

\begin{figure}[!ht]
\includegraphics[width=85mm]{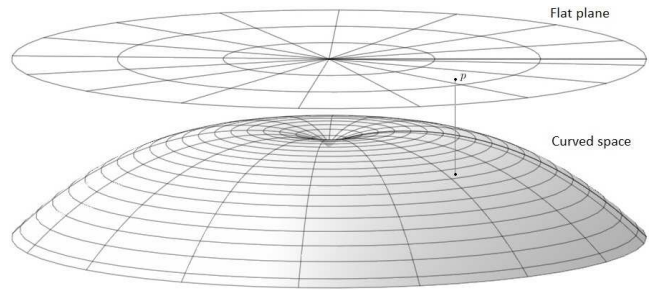}
\caption{An embedded curved surface, representing a slice of SdS space, and a flat plane. The curved surface represents the physical space while the plane above is useful in making flat diagrams. The correspondence between the points of the plane and the points of the curved surface is by direct projection, as depicted in the diagram.} \label{fig3}
\end{figure}

The overall shape of the embedded surface of metric \eqref{tp1e3} can be approximated by piecing together Flamm's paraboloid for small $r$ and the half shell for large $r$. This resulting surface, depicted in Figure \ref{fig3}, is a qualitative representation of the shape of the slice; its main use is in visualizing how the distances associated with the coordinates stretch due to intrinsic geometry. One may argue that to properly connect the surfaces of large $r$ and small $r$, the lower half of the spherical shell at large $r$ must be used, that is, $z$ must be taken negative in the transformation when ensuring bijection, however, for our purposes this is not important. This visualization will be an aid in qualitatively understanding how the system parameters $m$ and $\Lambda$ affect measurable angles.

Let us consider a static observer in the sub spacetime with metric \eqref{tp1e2} and constant coordinates ($r_{obs},\phi_{obs}$). Let the local frame of this observer be confined to this sub spacetime as well, that is $\theta=\frac{\pi}{2}$ and $\mathrm{d}\theta=0$. Since the direction of increasing proper time of the local frame of this observer coincides with the direction of increasing $t$, the space portion of the observer's frame coincides with a local patch around ($r_{obs},\phi_{obs}$) in the ($r,\phi$) surface with metric \eqref{tp1e3}. That is, the space, and curvature, around the static observer can be described by metric \eqref{tp1e3}, and can be visualized as a small patch on the isometrically embedded surface of Figure \ref{fig3}. This fact makes the special case of a static observer particularly useful in building understanding. However, outcomes of measurements generally depend on the motion of observers, and therefore a more detailed treatment is required for a complete description and establishment of practical relationships. As we progress to derive some general results, for arbitrary observers, we shall treat the case of a static observer at every step where observable angles are of interest. It will serve as a simple example of the physical phenomena at hand and as a specific case for others to be compared with.

Consider now two arbitrary curves on the flat ($r,\phi$) plane, intersecting at a point $p$. These curves may describe actual trajectories taking place in the curved physical space with metric \eqref{tp1e3}. The true (spatial) shape of these trajectories is fully determined only when projected from the flat plane onto the curved space, where the trajectories may physically exist. Consider a static observer at $p$ who makes a measurement of the intersection angle between the two curves. The situation is illustrated in Figure \ref{fig4} below. From the discussion in the previous paragraph, it is apparent that intersection angles measured by a static observer will be those on the embedded surface, which are sustained by the projected curves. Clearly, the Euclidean intersection angle, $\alpha_E$, appearing on the flat plane is different than the measurable intersection angle, $\alpha_M$, appearing on the embedded, curved, surface, see Figure \ref{fig4}. This is precisely the point we aim to make, and a fact that must be kept in mind when plotting curves that represent physical trajectories, on the flat plane.

\begin{figure}[!ht]
\includegraphics[width=85mm]{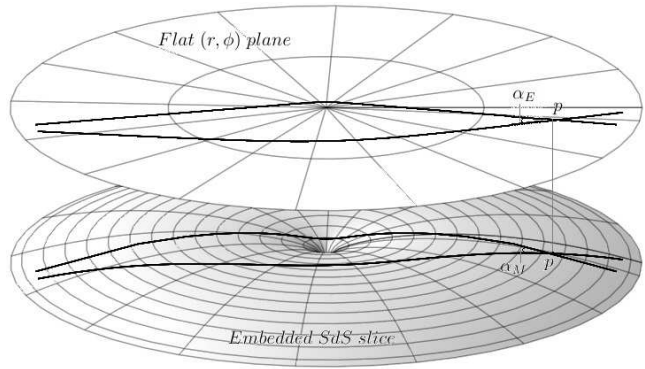}
\caption{Embedded SdS slice and a flat plane with intersecting trajectories. The intersection angle $\alpha_E$ is Euclidean and belongs to the flat plane. The intersection angle $\alpha_M$ takes place on the curved surface and measurable by a static observer.} \label{fig4}
\end{figure}

The difference between $\alpha_E$ and $\alpha_M$ comes only from the fact that the physical space is curved due to $m$ and $\Lambda$. It is already clear, qualitatively, that even if one of $m$ or $\Lambda$ were zero these angles would still not equal, and given the angle $\alpha_E$ one would need both $m$ and $\Lambda$ to find $\alpha_M$, and vice versa. Finally, we see that while the Euclidean angle, $\alpha_E$, depends only on the shape of the curves, the measurable angle, $\alpha_M$, depends on both the shape of the curves and the shape of the space itself, in which the true trajectories exist and intersect. The dependence of the shape of the space on the system parameters is clear and comes directly from the given metric. The dependence of the shape of curves on the parameters is determined in accordance to the particular situation being analyzed. Of course, the curves of central interest in the present work are the ones describing trajectories of light rays.

Let us restate the main conclusions of this section that are important to keep in mind in what follows. First, a clear distinction must be made between quantities that belong to the flat (Euclidean) plane on which diagrams are drawn, and quantities that are physically measurable. And second, to properly account for the various ways of influence when considering the dependence of measurable intersection angles between curves on the system parameters in general, one must consider both effects of the parameters on the curves and on the geometry of the space, where curves may physically exist and measurements may take place.

\section{Trajectories of light rays in SdS space and their dependence on $\Lambda$} \label{sec3}
Again we confine the motion to the plane $\theta=\frac{\pi}{2}$ without loss of generality, and use metric \eqref{tp1e2}. The two trivial Killing vectors ($\frac{\partial}{\partial t}$ and $\frac{\partial}{\partial \phi}$), along with the null condition, satisfied by trajectories of light, yield the following equations.
\begin{equation*}
\frac{d\phi}{d\lambda}=\frac{l}{r^2},\qquad \frac{dt}{d\lambda}=\frac{\gamma}{f(r)},
\end{equation*}
\begin{equation*}
-f(r)\left(\frac{dt}{d\lambda}\right)^2+\frac{\left(\frac{dr}{d\lambda}\right)^2}{f(r)}+r^2\left(\frac{d\phi}{d\lambda}\right)^2=0.
\end{equation*}
Here, $\lambda$ is an affine parameter, parametrizing the trajectory, and $l$ and $\gamma$ are constants of the motion. These equations can be combined to give the differential equation, satisfied by a curve in the ($r,\phi$) plane, describing the path of a light ray.
\begin{equation}
\left(\frac{dr}{d\phi}\right)^2=r^2\left[\left(\frac{1}{b^2}+\frac{\Lambda}{3}\right)r^2+\frac{2m}{r}-1\right], \label{tp2e1}
\end{equation}
where
\begin{equation*}
b=\frac{l}{\gamma}.
\end{equation*}
Solutions for this equation divide into a few categories and exhibit a number of interesting features. Although obtaining the exact solutions is not simple, they do exist in the literature, \cite{lake}, and can be used at any time to describe a path exactly or to test the validity of an approximation to any degree of accuracy. Fortunately, for realistic values of $m$ and $r$, the combination $\frac{m}{r}$ is very small, and approximations in the low orders of $\frac{m}{r}$ prove to be very accurate. Such approximations are most popular in the literature and textbooks on the subject, but it is comforting to know that exact solutions exist as well. The type of trajectories we shall mainly be interested in is the one for which there is an axis of symmetry along with other important features that we discuss in what follows. Such symmetric trajectories have a point of closest approach, with a minimum value of $r$, and extend to infinity (in the analytical sense, on the ($r,\phi$) plane) on both sides of the axis of symmetry. It can be shown that the value of $r$ for these trajectories does not go below $3m$. In regions where the value of $r$ is much larger than $m$, these trajectories exhibit asymptotic behaviour and can be described by straight lines, referred to as the asymptotes of the trajectory. The features listed here are well known and easy to establish analytically. We shall not cover all the mathematical details here but rather give an account of the key physical features and parameters that are important for what follows.

Concentrating on the symmetric trajectories with a point of closest approach, let the coordinates of this point be ($r_0,\phi_0$). At this point, the derivative $\frac{dr}{d\phi}$ is zero, and equation \eqref{tp2e1} gives
\begin{equation}
\frac{1}{b^2}+\frac{\Lambda}{3}=\frac{1}{r_0^2}-\frac{2m}{r_0^3}.  \label{tp2e1a}
\end{equation}
Let us also define a third parameter, $B$, as follows.
\begin{equation}
\frac{1}{B^2}=\frac{1}{r_0^2}-\frac{2m}{r_0^3}=\frac{1}{b^2}+\frac{\Lambda}{3}. \label{tp2e2}
\end{equation}
This allows us to write equation \eqref{tp2e1} in three ways using the three different parameters, $b$, $r_0$ and $B$. So in addition to \eqref{tp2e1} we also have, for convenience,
\begin{equation}
\left(\frac{dr}{d\phi}\right)^2=r^2\left[\left(\frac{1}{r_0^2}-\frac{2m}{r_0^3}\right)r^2+\frac{2m}{r}-1\right], \label{tp2e3}
\end{equation}
and
\begin{equation}
\left(\frac{dr}{d\phi}\right)^2=r^2\left[\frac{r^2}{B^2}+\frac{2m}{r}-1\right]. \label{tp2e4}
\end{equation}
Notice that only when the parameter $b$ is used in the governing differential equation does $\Lambda$ make an appearance. All three parameters will be discussed in considerable detail in the next section. Without any mathematical labour, we can assume that a required solution to equation \eqref{tp2e4} (as well as \eqref{tp2e3} and \eqref{tp2e1}) exists and can be written as follows, using either of the three parameters.
\begin{equation}
r=r(\phi,m,B,C), \label{tp2e5}
\end{equation}
\begin{equation}
r=r(\phi,m,\sqrt{\frac{1}{r_0^2}-\frac{2m}{r_0^3}},C)=r(\phi,m,r_0,C), \label{tp2e6}
\end{equation}
or
\begin{equation}
r=r(\phi,m,\sqrt{\frac{1}{b^2}+\frac{\Lambda}{3}},C)=r(\phi,m,\Lambda,b,C). \label{tp2e7}
\end{equation}
Here, $C$ is a constant of integration that is related to the orientation of the path. In each case there are two independent constants of motion to find in order to determine a specific trajectory in the subspace of interest, which is a particular set of points ($r,\phi$) through which the light ray passes. To this end, we must consider some boundary conditions. In what follows, four different sets of boundary conditions will be discussed in turn. We shall always assume that the value of $m$ is given in addition to any boundary conditions.

\subsubsection*{Set 1: Two known points through which the trajectory passes}
Let $p_1$ and $p_2$ be two points in space through which the light ray passes, with coordinates ($r_1,\phi_1$) and ($r_2,\phi_2$), respectively. Assume that the path of light connecting these points satisfies the conditions discussed above, i.e. point of closest approach, symmetry etc. Using the boundary conditions in \eqref{tp2e5} gives the following two equations with two unknowns.
\begin{equation*}
r_1= r(\phi_1,m,B,C), \qquad r_2=r(\phi_2,m,B,C).
\end{equation*}
It can be shown that the values of $B$ and $C$ are in general not unique for such boundary conditions; the possible values constitute a countable set, describing a family of curves connecting the two points. In this family each curve has a specific value of $r_0$, and there exists a unique trajectory with the largest value of $r_0$ connecting the two points. In practice, it is this trajectory which is usually of primary concern, and the one that is often approximated to various orders in $m$. Either way, it can be shown in general that for given two points in space connected by the path of light, a given mass $m$, and some additional restriction (which may be set by a requirement on the time-like interval or space-like distance of travel), it is possible to find unique values of $B(p_1,p_2,m)$ and $C(p_1,p_2,m)$, for which equation \eqref{tp2e5} will describe the required unique trajectory. Of course, an identical procedure can be followed by using equation \eqref{tp2e6} and the parameters $r_0$ and $C$ instead, leading to identical conclusions. Therefore, these considerations reveal that for such boundary conditions, the trajectory, which is a set of points on the ($r,\phi$) plane satisfying the governing equation, depends only on the mass $m$ and the two points in space $p_1$ and $p_2$ through which it passes; it is independent of $\Lambda$ in the simple sense that changing the value of $\Lambda$ will not alter the path satisfying these boundary conditions. In other words, with these boundary conditions the path of light in the subspace parametrized by $r$ and $\phi$ can be determined with or without knowledge of $\Lambda$.

\subsubsection*{Set 2: Known point of closest approach}
Let ($r_0,\phi_0$) be the coordinates of the point of closest approach of the trajectory. Since in this case $r_0$ is known from the start, we can use equation \eqref{tp2e3} as our first integral, for which all the parameters are known. Integrating this equation will give a solution of the form \eqref{tp2e6}, in which only the parameter $C$ remains to be determined from the boundary conditions. Plugging $r_0$ and $\phi_0$ in \eqref{tp2e6} gives an equation for $C$, which for a given choice of branch establishes a unique value of C($r_0,\phi_0,m$). That is, for these boundary conditions there is a unique path. The values of $B$ and $C$ are determined uniquely (up to a sign, which does not affect the shape of the path) from the values of $r_0$, $\phi_0$ and $m$. Again, we see that $\Lambda$ has no influence on the path in the same sense as for the previous set of boundary conditions. Varying the value of $\Lambda$ does not alter the path. A little investigation reveals that the parameter $B$, which depends only on $r_0$ and $m$, determines the overall shape of the path, while the parameter $C$ only determines the orientation (the direction of the axis of symmetry). Due to this fact and no loss of generality in setting orientation, it is often sufficient to use only the parameter $B$ to describe the path in many situations. These boundary conditions are particularly useful due to the uniqueness of the corresponding paths and the ability to find the parameter $B$ directly, without the need for integration or knowledge of $\Lambda$, from equation \eqref{tp2e2}.

\subsubsection*{Set 3: Known point on the path and direction of travel}
This set of boundary conditions can be considered as a generalization of the previous set. Let ($r_1,\phi_1$) be a known point on the path and $\alpha_E$ be the given Euclidean intersection angle on the flat diagram, sustained by the path of light under investigation and the radial path of light passing through ($r_1,\phi_1$). The situation is depicted in the following figure.

\begin{figure}[!ht]
\includegraphics[width=85mm]{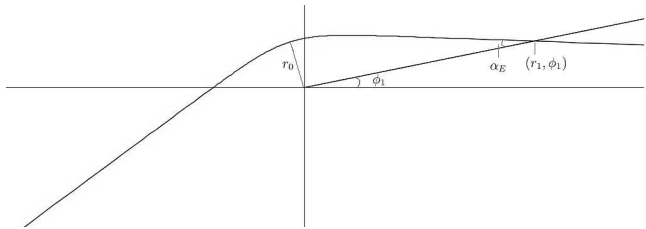}
\caption{A typical symmetric path of light on the flat $(r,\phi)$ plane, passing through a point with coordinates ($r_1,\phi_1$). The figure also shows the point of closest approach of this path, with $r=r_0$, and a radial path of light, which also passes through the point ($r_1,\phi_1$). The intersection angle between the two paths on this flat diagram is $\alpha_E$.} \label{fig5}
\end{figure}

In the flat, Euclidean, space of this diagram, the angle $\alpha_E$ is related to the differentials of the path at this point in the following way:
\begin{equation*}
\tan(\alpha_E)=r_1\left|\frac{d\phi}{dr}\right|_{(r_1,\phi_1)}.
\end{equation*}
This relationship can be easily formed by considering the local space around ($r_1,\phi_1$), and separating the radial and angular components of the tangent to the path. For simplicity let us drop the absolute value, and from now on assume that when there is a sign ambiguity it is the positive that is taken. The above can then be immediately rearranged to obtain $\frac{d\phi}{dr}$ as a function of $r_1$ and $\alpha_E$. Thus, boundary conditions which give a known point and a Euclidean intersection angle with a radial line at that point are equivalent to giving a known point and a derivative at that point. With these boundary conditions equations \eqref{tp2e3} and \eqref{tp2e4} can be used to find the parameters $r_0$ and $B$, either of which is sufficient to find the overall shape of the path, up to orientation. Upon integration, the parameter $C$ can be found as well by plugging the point ($r_1,\phi_1$) in the resulting relationship of the form \eqref{tp2e5} or \eqref{tp2e6}. Thus, with these boundary conditions the path is determined uniquely; the set of points ($r,\phi$) through which the light ray passes depends only on $m$, $r_1$, $\phi_1$ and $\alpha_E$. As in both previous cases, the trajectory does not depend on $\Lambda$. We see that this set of boundary conditions is in a sense equivalent to set 2, which may be considered as a special case. Whether it is set 1 that is initially given (with some condition to ensure uniqueness) or set 3, it may be convenient in each case to find the value of $r_0$ and classify the path according to this parameter, since its interpretation is intuitive and it is all that is needed for a complete description of the path, up to orientation. With this in mind, we shall always assume that a given trajectory of light, of the required type, may be uniquely described by a set of values $m$, $r_0$ and $\phi_0$, regardless of what Euclidean, or coordinate related, boundary conditions that are in the plane we initially start with.

\subsubsection*{Set 4: Known point on the path and a measurable intersection angle}
Let ($r_1,\phi_1$) be a known point on the path and $\alpha_M$ be the measurable intersection angle, at this point, between the trajectory of light under investigation and the radial trajectory of light, passing through ($r_1,\phi_1$), measured by an observer with 4-velocity $U$. This set of boundary conditions is different from the previous three sets in a fundamental way. It includes a directly measurable quantity as a boundary condition. Although the coordinates of the points $p_1$, $p_2$, ($r_0,\phi_0$), ($r_1,\phi_1$) and the derivative (or $\alpha_E$) of sets 1, 2 and 3 can, in principle, be determined through measurements, they are all Euclidean quantities that belong to the flat diagram. They may or may not have a physical interpretation as well, but their mathematical origin in the analysis has nothing to do with actual measurements. In contrast, the current set of boundary conditions includes a measurable angle, which may have a complicated relationship with the Euclidean quantities appearing on the plane that are needed to determine the path. Considering the discussion of the previous section and referring to Figure \ref{fig4}, we see that for the special case of a static observer, there can be constructed an intuitive relationship between the measurable angle $\alpha_M$ and the Euclidean angle $\alpha_E$. In this special case, which serves as a clear example, out of the parameters appearing in the relationship between $\alpha_M$ and $\alpha_E$ there will be both $m$ and $\Lambda$, since they both influence the geometry of the embedded space. In general, for an observer with arbitrary 4-velocity, $U$, the relationship between the angles will contain $m$, $\Lambda$, $r_1$, and the components of $U$ as parameters. Thus, to determine the path in the ($r,\phi$) plane with these boundary conditions one can find the Euclidean intersection angle, $\alpha_E$, from $\alpha_M$, $m$, $\Lambda$, $r_1$, and $U$, and use it along with the point ($r_1,\phi_1$) as in the case of set 3. Evidently, this set of boundary conditions is, in some sense, equivalent to set 3, both sets yield a unique path. With a given observer, for the current set, there is a one to one correspondence with the parameters of set 3, which can be used to convert from one set of boundary conditions to another. It is clear that the value of $\Lambda$ must be known in order to convert $\alpha_M$ of this set into $\alpha_E$ of set 3. In fact, without the knowledge of $\Lambda$ it is not possible to find the trajectory of light which satisfies the boundary conditions of the current set. Hence, with these boundary conditions the path is determined by $m$, $\Lambda$, $r_1$, $\phi_1$, $\alpha_M$, and $U$. We notice that $\Lambda$ does affect the path in this case, and, overall, it affects the path when certain (directly) measurable parameters are used as boundary conditions. It does not affect the path if all the boundary conditions are Euclidean or coordinate related, which appear on the flat diagram.\\

With the above examples in mind we see that, in contrast to the influence of $m$, the influence of $\Lambda$ on a path of light can come only from  uncommon boundary conditions that are usually associated with measurements. Since in most cases in the literature the boundary conditions are coordinate-like, or Euclidean, then in light of the above examples it may be loosely concluded that $\Lambda$ has no direct affect on the resulting paths. However, this common conclusion may be somewhat misleading if the assumptions on the boundary conditions are not stated explicitly. Indeed, it is important to keep in mind that no general conclusions should be made regarding the overall influence of $\Lambda$, which is sensitive to the particular situation being analyzed. As an additional example to set 4, which brings in $\Lambda$ through an observable quantity, consider a set of boundary conditions that contains two points on the path, one of them being the position of the source emitting the light ray; in the cosmological context this source could be a distant galaxy. Such a set is similar to set 1, it can be used in an identical way to establish the path of the ray, though it may have one important difference in regards to $\Lambda$. Given some astrophysical model, or tabulated data, which provides the position of the source, it could be the case that the position is a function of both time and $\Lambda$, and therefore, the appearance of $\Lambda$ once again will come from the boundary conditions but in a different way than it was for set 4. Thus, we stress that the influence of $\Lambda$ on a path of light and associated quantities of interest depends closely on the particular situation being analyzed, and in saying that a path is independent of $\Lambda$ one implicitly means that the path is subject to coordinate-like, or Euclidean, boundary conditions which do not depend on $\Lambda$ themselves.

Overall, it should be clear now in what way $\Lambda$ may influence a path of light, and how its influence is hidden in measurements, or rather, more generally, in boundary conditions. When analyzing a common setup, it may be straightforward to foresee whether $\Lambda$ will have an influence on results of interest or not. Let us consider a set of Euclidean boundary conditions, such as one of the first three sets discussed, and investigate the qualitative dependence of the resulting path of light in the ($r,\phi$) plane on the system parameters $m$ and $\Lambda$. As explained, it is convenient to convert any given set of Euclidean boundary conditions to the set of $r_0$, $\phi_0$, if it is not initially expressed as such. Further, without loss of generality, for illustration purposes we can orient the coordinates so that $\phi_0=\frac{\pi}{2}$. The following figures depict the dependence of the path on the parameters $m$ and $\Lambda$, for a set value of $r_0$.

\begin{figure}[!ht]
\includegraphics[width=85mm]{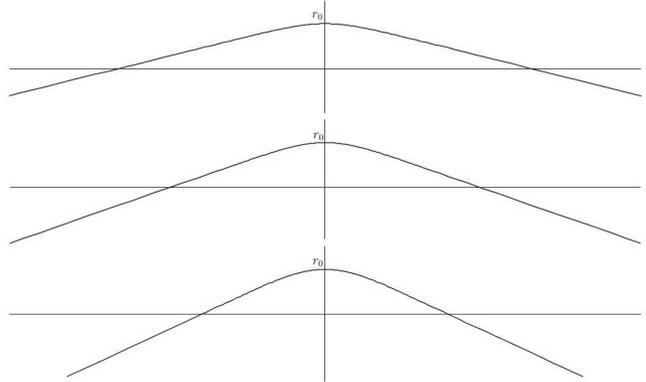}
\caption{A typical symmetric path of light on the flat $(r,\phi)$ plane, passing through a point with coordinates ($r_0,\frac{\pi}{2}$). The value of $m$ is successively increasing, starting from the top, and its influence is illustrated through the three diagrams. The value of $\Lambda$ is kept constant and it is assumed that the outer horizon is too far to be shown on the graphs.} \label{fig6}
\end{figure}

\begin{figure}[!ht]
\includegraphics[width=85mm]{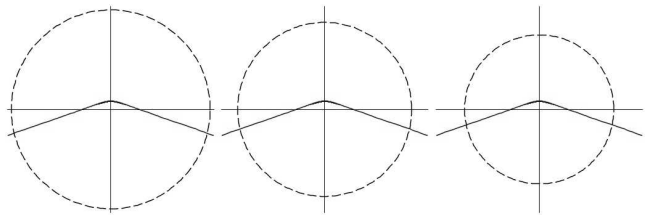}
\caption{A typical symmetric path of light on the flat $(r,\phi)$ plane, passing through a point with coordinates ($r_0,\frac{\pi}{2}$). While the value of $m$ is kept constant, the value of $\Lambda$ is successively increasing, starting from the left, and its lack of influence on the path is illustrated through the three diagrams. The outer horizon is also shown on the three diagrams as the dashed circle. Although the shape of the path does not change with varying $\Lambda$, the geometry of the underlying space as well as the location of the outer horizon both change.} \label{fig7}
\end{figure}

These figures make it clear that, in the region between the horizons, for typical Euclidean boundary conditions, only when varying $m$ the path of light changes. Varying $\Lambda$ only changes the location of the outer horizon on the diagram. But although the path itself may be independent of $\Lambda$, we shall make it abundantly clear that there is an influence of $\Lambda$ on measurements of intersection angles of light rays, and as one may expect this influence near the outer horizon may be quite significant.\\

The fact that, while both $m$ and $\Lambda$ appear in the metric, but only $m$ has an effect on paths of light in space deserves further attention. It is illuminating to study the paths in de Sitter space, for the case $m=0$ in equations \eqref{tp2e1}, \eqref{tp2e3}, \eqref{tp2e4}. The three equations are then
\begin{equation*}
\left(\frac{dr}{d\phi}\right)^2=r^2\left[\left(\frac{1}{b^2}+\frac{\Lambda}{3}\right)r^2-1\right],
\end{equation*}
\begin{equation*}
\left(\frac{dr}{d\phi}\right)^2=r^2\left[\frac{r^2}{B^2}-1\right],
\end{equation*}
and
\begin{equation*}
\left(\frac{dr}{d\phi}\right)^2=r^2\left[\frac{r^2}{r_0^2}-1\right].
\end{equation*}
We immediately recognize that for $m=0$ the paths are straight lines with a point of closest approach at $r=r_0$. Notice that $B=r_0$ in this case and, as before, either of these two parameters can be found from Euclidean boundary conditions without the need for $\Lambda$, and conveniently describe the entire path up to orientation. The parameter $b$, on the other hand, has no independent interpretation in this case; it is determined through its relation to $r_0$, and can only be found given knowledge of $\Lambda$. Thus, paths of light in de Sitter space are straight lines, and are independent of $\Lambda$ for given Euclidean boundary conditions. In other words, the set of points that lay on the path of a light ray in the ($r,\phi$) plane that connects two given points is independent of the value of $\Lambda$. Intuitively, in defining a bending angle for paths of light, the value of such an angle should be zero for a path which is a straight line. This is an intuitive and important requirement to keep in mind when considering bending angles in SdS space.

It is also interesting to further investigate the non influence of $\Lambda$ on paths of light from the following mathematical perspective. Evidently, the way in which $\Lambda$ appears in the first order differential equation, \eqref{tp2e1}, makes it 'entangled', in some sense, with the parameter $b$, allowing for the complete absorption of $\Lambda$ by transforming to a new parameter, for example $B$ or $r_0$. For the sake of curiosity, let us consider a more general coefficient of $\frac{\Lambda}{3}$ in the metric \eqref{tp1e1}, changing $r^2$ to $r^n$ in $f(r)$, for some $n$. Proceeding as before to obtain the first order equation of the path, we find
\begin{equation}
\left(\frac{dr}{d\phi}\right)^2=r^2\left[\left(\frac{r^2}{b^2}+\frac{\Lambda}{3}r^n\right)+\frac{2m}{r}-1\right]. \label{tp2e8}
\end{equation}
Again, restricting to symmetric trajectories with a point of closest approach, setting $\frac{dr}{d\phi}=0$ at $r_0$ gives
\begin{equation*}
\frac{1}{b^2}=\frac{1}{r_0^2}-\frac{2m}{r_0^3}-\frac{\Lambda}{3}r_0^{n-2},
\end{equation*}
which can be used to rewrite \eqref{tp2e8} in terms of the coordinate distance of closest approach, $r_0$:
\begin{align*}
\left(\frac{dr}{d\phi}\right)^2=r^2 \left[ \left(\frac{1}{r_0^2} \right. - \frac{2m}{r_0^3} - \frac{\Lambda}{3}r_0^{n-2} + \frac{\Lambda}{3}r^{n-2}\right)r^2&\\
+ \left. \frac{2m}{r}-1\right]&.
\end{align*}
The value of $r_0$ can be set by boundary conditions in a given setup, making the effect of $\Lambda$ on the path clear for a given value of $n$. Interestingly, only when $n=2$ does the effect of $\Lambda$ on the path vanish. Then $\Lambda$ completely disappears from the equation, leaving $r_0$ and $m$ the only parameters. It is this specific value of $n$ that happens to occur in the SdS (and de Sitter) metric, making it the only special case in which $\Lambda$ has no affect on paths of light in space. Thus, the power of 2 appearing in the $r$ coefficient of $\Lambda$ reveals much about its geometric characteristics and its apparent influence on paths of light.

\section{Discussion of parameters and additional definitions} \label{sec4}
\subsection{Constants of motion}
Going back to equations \eqref{tp2e1}, \eqref{tp2e3} and \eqref{tp2e4}, we wish to make a clear distinction between the three parameters $B$, $r_0$ and $b$, and gain clear mathematical and physical interpretations for each. As discussed, the parameter $r_0$ is particularly useful; it gives the shape of a unique path up to orientation. Given $r_0$, all the important features of a trajectory can be found without knowledge of $\Lambda$. Given any other complete set of Euclidean boundary conditions, $r_0$ can be found and used to describe the path on its own. An important question is whether $r_0$ is measurable. In principle, a static observer in a spherically symmetric, static spacetime can find its radial coordinate through measurements. For example, the measurable circumference of a stationary ring centred around the origin is $2\pi r$. By slowly moving around the circumference or setting an array of observers, the corresponding length can be found and $r$ can be determined. Similarly, by dividing the ring into sections, angular separations can be set. See \cite{rindler} (chapter 9) for remarkably clear and illuminating discussions related to such measurements. Thus, in principle, the coordinates of a given static point ($r,\phi$) in the space slice can be found through measurements by observers in that space. In particular, the coordinates of any point through which a given, fixed, light ray passes can be found by means of measurements, including ($r_0,\phi_0$). The method in this example may not be practical but it is meant to make a clear illustration of the fact that it is possible, in principle, to determine the value of $r_0$ through measurements without knowledge of $\Lambda$, or even $m$. Clearly, it is possible to convert $r_0$ to $B$ and vice versa, for values of $r_0>3m$, without the knowledge of $\Lambda$, see equation \eqref{tp2e2}. Therefore, as far as the mathematical description of the path is concerned, the two parameters are equivalent for $r_0>3m$. Since $B$ can be found from $r_0$, which can be found from measurements, we conclude that $B$ can be found, indirectly, from measurements as well, without the need for $\Lambda$. We shall see that $B$ happens to be the impact parameter, to be defined more precisely in what follows. Finally, given the parameter $B$, the shape of the path can be described up to orientation, without the need for $\Lambda$.

The remaining parameter to discuss is $b$, which is unfortunately the least useful and most popular of the three. It is immediately evident that given a fixed path, for which $r_0$ and $B$ can be determined, the value of the parameter $b$ can only be found with the knowledge of $\Lambda$ from equations \eqref{tp2e2}. Therefore, for a given value of $b$, one needs the value of $\Lambda$ to determine the shape of the path, up to orientation. Of course, in a situation where $b$ is given a priori, one may conclude that $\Lambda$ influences the shape of the path. However, $b$ should not be treated as a boundary condition, but rather as a parameter to be determined from boundary conditions, in the same way as $B$ and $C$ of equations \eqref{tp2e5}-\eqref{tp2e7}. Further, considering the relationship between $b$ and $B$ in equations \eqref{tp2e2} leads to the following question. Which of the two parameters is independent of $\Lambda$, if any, and which is dependent? At this point, the answer to this question is somewhat straightforward. For a path with typical Euclidean boundary conditions the value of $B$ can be determined independently of $\Lambda$. Therefore $B$ can be viewed as a parameter of the trajectory that is independent of $\Lambda$. In fact, $B$ can be used as a boundary condition since it is in one to one correspondence with $r_0$, for a given $m$ and $r_0>3m$. This leaves the parameter $b$ as the parameter that depends on the values of $B$ and $\Lambda$ in the relationship given by equations \eqref{tp2e2}. Thus, for a given trajectory, $b$ should never be treated as a parameter that is independent of $\Lambda$, especially when studying the effects of $\Lambda$. Technically, we could even throw Avogadro's number, say $N_A$, into the sum containing $b$ and $\Lambda$, that is: $\left(\frac{1}{b^2}+\frac{\Lambda}{3}\right) \rightarrow \left(\frac{1}{b^2}+\frac{\Lambda}{3}+N_A\right)$, and the situation would not change, since the boundary conditions will determine the value of the whole sum in the brackets. It is the value of $B$ (represented by this sum) that sets the shape of the path, while the value of $b$ shifts to compensate for $\Lambda$, or whatever else you throw at it, like Avogadro's number or any other imaginable constant. In other words, the boundary conditions will set the value in the brackets above, which is a constant of the path that does not depend on $\Lambda$, shifting the value of $\Lambda$ or adding anything new into the brackets will result in a shift of the value of $b$ so that the total value of the brackets remains the same. Although the physical interpretation of the parameter $b$ is not yet clear, these considerations clarify the mathematical role of $b$ in a typical situation. An important question now is whether it is theoretically possible to measure $b$ directly or, rather, find it from measurable quantities without knowledge of $\Lambda$. If possible, this could lead to a way of finding $\Lambda$ experimentally (by determining $b$ and $B$ independently), and allow for situations in which the parameter $b$ can be known a priori, which would force us to reconsider it as a possible boundary condition.

Let us investigate the above question in detail. At a given point in the ($r,\phi$) plane through which a ray of light passes, the possible measurements that can be made by an observer on the ray are the energy of the photons and the angle the ray makes with a given reference direction. Of course, for light consisting of a bundle of rays there may be more possible measurements to make, for example the size of the visible solid angle associated with the bundle. Such measurements we study in detail in \cite{ll}, but these are of no major consequence in the current discussion; more on this in the next section. In realistic situations, the deflecting mass is a luminous object, making radial light rays a good reference. As previously discussed, the coordinate parameters $r$ and $\phi$ can be found, in principle, through measurements independent of $\Lambda$. If we consider an extended frame around the observer, large enough to contain a sufficient amount of points through which the light passes to make accurate measurements, and if the coordinates ($r,\phi$) of each point are found as well, then the change in $r$ can be compared to the change in $\phi$ of this ray, making the derivative $\frac{dr}{d\phi}$ an indirectly measurable quantity. Also, if the proper time in the observer's frame is given by $\tau$, the changes in $r$ and $\phi$ can be compared to the change in time, making the quantities $\frac{dr}{d\tau}$ and $\frac{d\phi}{d\tau}$ indirectly measurable as well. With this in mind we proceed. For simplicity let us first consider the extended frame of a static observer (or, rather, multiple neighbouring static observers).

To be able to determine the value of $b$ through measurements, for a given light ray in the ($r,\phi$) plane, one must find a relationship between $b$ and directly measurable quantities. By definition, $b=\frac{l}{\gamma}$, where $l=r^2\frac{d\phi}{d\lambda}$ and $\gamma=f(r)\frac{dt}{d\lambda}$, for an affine parameter $\lambda$. Let $E$ be the measurable energy of the photons, and $\alpha_M$ be the measurable angle between the light ray under consideration and a radial light ray passing through this point. Let $\alpha_E$ be the Euclidean intersection angle, corresponding to $\alpha_M$, see Figures \ref{fig4} and \ref{fig5} for an illustration of the situation. Let $U$ be the 4-velocity of the observer and $K$ be the 4-momentum of the ray of light under investigation. With the proper time $\tau$, and an appropriate choice of $\lambda$, $U$ and $K$ can be expressed as
\begin{eqnarray*}
U^\alpha=(U^t,U^r,U^\phi,U^\theta)=\left( \frac{dt_{(U)}}{d\tau},\frac{dr_{(U)}}{d\tau},\frac{d\phi_{(U)}}{d\tau},\frac{d\theta_{(U)}}{d\tau} \right),\\
K^\alpha=(K^t,K^r,K^\phi,K^\theta)=\left( \frac{dt_{(K)}}{d\lambda},\frac{dr_{(K)}}{d\lambda},\frac{d\phi_{(K)}}{d\lambda},\frac{d\theta_{(K)}}{d\lambda} \right).
\end{eqnarray*}
The subscripts $U$ and $K$ in the coordinates above are introduced for clarity, and will be dropped when there is no room for ambiguity; clearly we are free to set $dt_{(U)}=dt_{(K)}$. With $g_{\alpha \beta}$ the metric tensor, the measurable energy, $E$, can then be expressed in terms of the inner product
\begin{equation}
E=-g_{\alpha \beta}U^\alpha K^\beta. \label{energy}
\end{equation}
For the case of a static observer then, where $U^r=U^\phi=U^\theta=0$, it is trivial to find $U^t=\frac{1}{\sqrt{f(r)}}$ from the required condition $U \cdot U=-1$. Thus, we have
\begin{align}
\nonumber E&=-g_{\alpha \beta}U^\alpha K^\beta=f(r)U^t K^t\\
&=f(r) \left(\frac{1}{\sqrt{f(r)}}\right) \left(\frac{\gamma}{f(r)}\right)=\frac{\gamma}{\sqrt{f(r)}}.
\end{align}
\begin{equation}
\Rightarrow \; \gamma=E\sqrt{f(r)}.
\end{equation}
Here, $E$ is a measurable quantity, by definition; the constant of motion $\gamma$ can be determined, from the measurement of $E$, only if both $m$ and $\Lambda$ are known. Further,
\begin{align}
\nonumber l &=r^2\frac{d\phi_{(K)}}{d\lambda}=r^2 \frac{d\phi_{(K)}}{d\tau} \frac{d\tau}{dt_{(U)}} \frac{dt_{(K)}}{d\lambda}\\
&=r^2 \frac{d\phi_{(K)}}{d\tau} \sqrt{f(r)} \frac{\gamma}{f(r)}=r^2 \frac{d\phi_{(K)}}{d\tau} E.
\end{align}
Hence, the constant of motion $l$ can be expressed entirely in terms of measurable quantities (in this case, measurable by a static observer), and can be determined without knowledge of $m$ or $\Lambda$. With the above relationships the parameter $b$ can be expressed as follows:
\begin{equation}
b=\frac{l}{\gamma}=r^2 \left(\frac{d\phi}{d\tau}\right) \frac{1}{\sqrt{f(r)}}. \label{tp3e1}
\end{equation}
And again, we see that this equation cannot be used to determine $b$ from measurable quantities without a prior knowledge of the value of $\Lambda$, and in this case $m$ as well. Since the derivative $\frac{dr}{d\phi}$, as well as $\frac{d\phi}{d\tau}$, can be found at the point of intersection, as discussed, it is possible to determine $\alpha_E$ through
\begin{equation}
\tan(\alpha_E)=r\frac{d\phi}{dr}. \label{tp3e2}
\end{equation}
And since for the stationary observer, as for any other, the ray moves at the speed of light, set to unity in our coordinates, we have
\begin{align*}
1&=\frac{1}{f(r)}\left(\frac{dr}{d\tau}\right)^2+r^2\left(\frac{d\phi}{d\tau}\right)^2\\
&=\left(\frac{1}{f(r)}\left(\frac{dr}{d\phi}\right)^2+r^2\right)\left(\frac{d\phi}{d\tau}\right)^2.
\end{align*}
Using \eqref{tp3e2} in the above gives
\begin{equation}
\left(\frac{d\phi}{d\tau}\right)^2=\left(\frac{r^2}{f(r)\tan^2(\alpha_E)}+r^2\right)^{-1},
\end{equation}
which can be used in \eqref{tp3e1} to re-express $b$ in terms of $\alpha_E$, then
\begin{equation}
b=\frac{r\tan(\alpha_E)}{\sqrt{1+f(r)\tan^2(\alpha_E)}}.  \label{tp3e3}
\end{equation}
Thus, even if the angle $\alpha_E$ can be determined through measurements in an extended frame, one still needs the values of $m$ and $\Lambda$ to calculate $b$. Consider now the measurable angle $\alpha_M$ at the point of intersection, measured by a static observer, which is obviously different from the Euclidean angle $\alpha_E$, as discussed. The relationship between these angles, derived in the next section, turns out to be
\begin{equation}
\tan(\alpha_M)=\sqrt{f(r)}\tan(\alpha_E). \label{tp3e4}
\end{equation}
In contrast to $\alpha_E$, the angle $\alpha_M$ can be determined through a direct measurement at a single point by a single observer. To determine $\alpha_E$ an extended frame is needed, which for theoretical reasons is important to consider but may not be practical. Equation \eqref{tp3e4} can be used to replace $\alpha_E$ by the measurable angle $\alpha_M$ in the last expression of $b$, \eqref{tp3e3}, giving
\begin{equation}
b=\frac{r\tan(\alpha_M)}{\sqrt{f(r)}\sqrt{1+\tan^2(\alpha_M)}}=\frac{r}{\sqrt{f(r)}}\sin(\alpha_M).
\end{equation}
The above relationship is of simple form and allows finding $b$ from the measurable intersection angle $\alpha_M$. In fact, this equation can be used to recover the relationship between $b$ and $r_0$, \eqref{tp2e2}, by setting $r=r_0$ at $\alpha_M=\frac{\pi}{2}$, and may be of use in certain applications. However, once again we see that without the knowledge of $\Lambda$ (and $m$) the value of $b$ cannot be established. In summary, we found that out of the three, related, constants of motion $\gamma$, $l$ and $b$, it is only the value of $l$ that can be established without prior knowledge of $\Lambda$ form the possible measurements discussed here. In particular, the value of $b$ cannot be found without prior knowledge of $\Lambda$ from such measurements. These conclusions remain true when considering measurements done by any observer. The special case of a static observer was considered here only for a simple illustration of the situation. Thus, the answer to the previous question concerning the determination of $b$ is in the negative. The value of $b$ cannot be established without knowledge of $\Lambda$, $b$ cannot be used as a realistic boundary condition, and finally, due to its dependence on $\Lambda$, its use can be misleading when investigating the influence of $\Lambda$ on other quantities.

We notice that, since it is theoretically possible to measure $\alpha_M$ and determine $\alpha_E$ from measurements, equation \eqref{tp3e4} can be used to express $\Lambda$ in terms of measurable quantities. Hence, this suggests one theoretically possible, although maybe not practical, method to find $\Lambda$ experimentally. This method of finding $\Lambda$ is somewhat equivalent to determining the parameter distance and the measurable distance between two points, and using a relationship between the two quantities, similar to equation \eqref{tp3e4}, to establish the value of $\Lambda$. These effects are a result of the curvature induced by $\Lambda$, and can be viewed as the effect of $\Lambda$ on the embedded surface of Figure \ref{fig3}. $\Lambda$ affects the relationships between measurable quantities and corresponding Euclidean (or coordinate) quantities. A visual example of the influence of $\Lambda$ on such relationships can be seen in Figure \ref{fig4}, which is a particularly good illustration when considering measurements made by static observers. We state again, the possible measurements discussed in this section are for theoretical purposes only, whether or not they are practical is of no concern.

The main goal of this section is to interpret and discuss the parameters $r_0$, $B$ and $b$, and determine which of these can be found through measurements without knowledge of $\Lambda$. We have established the mathematical roles of all three, and found that for a given path of light only $b$ depends on $\Lambda$; its value cannot be determined without it. While the geometrical interpretation of $r_0$ is clear from its definition, the geometrical interpretation of $B$ requires a little more analysis, to be done shortly, which will reveal that $B$ is the impact parameter of the trajectory. As for the parameter $b$, there is no clear geometrical interpretation in the general case of $m \neq 0$ and $ \Lambda\neq 0$. In the special case where $\Lambda=0$, $b$ is the impact parameter, since $b=B$. But even when $m=0$ and $\Lambda \neq 0$, $b$ loses its geometrical meaning and gains dependence on $\Lambda$. Thus, in Schwarzschild space, the usefulness of $b$ comes only from the fact that $b=B$. In SdS space, the parameter $b$ loses its worth.

\subsection{Supplementary definitions} \label{sec4b}
When discussing some of the recent papers on the topic, we shall have clear definitions of the important quantities in mind. Much of the disagreement in the literature seems to emerge from misunderstanding conclusions due to lack of clarity and ambiguity. In many cases, parameters that are defined and often used in analyzing trajectories of light in Schwarzschild space are imported to the analysis in SdS space without mentioning their exact definitions or discussing if they remain appropriate to use. Furthermore, even in cases where these imported parameters do remain appropriate to use in SdS space, their interpretations may change considerably, which should be noted to avoid confusion. For the sake of clarity, we present a few definitions in what follows. Although the manner in which $\Lambda$ influences measurements while not having an influence on paths of light should be clear by now from the previous sections, the definitions presented in this section are meant to clarify some of the terminology in the current literature on the topic. The parameters discussed may or may not be of much practical or theoretical use, however, they encompass some of the popular quantities used in the literature and can aid in making it simple and systematic to understand the results and conclusions of some recent papers.

\subsubsection{Impact parameter}
In the general context, the impact parameter is defined for a trajectory in a radially dependent potential field, whose first derivative vanishes at large values of the radial coordinate, as the perpendicular distance between an asymptote of the trajectory and the origin. In such a potential field, trajectories that go to infinity can be approximated by straight lines at large radial coordinate, $r$, and for our purposes we also assume that these trajectories have a point of closest approach to the origin with a minimum value of $r$. See the following figure.

\begin{figure}[!ht]
\includegraphics[width=85mm]{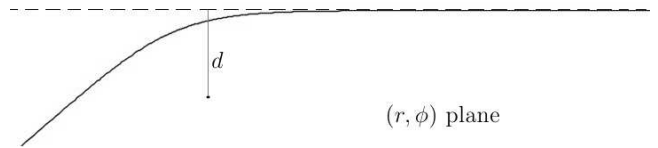}
\caption{A trajectory on the flat ($r,\phi$) plane under the influence of a radially dependent potential field. The solid curve represents the trajectory of interest, the dashed line represents one of its asymptotes. The impact parameter is the distance $d$ appearing on the diagram.} \label{ip1}
\end{figure}

In the context of general relativity, specifically for trajectories of light in Schwarzschild space, a second definition, equivalent to the first, is used in many books. In this context, the impact parameter is defined as the perpendicular distance between the path and the radial line, that is parallel to an asymptote of the path, at large values of $r$. More exactly, it is the limit that this distance approaches as $r$ goes to infinity. The next figure will make this definition clear.

\begin{figure}[!ht]
\includegraphics[width=85mm]{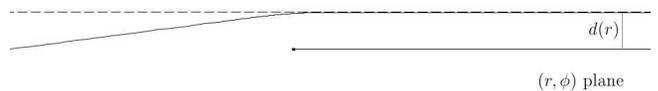}
\caption{A symmetric path of light on the flat ($r,\phi$) plane with a point of closest approach. The solid curve represents the path of interest, the dashed line represents one of its asymptotes, and the solid line is a radial line parallel to the asymptote. The impact parameter equals the limit of the distance $d(r)$, appearing on the diagram, as $r \rightarrow \infty$.} \label{ip2}
\end{figure}

We see that this second definition suggests an experimental method to find the impact parameter for a given, fixed, path of light. For example, in Schwarzschild space, which is asymptotically flat, radial lines can, in principle, be identified, and the required distance corresponding to the impact parameter of Figure \ref{ip2} can, theoretically, be measured directly by static observers. Thus, in addition to the fact that the impact parameter can be calculated from some boundary conditions, in Schwarzschild space it can also be found from direct measurements as well. When extending the concept of the impact parameter to trajectories of light in SdS space, which are mathematically the same as in Schwarzschild space, both of the two common definitions remain valid. However, in this case, the second definition no longer suggests a method to measure the impact parameter directly, as it does for Schwarzschild space, since the space is no longer asymptotically flat. Thus, the impact parameter of trajectories of light in SdS space can still be calculated from some boundary conditions, which determine the trajectory, but can no longer be measured directly. The impact parameter can be found analytically as follows. First, let us refer to Figure \ref{ip2} and orient the angular coordinate, $\phi$, so that the radial line will corresponds to $\phi=0$. Far from the origin, the perpendicular coordinate distance between a point on the path under investigation and the radial line is $d(r)=r \sin(\phi)$, where the coordinates $r$ and $\phi$ are of a point on the path (with $\phi<\frac{\pi}{2}$). Therefore, the impact parameter is the limit of $r \sin(\phi)$ as $r$ approaches infinity. This limit can be easily found with the aid of equations \eqref{tp3e2} and \eqref{tp2e4}. $\alpha_E$ of equation \eqref{tp3e2} is the Euclidean intersection angle at a point on the path under investigation sustained by the path and the radial line through this point. $B$ of equation \eqref{tp2e4} is assumed to be a fixed parameter for this particular path.
\begin{align*}
\text{Impact parameter} &= \lim_{r \to \infty} d(r)\\
&= \lim_{r \to \infty} \left(r\sin(\phi)\right)\\
&= \lim_{r \to \infty} \left(r \tan(\phi)\right) \lim_{r \to \infty} \cos(\phi)\\
&= \lim_{r \to \infty} \left(r \tan(\alpha_E)\right)\\
&= \lim_{r \to \infty} \left(r^2 \frac{d\phi}{dr}\right)\\
&= \lim_{r \to \infty} \frac{r}{\sqrt{\frac{r^2}{B^2}+\frac{2m}{r}-1}}\\
&= \lim_{r \to \infty} \frac{B}{\sqrt{1+\frac{2mB^2}{r^3}-\frac{B^2}{r^2}}}\\
&= B.
\end{align*}
And so we find that the impact parameter of a given fixed trajectory is the constant of motion $B$. This gives us the geometrical significance of $B$, but again, since SdS space is not asymptotically flat, the value of $B$ cannot be measured directly, though it could easily be found analytically from boundary conditions. The facts to keep in mind when bringing up the concept of the impact parameter in the context of trajectories of light in SdS space are the following: in the special case of $\Lambda=0$, the space is asymptotically flat and we have $b=B$, so not only does $b$ become the impact parameter, but also the impact parameter becomes directly measurable at large distances. However, these two features do not remain true for $\Lambda \neq 0$. In general, the impact parameter, $B$, always maintains its mathematical role and geometrical meaning for any value of $\Lambda$, but can not always be interpreted as a physical distance. The parameter $b$, on the other hand, loses its mathematical and geometrical meanings when a non-zero $\Lambda$ is introduced. Overall, the impact parameter is a Euclidean quantity that belongs to diagrams on the flat ($r,\phi$) plane, it only gains a physical (measurable) significance in a special situation.\\

The following two figures will be referred to in the subsequent definitions. They depict a typical path of the kind we are interested in, with a few important features.

\begin{figure}[!ht]
\includegraphics[width=77mm]{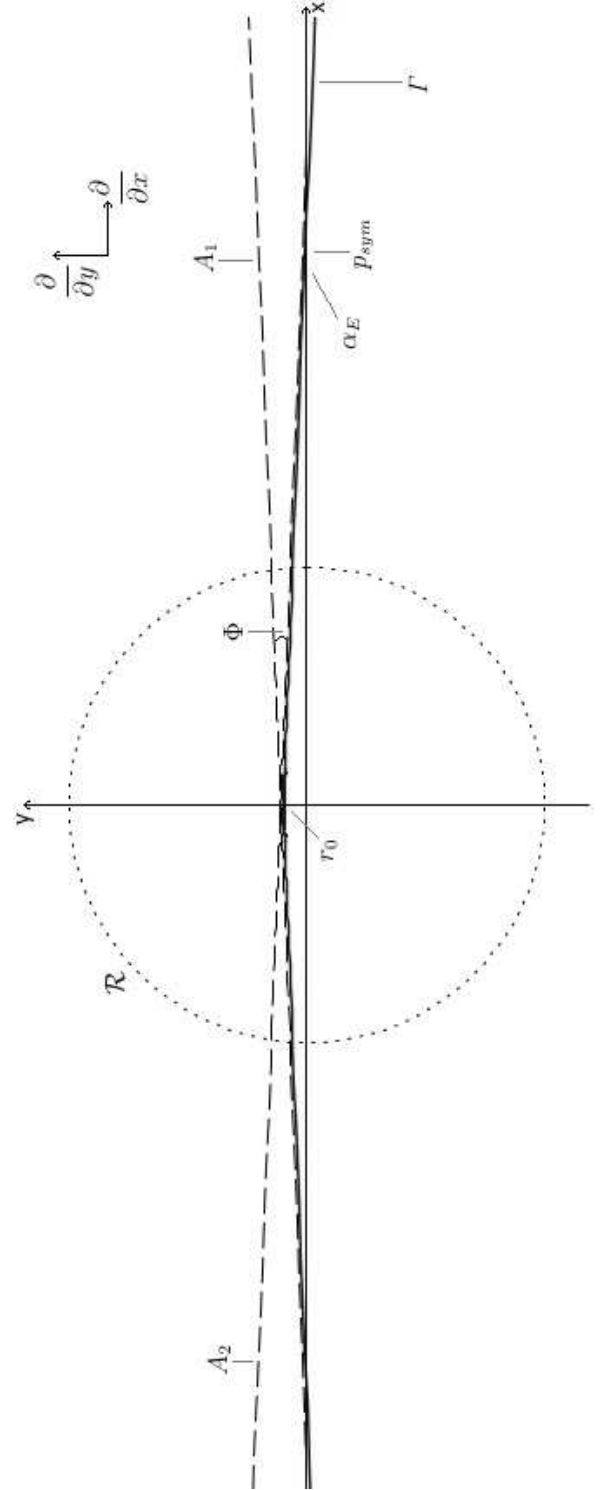}
\caption{A curve on the flat ($r,\phi$) plane, representing a typical symmetric path of light with a point of closest approach at ($r_0,\frac{\pi}{2}$). The features and parameters appearing on this figure are defined and discussed below.} \label{split1}
\end{figure}

\begin{figure}
\includegraphics[width=85mm]{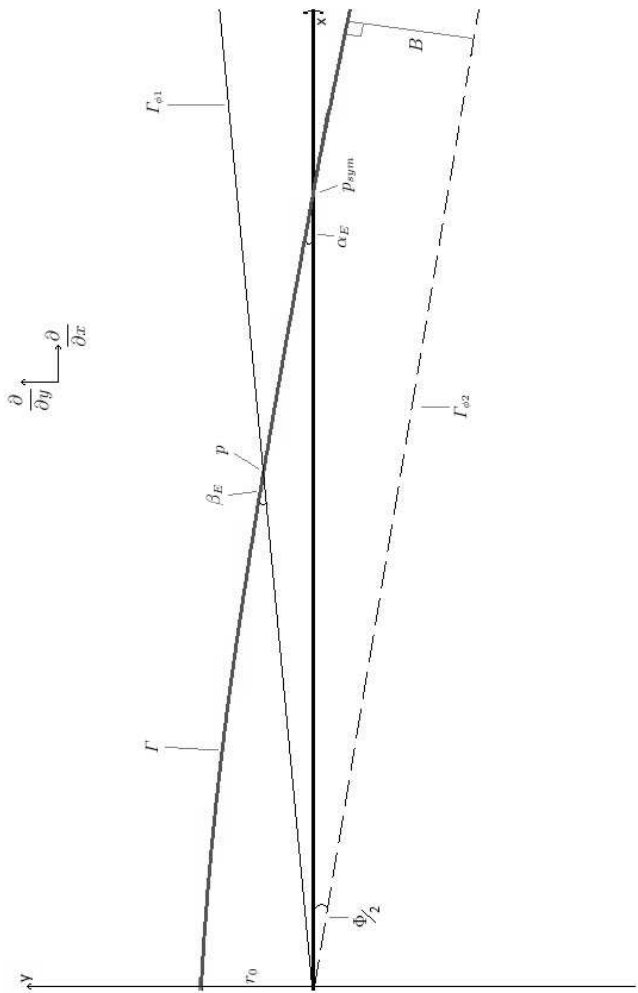}
\caption{The above is the right half of the previous figure, with a few additions. It is stretched in the $y$ direction for clarity. The features and parameters appearing on this figure are defined and discussed below.}  \label{split2}
\end{figure}

\bigskip

In these diagrams, the path of light under investigation is the curve represented by $\varGamma$. For the chosen orientation, the shape of $\varGamma$ is entirely determined by $r_0$, or equivalently $B$, which both appear on the diagrams. The straight (dashed) lines $A_1$ and $A_2$ are the asymptotes of the path, which approximate the path well at sufficiently large values of $r$. The (dotted) circle $\mathcal R$ represents a region outside of which the effects of $m$ are negligible on, both, paths of light and curvature of space. It is outside of this region that $r$ is considered to be sufficiently large, where the path is straight and Euclidean quantities are not distorted by $m$. Of course, the position of $\mathcal R$ will ultimately depend on the sensitivity of instruments and the desired accuracy. However, it is usually assumed that the intersection of $\varGamma$ with the $x$ axis (in the diagrams) occurs well beyond this circle. The Cartesian coordinates ($x,y$) are related to the polar coordinates ($r,\phi$) in the usual way, $x=r \cos(\phi)$ and $y=r \sin(\phi)$. This makes the vectors $\frac{\partial}{\partial x}$ and $\frac{\partial}{\partial y}$ well defined at every point on the plane. In the orientation of these diagrams, $\varGamma$ is symmetric about the $y$ axis, and the point $p_{sym}$ on $\varGamma$ is symmetric about the origin as well. Let us refer to this point as the point of symmetry, which in this case is the point of intersection of $\varGamma$ with the $x$ axis. At this point, the Euclidean intersection angle appearing on the diagrams between $\varGamma$ and the $x$ axis is $\alpha_E$. This angle (when very small) is approximately half the magnitude of the angle between $A_1$ and $A_2$, the asymptotes of the path, which is given by $\Phi$. The curves $\varGamma_{\phi 1}$ and $\varGamma_{\phi 2}$ represent radial rays of light, which are straight lines, with constant angular coordinate $\phi=\phi_1=\phi_p$ and $\phi=\phi_2=\frac{\Phi}{2}$, respectively. The purpose of $\varGamma_{\phi 2}$ is for the illustration of the impact parameter, $B$, while the purpose of $\varGamma_{\phi 1}$ is to serve as a reference direction at a point $p$ on the path. Although the de Sitter horizon is assumed to be outside the range of these diagrams and $\Lambda$ has no affect on the illustrated path, the possible influence of $\Lambda$ on measurements through the curvature of space should not be neglected. Let $\alpha_M$ be the measurable intersection angle by a static observer at $p_{sym}$ corresponding to the Euclidean angle $\alpha_E$.

\subsubsection{Bending angle}
The bending angle is originally defined for paths of light in Schwarzschild space and is also referred to as the total bending angle, the deflection angle, and the total deflection angle by some authors. In certain cases definitions differ by a factor of 2, and the word ``total" is used to make the distinction for clarity. Extending this concept to paths of light in SdS space can give rise to some ambiguity and confusion, so we shall do it carefully. Since the curve $\varGamma$ (and its associated Euclidean quantities) in the above figures does not depend on $\Lambda$, as should be presently clear, such curves may be used in modelling paths of light in either Schwarzschild or SdS space. In the context of Schwarzschild space, the bending angle is usually defined, in most textbooks, in one of the following two equivalent ways.\\
\\
\textbf{Definition 1:} The bending angle of a symmetric path of light in Schwarzschild space is the (small) angle between the two asymptotes of the path.\\
\\
In reference to Figures \ref{split1} and \ref{split2}, the bending angle is the Euclidean angle $\Phi$, between $A_1$ and $A_2$. This definition is purely mathematical in the sense that there is no reference to any measurements. The definition suggests that the bending angle can be found by determining the path from some boundary conditions and finding the bending angle through its asymptotic behaviour.\\
\\
\textbf{Definition 2:} The bending angle of a symmetric path of light in Schwarzschild space is double the (small) measurable intersection angle by a static observer between the path and a radial ray at the point of symmetry, far from the origin.\\
\\
According to this definition, referring to Figures \ref{split1} and \ref{split2}, the bending angle is double the measurable intersection angle $\alpha_M$, which corresponds to the Euclidean angle $\alpha_E$. The assumption made in the figure that the point of symmetry, $p_{sym}$,  is outside the circle $\mathcal R$, where the affects of $m$ are negligible, is what's meant by being far from the origin in the definition. Hence, in the asymptotically flat Schwarzschild space, the measurable angle by a static observer at the point of symmetry $\alpha_M$ is the same as the Euclidean angle $\alpha_E$ appearing on the flat diagram. It is also clear that $\alpha_E \approx \frac{\Phi}{2}$, since the path is already exhibiting its asymptotic behaviour at $p_{sym}$. Therefore we see that, in the context of Schwarzschild space, the two definitions are equivalent.

The second definition suggests that the bending angle is a quantity that can be directly measured. Similar to the impact parameter, in Schwarzschild space, the bending angle can be found from some boundary conditions that determine the path as well as measured directly at a distant point. However, in contrast to the case of the impact parameter, when extending the concept of bending angle to SdS space the two common definitions of the parameter given here are no longer equivalent. Since $\Lambda$ will affect the geometry at $p_{sym}$, the measurable intersection angle, $\alpha_M$, will be different than the Euclidean angle, $\alpha_E$.

In extending the concept of the bending angle to SdS we shall build on both of the above definitions and define two kinds of angular quantities, purely mathematical and measurable, concerned with symmetric paths of light. First, by restricting to definition 1 of the bending angle in Schwarzschild space, let us explicitly state what will be referred to as the bending angle of a symmetric path of light in SdS space.\\
\\
\textbf{Definition:} The bending angle of a symmetric path of light in SdS space is the (small) angle between the two asymptotes of the path.\\
\\
Although measurements by observers are important to consider, the bending angle is a measure of how much the entire path is bent, and should be independent of observers. For this reason we extend the concept of the bending angle to SdS space in accordance with definition 1 (of Schwarzschild space) and reserve definition 2 for a different quantity that is measurable. In reference to Figures \ref{split1} and \ref{split2}, according to the above definition, the bending angle is $\Phi$. With this definition for the bending angle in SdS we see again a similarity with the case of the impact parameter. The bending angle can be found from some boundary conditions that determine the path, and therefore can be determined from measurable quantities, but can no longer be measured directly. In particular, the bending angle can be found by taking the limit as $r$ goes to infinity in the solution for the orientation in Figure \ref{split1}, and since the path does not depend on $\Lambda$ the bending angle does not depend on $\Lambda$ either. It is clear from the symmetry that the bending angle should only depend on $m$ and $r_0$, and since these parameters only appear as the combination $\frac{m}{r_0}$ in the analysis, the bending angle will be a function of $\frac{m}{r_0}$. It is easily found that for a small bending angle, $\Phi$, to first order in $\frac{m}{r_0}$, we have
\begin{equation}
\Phi=\dfrac{4m}{r_0}.
\end{equation}
Also, to this order in $\frac{m}{r_0}$, equation \eqref{tp2e2} gives
\begin{equation}
\frac{m}{B}=\frac{m}{r_0}.
\end{equation}
Therefore,
\begin{equation}
\Phi=\dfrac{4m}{r_0}=\dfrac{4m}{B}.
\end{equation}
Equation \eqref{tp2e2} can also be used to replace $B$ in the above equation and express $\Phi$ in terms of $m$, $b$ and $\Lambda$. But, given the discussion of the parameter $b$ in this section, we see that this relationship will be of little use and, in a way, misleading. Finally, it is important to keep in mind that, in the case of SdS space, the bending angle should be interpreted only as a Euclidean quantity, which belongs to the flat ($r,\phi$) plane. Since paths of light are independent of $\Lambda$, extending the bending angle to SdS space in such a way does not affect its mathematical interpretation. Now, however, only in the special case of $\Lambda=0$ the bending angle gains a physical significance as well by becoming equivalent to a measurable quantity.

\subsubsection{Measurable deflection angle at the point of symmetry by a static observer}
In light of the definition 2 of the bending angle in Schwarzschild space, we define a similar angular quantity for a path of light in SdS space, which refers to an actual measurement. In reference to Figures \ref{split1} and \ref{split2} and the paragraph following it, let the \textit{measurable deflection angle at the point of symmetry by a static observer} be defined as the angle $\alpha_M$, which corresponds to the Euclidean angle $\alpha_E$. For concreteness, rather than taking $\alpha_M$ to be the measurable intersection between the path of light $\varGamma$ and the $x$ axis, which leaves room for ambiguity, we can define it to be the measurable intersection angle between the path of light $\varGamma$ and the radial light ray going through $p_{sym}$. Notice that for this definition, of a measurable angular quantity, we only consider the one sided intersection angle (in contrast to the double of definition 2 above), since it is the measurement that is significant here rather than the overall shape of the path. To further distinguish this measurable, one sided, angle from the Euclidean bending angle, we refer to it as a measurable \textit{deflection} angle. The way in which the measurable angle, $\alpha_M$, is related to its corresponding Euclidean angle, $\alpha_E$, is illustrated in Figure \ref{fig4}; $\alpha_M$ is the projection of $\alpha_E$ onto the embedded surface discussed in section \ref{sec2}. The angle $\alpha_M$ is physically measurable by using the radial ray at $p_{sym}$ as a reference, which in the Euclidean sense is parallel to the direction of the path at $r_0$, and for this reason it is a measure of the deflection of the path as it goes from ($r_0,\phi_0$) to $p_{sym}$. If the mass at the centre of coordinates is luminous, as is usually the case in practice, then radial reference rays are available at all points to all observers. Since the observer and the point of measurement are set in the definition, the measurable angle $\alpha_M$ can be considered as a function of $r_0$ only, in addition to $m$ and $\Lambda$, of course. Clearly, for a fixed path, this measurable deflection angle will depend on $\Lambda$, in the simple sense that changing $\Lambda$ while keeping the boundary conditions will alter the measurement. By means of equation \eqref{tp3e4}, which will be derived in the next section, we can explicitly write the relationship between the measurable angle $\alpha_M$ and the corresponding Euclidean angle $\alpha_E$.
\begin{equation*}
\tan(\alpha_M)=\sqrt{f(r_{p_{sym}})}\tan(\alpha_E).
\end{equation*}
Since $\tan(\alpha_E)=r_{p_{sym}} \frac{d\phi}{dr}|_{p_{sym}}$, for convenience $\alpha_M$ can also be expressed in terms of $r_0$ and $r_{p_{sym}}$, or $B$ and $r_{p_{sym}}$. And of course, $r_{p_{sym}}$ can be determined from $r_0$ and $m$ regardless of orientation.\\

With the last two definitions of the bending angle and the measurable deflection angle at the point of symmetry, we have sufficiently extended the usual concept of the bending angle for paths of light in Schwarzschild space to SdS space. It is now important to mention that in doing so, it is intuitive to expect, or rather require, the following conditions. First, for $\Lambda=0$ any defined angular quantity, which is a measure of the deflection of the path, should reduce to the usual bending angle of Schwarzschild space, so that it can be interpreted as a proper generalization. And second, for $m=0$, in which case the path is a straight line on the ($r,\phi$) plane, the defined angular quantity should equal zero. It is easily verified that the definitions we made above meet these two conditions. For $\Lambda=0$ the new definitions simply reduce to the original definitions 1 and 2. For $m=0$ we must deal with a limit and some assumptions on $\Lambda$ to show that the condition is met for $\alpha_M$ (or rather use the generalization of this angle, $\beta_M$ below, for a more intuitive approach).

Next, we generalize the last definition of the measurable deflection angle, $\alpha_M$, to an arbitrary point on the path and an arbitrary observer in the following two definitions. The following figure is a magnification of the area around the point $p$ on Figure \ref{split2}. It represents a small patch on the ($r,\phi$) surface containing $p$.

\begin{figure}[!ht]
\includegraphics[width=85mm]{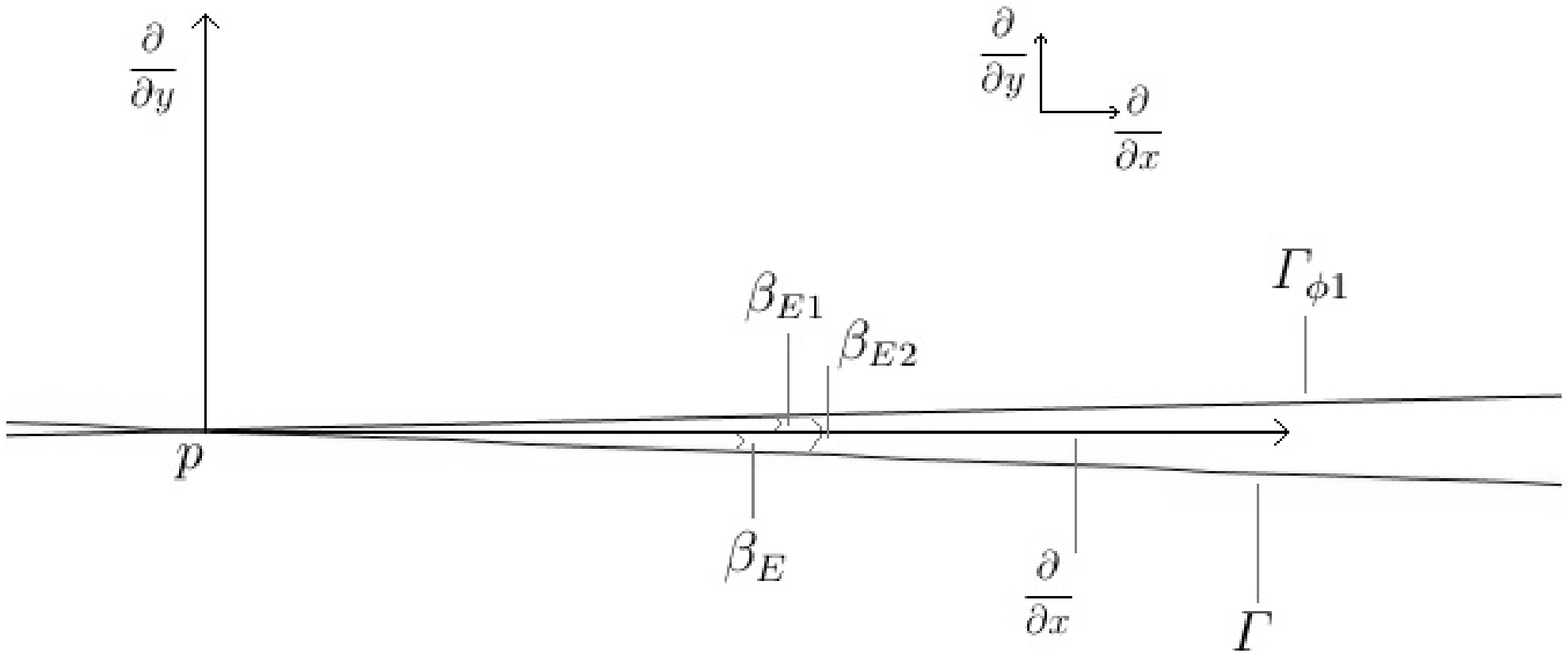}
\caption{Complementing Figure \ref{split2}, this is the area around the point $p$ where $\varGamma$ and $\varGamma_{\phi 1}$ intersect. The vectors and angles appearing on the figure are defined below.} \label{fig10a}
\end{figure}

At any point on Figure \ref{split2}, including $p$, the vectors $\frac{\partial}{\partial x}$ and $\frac{\partial}{\partial y}$ are well defined, and shown on the above diagram. $\beta_E$ is the Euclidean angle on this flat diagram sustained by the vector $\frac{\partial}{\partial x}$ and the tangent vector of $\varGamma$ at this point. $\beta_{E1}$ is the Euclidean angle between $\varGamma_{\phi 1}$ and $\frac{\partial}{\partial x}$. $\beta_{E2}$ is the Euclidean intersection angle between $\varGamma_{\phi 1}$ and $\varGamma$, so that $\beta_E=\beta_{E2}-\beta_{E1}$. Let $\beta_M$ be a measurable angle by a static observer at $p$, which corresponds to the Euclidean angle $\beta_E$, in the sense of the projection onto the embedded surface of section \ref{sec2}. Let $\beta_{M1}$ and $\beta_{M2}$ be the measurable angles by a static observer at $p$ corresponding to the Euclidean angles $\beta_{E1}$ and $\beta_{E2}$, respectively.

\subsubsection{Measurable deflection angle at any point by a static observer}
We generalize the previous definition of the measurable deflection angle $\alpha_M$ as follows. In reference to Figure \ref{split2} and the paragraph following it, let the \textit{measurable deflection angle by a static observer at a point $p$} be defined as the angle $\beta_M$, which corresponds to the Euclidean angle $\beta_E$. $\beta_M$ is equal to the projection of $\beta_E$ onto the embedded surface discussed in section \ref{sec2}, and therefore it depends on both $m$ and $\Lambda$. Due to symmetry, for given values $r_p$ and $r_0$, $\beta_M$ can be found analytically independent of orientation, assuming that $\phi_p$ at this $r_p$ satisfies the solution. The reference direction used to determine $\beta_M$ is the direction parallel to the $x$ axis in the setup of Figures \ref{split1} and \ref{split2}, which in the Euclidean sense is parallel to the direction of the path at $r_0$, and for this reason $\beta_M(r_0,r_p)$ is a measure of the deflection of the path as it goes from $r_0$ to $r_p$. For the standard transformation between the polar and Cartesian coordinates in the plane, the direction of increasing $x$ is well defined. A vector in this direction in Cartesian coordinates is $\frac{\partial}{\partial x}$ which can be transformed to polar coordinates at any point on the plane through $\frac{\partial}{\partial x}=\frac{\partial r}{\partial x}\frac{\partial}{\partial r}+\frac{\partial \phi}{\partial x}\frac{\partial}{\partial \phi}$. For concreteness, in reference to Figures \ref{split1} and \ref{split2} (and \ref{fig10a}), we can define the bending angle $\beta_M$ to be the measurable intersection angle between the path of light under investigation and the path of light whose tangent is parallel to the vector $\frac{\partial}{\partial x}$ at the point of measurement on the ($r,\phi$) plane. This angle is well defined, but unlike the special case of $r=r_{p_{sym}}$, for which a radial light ray could serve as the reference direction, in this general case the available radial light ray, $\varGamma_{\phi 1}$, is not going in the required $\frac{\partial}{\partial x}$ direction. Analytically, this angle can be found by referring to Figures \ref{split1}, \ref{split2} and \ref{fig10a} and using the measurable angles $\beta_{M1}$ and $\beta_{M2}$ corresponding to $\beta_{E1}$ and $\beta_{E2}$, respectively. Then, $\beta_{M}=\beta_{M2}-\beta_{M1}$, where both $\beta_{M1}$ and $\beta_{M2}$ refer to angles measured in reference to the radial light ray, and therefore both will satisfy a relationship of the form \eqref{tp3e4}, which allows for expressing $\beta_{M}$ in terms of $\beta_{E1}$ and $\beta_{E2}$. Clearly, $\beta_{E1}$ equals the value of $\phi_p$ at the point on the path, given the orientation of Figures \ref{split1} and \ref{split2}. The angle $\beta_{E2}$ is the Euclidean intersection angle between the path of light under investigation, $\varGamma$, and the radial light ray at the point $r_p$ on the path, and therefore can be expressed in terms of $r_0$, $r_p$ and $m$. As it is for the measurable deflection angle $\alpha_M$, this measurable deflection angle, $\beta_M$, also depends on both $\Lambda$ and $m$. The angle can be physically measured if the required reference light ray exists. Although not practical, but of theoretical significance, it is worth mentioning that a reference light ray for the required measurement can be produced in an experiment, even without the knowledge of $\Lambda$. Notice that this definition reduces to the bending angle of Schwarzschild space (if doubled) when $\Lambda$ is taken to be zero, assuming that $p$ is in the asymptotically flat region, outside $\mathcal R$ as in Figure \ref{split1}. In addition, for the case of $m=0$, the paths are straight lines, and the deflection angle $\beta_M$ equals zero at any point on a path, as expected.\\

\subsubsection{Measurable deflection angle at any point by any observer}
We generalize the previous definition of the measurable deflection angle $\beta_M$ even further as follows. Given the details of the previous definition of the bending angle $\beta_M$, let $K$ and $W$ represent the 4-vectors of the intersecting null geodesics at $p$ corresponding to the path of light under investigation, $\varGamma$, and the path of light whose tangent is parallel to $\frac{\partial}{\partial x}$ at $p$, respectively. For analytical purposes, $K$ can be found, up to an overall factor, from the derivative of the path given by the governing differential equation, \eqref{tp2e3}, at the point of intersection and the null condition. $W$ can be expressed in Kottler coordinates, up to an overall factor, by converting $\frac{\partial}{\partial x}$ to polar coordinates at the point of measurement and using the null condition. Let the \textit{measurable deflection angle by a given observer at a point $p$} be defined as the measurable angle between $K$ and $W$ by an observer with 4-velocity $U$ at $p$.

Let us designate this measurable deflection angle by $\bar{\beta}_M$. For the three 4-vectors $K$, $W$ and $U$, the angle  $\bar{\beta}_M$ is well defined. It may not yet be clear, though will be in the next section, how this angle can be found analytically directly from these vectors. In principle, with reference to a static observer, we can find $\beta_M$ and use the aberration equation to relate $\bar{\beta}_M$ to $\beta_M$, thereby expressing $\bar{\beta}_M$ in terms of the parameters of the setup, including the relative speed between the observers. Since the vector $\frac{\partial}{\partial x}$ is used as a reference direction in the definition of $\bar{\beta}_M$, we see that $\bar{\beta}_M$, as $\beta_M$, is a measure of the deflection of $\varGamma$ as it goes from ($r_0,\phi_0$) to $p$. Clearly, for a static observer $\bar{\beta}_M$ reduces to $\beta_M$, and satisfies all the expected limits of the definition. As before, this angle can be physically measured if a reference ray, with the required 4-vector $W$, exists, and its value depends on $\Lambda$ in addition to $m$.\\

We conclude as follows. In order to properly extend the concept of a bending angle for a trajectory of light to SdS space, we restricted the original definition of the bending angle in Schwarzschild space to the geometrical definition (definition 1), for which no measurements are considered, and further defined three additional measurable angular quantities. The use of the bending angle, $\Phi(r_0)$, in SdS space is clear from its definition and geometrical interpretation. It gives a quantitative measure of an important geometrical characteristic of the path in the ($r,\phi$) plane. Any path of the required type can be classified by its bending angle, which is in one to one correspondence to $r_0$. Although it is a Euclidean angle in nature, its value gives a visual and intuitive quality of the path, which makes it a useful parameter as well. On the other hand, the measurable deflection angles, $\alpha_M(r_0)$, $\beta_M(r_0,r)$ and $\bar{\beta}_M(r_0,r,U)$, give a type of observable measure of the deflection of the path as it goes from ($r_0,\phi_0$) to the observer, the practical use of which is not obvious. Although measurements are important to consider, these measurable quantities are observer dependent and are not informative in describing the behaviour of the path in the ($r,\phi$) plane, where the concept of the bending angle has originated. Nevertheless, these four definitions encompass many of the recent attempts to extend the bending angle to SdS space. With these definitions in mind it is straightforward to understand and compare the conclusions of many recent papers on the topic. Much of the disagreement in the recent literature seems to originate from lack of proper definitions, resulting in a mix-up of distinct quantities and improper comparison of results.\\

In closing this section it is worth noting again the following important conclusions about the commonly used parameters. Of the parameters $B$, $r_0$ and $b$, given typical boundary conditions, $B$ and $r_0$ do not depend on $\Lambda$ and can be determined from measurable quantities without its knowledge. The parameter $b$ does depend on $\Lambda$, it cannot be known a priori in an experiment and cannot be used as a boundary condition. It can only be found if the value of $\Lambda$ is known. Out of the angles $\Phi(r_0)$, $\alpha_M(r_0)$, $\beta_M(r_0,r)$ and $\bar{\beta}_M(r_0,r,U)$, only the bending angle $\Phi(r_0)$ does not depend on $\Lambda$. The other three angles are progressive generalizations, they are measurable and all depend on $\Lambda$, as should be expected given the previous discussions. We have considered creating a table that summarizes the above mentioned parameters and categorizes them based on their dependence on $\Lambda$ and way of measurement, but decided against it for the following reasons. The three parameters $B$, $r_0$ and $b$ are discussed to exhaustion, and the angular quantities are mentioned here for the purpose of their exact definition that is accompanied by an adequate discussion. Simply put, the parameters $B$, $r_0$ and $b$ have important physical and geometrical interpretations that should be abundantly clear by now and kept in mind throughout the rest of our presentation, and when we later refer to the angular parameters $\Phi(r_0)$, $\alpha_M(r_0)$, $\beta_M(r_0,r)$ and $\bar{\beta}_M(r_0,r,U)$ it is our intention that their exact definitions and our discussion of them will be read and understood. Finally, it is interesting to note that in the case where $r_0$ is the \textit{radius} of a static star, and is determined from other theoretical considerations, its value still does not depend on $\Lambda$, \cite{lake}.

\section{Measurable intersection angles} \label{sec5}
The main goal of this section is to find an expression for a measurable intersection angle for a given observer associated with two null geodesics in terms of the three 4-vectors representing the 4-velocity of the observer and the two tangent 4-vectors of the null geodesics evaluated at the point of intersection. Of course, since the motivation for the derivation of such an expression sprang out of the investigation of light rays in SdS space, being the central theme of this work, we shall attempt to keep our attention on it throughout this section. However, some of the results derived here are far reaching in their applicability, and may themselves be of much greater significance than their application to SdS space. Before considering the case of a general observer in SdS space, as a warm up, we derive the expression of a measurable intersection angle by a static observer in terms of the Euclidean intersection angle appearing on the flat ($r,\phi$) plane. This derivation is particularly informative and serves as an intuitive way to illustrate the influence of $\Lambda$ on measurable angles.
\subsection{Static observer in SdS space} \label{sec5a}
Although we should always assume that the background spacetime is SdS, it is noteworthy that the following derivation only assumes a spherically symmetric, static metric that is locally Minkowski. In spherical coordinates ($t,r,\phi,\theta$), we may assume that $r$ is the areal radius, and the metric can be written as in equation \eqref{tp1e1}. Further, the coordinate $r$ is restricted to a region where $f(r)$ in \eqref{tp1e1} is positive, and without loss of generality we restrict all motion and measurements to the slice $\theta=\frac{\pi}{2}$. The 4-velocity, $U$, of a static observer in these coordinates is defined by the requirement that $U^r=U^\phi=U^\theta=0$. With the condition $U \cdot U=-1$, we have
\begin{equation}
U=U^t\frac{\partial}{\partial t}, \qquad U^t=\frac{1}{\sqrt{f(r)}}. \label{32a}
\end{equation}
Let the space-like coordinates in the local Minkowski spacetime of the observer be $x$ and $y$. Since the 4-velocity vector of the static observer is parallel to $\frac{\partial}{\partial t}$, the local ($x,y$) plane corresponds to a small neighbourhood in the ($r,\phi$) plane around the location of the observer. We can orient the coordinates $x$ and $y$ without loss of generality such that $\frac{\partial}{\partial x}$ and $\frac{\partial}{\partial y}$ are parallel to $\frac{\partial}{\partial r}$ and $\frac{\partial}{\partial \phi}$, respectively, at the location of the observer. The metric of the local space around the observer can be written in terms of the coordinates $r$ and $\phi$, as given by equation \eqref{tp1e3}, or in terms of $x$ and $y$, as the flat metric,
\begin{equation}
\mathrm{d}s^2 =\mathrm{d}x^2+\mathrm{d}y^2. \label{33}
\end{equation}
The Minkowski coordinates $x$ and $y$ serve as real distance measurements of a static observer at the given point. Let $W$ and $K$ be the 4-vectors of two intersecting null geodesics at the point of the observer. For simplicity, we first assume that the path associated with $W$ is radial. Let $p$ be the point of intersection on the ($r,\phi$) plane, with coordinates ($r_p,\phi_p$). Let the point $p_1$ be a neighbouring point to $p$ lying on the path associated with $K$, with coordinates ($r_p-dr,\phi_p+d\phi$). Let $p_2$ be a neighbouring point on the radial path with coordinates ($r_p-dr,\phi_p$), where the $dr$ is the same as for $p_1$.

\begin{figure}[!ht]
\includegraphics[width=85mm]{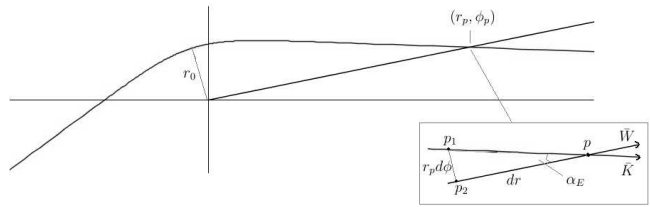}
\caption{A typical symmetric path of light on the flat $(r,\phi)$ plane, passing through a point with coordinates ($r_p,\phi_p$). The figure also shows the point of closest approach of this path, with $r=r_0$, and a radial path of light, which also passes through the point ($r_p,\phi_p$). The boxed diagram is of a small neighbourhood around the point ($r_p,\phi_p$), it can be thought of as a magnification of this point. The points $p_1$ and $p_2$ are points in this neighbourhood lying on the paths corresponding to $K$ and $W$, respectively. The space-like vectors $\overline{K}$ and $\overline{W}$ at $p$ on the diagram are the projections of $K$ and $W$ onto the ($r,\phi$) space, respectively. The intersection angle between the two paths on this flat diagram, which is the angle between $\overline{K}$ and $\overline{W}$, is $\alpha_E$.} \label{fig11}
\end{figure}

Assuming that the points $p_1$ and $p_2$ are in the immediate vicinity of the observer, let the distance measured form $p$ to $p_2$ be $dx$ and the distance from $p_1$ to $p_2$ be $dy$. The measurable intersection angle by the static observer, $\alpha_M$, corresponding to $\alpha_E$ on the figure above, can then be expressed as
\begin{equation}
\tan(\alpha_M)=\frac{dy}{dx}. \label{34}
\end{equation}
Since $\frac{\partial}{\partial x}$ and $\frac{\partial}{\partial y}$ are parallel with $\frac{\partial}{\partial r}$ and $\frac{\partial}{\partial \phi}$, respectively, we immediately see that
\begin{equation}
dx=\frac{dr}{\sqrt{f(r_p)}} \quad  and \quad dy=r_p d\phi. \label{35}
\end{equation}
Using \eqref{35} in \eqref{34},
\begin{equation}
\tan(\alpha_M)=\sqrt{f(r_p)}r_p\frac{d\phi}{dr}. \label{36}
\end{equation}
The angle $\alpha_E$ in the figure can be expressed as
\begin{equation}
\tan(\alpha_E)=r_p\frac{d\phi}{dr}. \label{37}
\end{equation}
Using \eqref{37} in \eqref{36}, dropping the subscript $p$, we find
\begin{equation}
\tan(\alpha_M)=\sqrt{f(r)}\tan(\alpha_E). \label{38}
\end{equation}
This is the first equation we were seeking: it relates the measurable intersection angle $\alpha_M$ to the Euclidean angle $\alpha_E$. In SdS space, $f(r)$ will depend on both $m$ and $\Lambda$, which is how $\Lambda$ has an influence on such measurements. The source of this influence can be viewed as the stretching of space due to $\Lambda$, quantitatively entering the analysis through the first of equations \eqref{35}. We can also express the measurable angle, $\alpha_M$, in terms of $r$ and $r_0$. Using \eqref{tp2e3} in \eqref{36} gives
\begin{align}
\nonumber \tan(\alpha_M) &= \frac{\sqrt{f(r)}}{\sqrt{(\frac{1}{r_0^2}-\frac{2m}{r_0^3})r^2+\frac{2m}{r}-1}}\\
&= \frac{\sqrt{\frac{f(r)}{r^2}}}{\sqrt{\frac{f(r_0)}{r_0^2}-\frac{f(r)}{r^2}}}. \label{38a}
\end{align}
If $r_0$ can be found from some boundary conditions, then the last equation above is particularly useful.  Although equation \eqref{38} was derived under the assumption that one of the paths is radial, it is useful due to its simple form. It illustrated the role of $\Lambda$ in a most simple and intuitive way, and it can be used as a starting point to establish a more general relationship.

Consider now the situation where neither $K$ nor $W$ is associated with the radial trajectory. Again, let $\overline{K}$ and $\overline{W}$ be the projections of the vectors onto the slice ($r,\phi$) and $\alpha_E$ be the Euclidean angle, appearing on the flat plane, between them. The following figure is of a neighbourhood around the point of intersection on the ($r,\phi$) plane, much like the boxed part of the last figure, however this time both $W$ and $K$ are in arbitrary directions.

\begin{figure}[!ht]
\includegraphics[width=85mm]{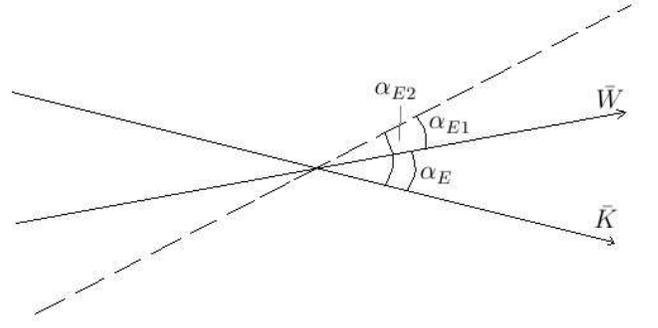}
\caption{Similar to the boxed part of Figure \ref{fig11}, this figure is of a small neighbourhood around an intersection point on the ($r,\phi$) plane. The vectors $\overline{K}$ and $\overline{W}$ are the tangent vectors of the intersecting paths on the plane, the dashed line represents the radial direction at this point. The angles $\alpha_{E1}$ and $\alpha_{E2}$ are Euclidean angles on the flat plane between the radial direction and the vectors $\overline{W}$ and $\overline{K}$, respectively.} \label{fig12}
\end{figure}

What we are after is the measurable intersection angle $\alpha_M$, corresponding to $\alpha_E$ on the above figure. Let $\alpha_{M1}$ and $\alpha_{M2}$ be the measurable angles corresponding to $\alpha_{E1}$ and $\alpha_{E2}$, respectively. Clearly,
\begin{equation}
\alpha_M=\alpha_{M2}-\alpha_{M1}. \label{39}
\end{equation}
Since the angles $\alpha_{M1}$ and $\alpha_{M2}$ are sustained with the radial direction, they can be expressed in terms of $\alpha_{E1}$ and $\alpha_{E2}$ according to equation \eqref{38}. This outlines a method of finding the expression for the measurable angle in an arbitrary orientation. The influence of $\Lambda$ in this case is clear and comes from the reference to equation \eqref{38}.

Let us consider a different approach to the problem. We have already established that the local space around the static observer can be represented by metric \eqref{tp1e3} in the coordinates ($r,\phi$), as well as the flat metric \eqref{33} in local Minkowski coordinates. The intersection angle measured between two trajectories of light with 4-velocities $K$ and $W$ is the angle sustained by the projections of $K$ and $W$ onto the local space of the observer. For a static observer, these projections are just projections onto the local ($r,\phi$) space, since it corresponds to the local ($x,y$) space. With $\overline{K}$ and $\overline{W}$ being the projections of $K$ and $W$, without restricting to any particular orientation, we have in general
\begin{equation}
\cos(\alpha_M)=\frac{\overline{K} \cdot \overline{W}}{|\overline{K}||\overline{W}|}. \label{40}
\end{equation}
Intuitively, since the angle $\alpha_M$ belongs to the Minkowski space of the observer, we may consider the vectors in \eqref{40} to exist in this Minkowski space and to be written in the $x$ and $y$ coordinates, with the inner products taking place in the ($x,y$) plane. However, since inner products are coordinate independent, there is no need to use any coordinates other than the given ($r,\phi$) to establish a relationship from \eqref{40} for a particular situation. Equation \eqref{40} can be used for any given 4-vectors $K$ and $W$, the projected vectors $\overline{K}$ and $\overline{W}$ can be found, for a static observer, simply by eliminating the $t$ component in each of the 4-vectors. For the special case where $W$ is associated with a radial trajectory, we find
\begin{equation}
\overline{K} \cdot \overline{W}=\frac{1}{f(r)} K^r W^r, \label{41}
\end{equation}
\begin{equation}
|\overline{K}|=\sqrt{\frac{K^{r2}}{f(r)}+r^2 K^{\phi 2}}, \label{42}
\end{equation}
\begin{equation}
|\overline{W}|=\frac{W^r}{\sqrt{f(r)}}. \label{43}
\end{equation}
Using \eqref{41}, \eqref{42}, and \eqref{43} in \eqref{40}
\begin{align}
\nonumber \cos(\alpha_M) &= \frac{K^r}{f(r)\sqrt{\frac{K^{r2}}{f^2(r)}+r^2 \frac{K^{\phi 2}}{f(r)}}}\\
&=\frac{K^r}{\sqrt{K^{r2}+r^2 K^{\phi 2} f(r)}}. \label{44}
\end{align}
From the trigonometric identity
\begin{equation}
\tan^2(\alpha)=\frac{1}{\cos^2(\alpha)}-1, \label{45}
\end{equation}
we find
\begin{equation}
\tan(\alpha_M)=\sqrt{1+f(r) r^2 \frac{K^{\phi 2}}{K^{r2}}-1}=\sqrt{f(r)} r \frac{K^{\phi}}{K^r}. \label{46}
\end{equation}
Since $\frac{K^{\phi}}{K^r}=\frac{\left( \frac{d \phi}{d \lambda} \right)}{\left( \frac{dr}{d \lambda}\right)}=\frac{d \phi}{dr}$, we have
\begin{equation}
\tan(\alpha_M)=\sqrt{f(r)} r \frac{d \phi}{dr}=\sqrt{f(r)} \tan(\alpha_E), \label{48}
\end{equation}
where $\alpha_E$ is the corresponding Euclidean intersection angle on the flat ($r,\phi$) plane. Thus, we have derived equation \eqref{38} as a special case of equation \eqref{40}. A similar reasoning to the one used in establishing equation \eqref{40} will be employed in the derivation of the general relation, applicable to any observer, which is the main goal of this section. In the meanwhile, we notice that with an established relationship for a static observer one can construct a relationship for any other observer by using the aberration equation to relate the measurable angles. This fact is important and may be of practical use, however, it may be inconvenient in some cases to refer to a static observer that is not part of the setup. Moreover, it is of mathematical curiosity to establish a relationship between a measurable angle and the associated three 4-vectors from first principles, with no reference to a proxy observer.

\subsection{Derivation of a general formula for measurable intersection angles} \label{sec5b}
The following is a derivation of the general formula for the measurable intersection angle by any observer. The final result is the main goal of this section and, perhaps, the result of most importance in this paper. For the sake of generality, we make no assumptions except that the metric of the space where the event occurs is locally Minkowski. For simplicity, in the derivation we consider the spacetime to be four dimensional, and we stick with the convention of positive signature. The generalization of the derivation to any higher dimension is trivial and the final relationship is true for all dimensions.

\begin{figure}[!ht]
\includegraphics[width=85mm]{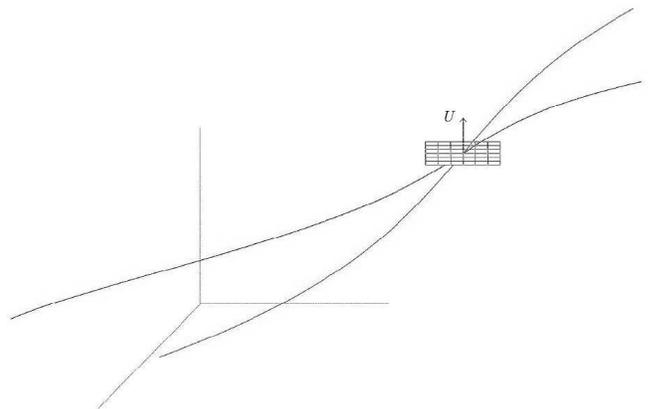}
\caption{Two intersecting arbitrary trajectories. One of the space-like dimensions is suppressed. The 4-velocity vector of the observer at the point of intersection is shown. The local flat space of the observer, in which measurements take place and to which the 4-velocity, $U$, is normal is shown as well. This local space together with the 4-velocity vector constitute the local Minkowski spacetime of the observer at the event of intersection.} \label{fig13}
\end{figure}

\begin{figure}[!ht]
\includegraphics[width=75mm]{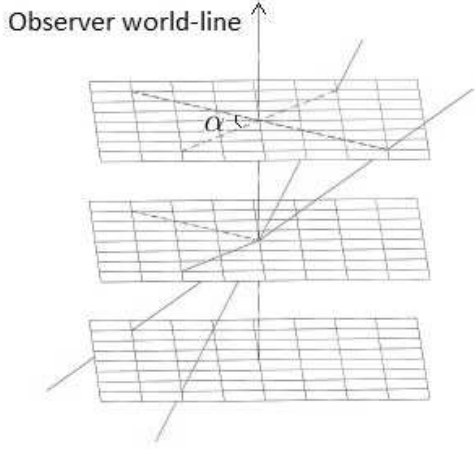}
\caption{An enlargement of the event of intersection in the previous figure. The world-line of the observer around this event is the vertical line, the two intersecting trajectories appear locally straight. As the trajectories pass through the neighbourhood of the observer, they trace paths in its local space. The dashed lines are the paths in the local space of the observer, traced by the two trajectories. The space slices depicted are successive instances of proper time $\tau$. The measurable intersection angle, $\alpha$, is the angle sustained by the traced paths.} \label{fig13a}
\end{figure}

Let $U$ be the 4-velocity of the observer at the event of intersection. Let $K$ and $W$ be the 4-vectors of the intersecting trajectories. At this point we do not make any assumptions on $K$ and $W$. The trajectories can be time-like, null or space-like. An example of a space-like trajectory is a simultaneous chain of events in some extended rigid frame. Consider a rigid line in the frame of which clocks at different locations are synchronized and simultaneity is well defined. Consider now a flash, or rather a brief change in colour, taking place simultaneously at each point on the line. In a different extended frame, in relative motion to the frame of the line, the chain of events will not be simultaneous. Rather, in the second frame the flash, or change in colour, travels along the points of the line faster than the speed of light, appearing as a traced path. This is a space-like trajectory, projected onto the second frame. Clearly, time-like and null trajectories represent paths of massive objects and rays of light, respectively. Let ($w^0,w^1,w^2,w^3$) be a given set of coordinates in a patch of the underlying spacetime, and let ($\tau,x,y,z$) be the local Minkowski coordinates of the observer at the point of intersection, with $\tau$ being the proper time. At the point of intersection, during a short interval of proper time around the event, the trajectories pass through the frame of the observer, tracing paths in the local space, ($x,y,z$), of the observer. This is illustrated in Figures \ref{fig13} and \ref{fig13a}.

\begin{figure}[!ht]
\includegraphics[width=70mm]{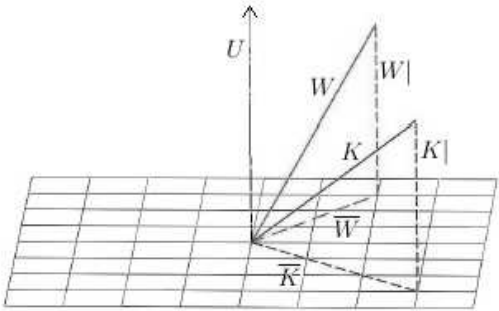}
\caption{The figure shows the 4-velocity $U$ and the perpendicular local space (the laboratory space), the 4-vectors $K$ and $W$ and their projections $\overline{K}$ and $\overline{W}$ onto the local space. Also shown, the components of the vectors parallel to U, $K|$ and $W|$. The measurable intersection angle $\alpha$, is the angle sustained by the two projected vectors $\overline{K}$ and $\overline{W}$ on this diagram.} \label{fig14}
\end{figure}

The measurable intersection angle by the observer is the angle between the traced paths in the observer's space, $\alpha$. This angle is determined by the tangent vectors of the projected paths in space. These tangent vectors are the projections of the 4-vectors $K$ and $W$ onto the space of the observer. Let $\overline{K}$ and $\overline{W}$ be the projections of $K$ and $W$ onto the space of the observer, respectively. Let $K|$ and $W|$ be components of $K$ and $W$, respectively, parallel to $U$. We have
\begin{equation}
K=K|+\overline{K}, \qquad W=W|+\overline{W}. \label{49}
\end{equation}
See Figure \ref{fig14}.

The measurable intersection angle $\alpha$, in the observer's frame, can therefore be expressed as usual.
\begin{equation}
\cos(\alpha)=\frac{\overline{K} \cdot \overline{W}}{|\overline{K}||\overline{W}|}. \label{50}
\end{equation}
The task now is to find $\overline{K}$ and $\overline{W}$ given the three 4-vectors $U$, $K$ and $W$, and take the inner product in accordance with the metric. We shall be abundantly clear in the following derivation. In the local Minkowski coordinates of the observer, the 4-vectors $K$ and $W$ can be expressed as
\begin{equation}
K=K^\tau \frac{\partial}{\partial \tau} + K^x \frac{\partial}{\partial x} + K^y \frac{\partial}{\partial y} + K^z \frac{\partial}{\partial z}, \label{51}
\end{equation}
\begin{equation}
W=W^\tau \frac{\partial}{\partial \tau} + W^x \frac{\partial}{\partial x} + W^y \frac{\partial}{\partial y} + W^z \frac{\partial}{\partial z}. \label{52}
\end{equation}
And, clearly,
\begin{equation}
\overline{K}=K^x \frac{\partial}{\partial x} + K^y \frac{\partial}{\partial y} + K^z \frac{\partial}{\partial z}, \qquad K|=K^τ \frac{\partial}{\partial \tau}, \label{53}
\end{equation}
\begin{equation}
\overline{W}=W^x \frac{\partial}{\partial x} + W^y \frac{\partial}{\partial y} + W^z \frac{\partial}{\partial z}, \qquad W|=W^τ \frac{\partial}{\partial \tau}. \label{54}
\end{equation}
In the coordinates $w^\alpha$, we have
\begin{equation}
K=K^\alpha \frac{\partial}{\partial w^\alpha}, \qquad  W=W^\alpha \frac{\partial}{\partial w^\alpha}, \label{55}
\end{equation}
\begin{equation}
\overline{K}=\overline{K}^\alpha \frac{\partial}{\partial w^\alpha}, \qquad  \overline{W}=\overline{W}^\alpha \frac{\partial}{\partial w^\alpha}, \label{56}
\end{equation}
\begin{equation}
K|=K|^\alpha \frac{\partial}{\partial w^\alpha}, \qquad  W|=W|^\alpha \frac{\partial}{\partial w^\alpha}. \label{57}
\end{equation}
In general, for any set of independent coordinates $q^\alpha$, and a vector $A$, such that $A=A^\alpha \frac{\partial}{\partial q^\alpha}$, the quantities $A^\alpha$ are determined by $A^\alpha=dq^\alpha(A)$, where $dq^\alpha$ is the differential 1-form corresponding to the coordinate $q^\alpha$, for a given value of the index $\alpha$. Therefore,
\begin{equation}
K^\tau=d\tau(K)=d\tau(K^\alpha \frac{\partial}{\partial w^\alpha})=K^\alpha \frac{\partial \tau}{\partial w^\alpha}, \label{58}
\end{equation}
\begin{equation}
W^\tau=d\tau(W)=d\tau(W^\alpha \frac{\partial}{\partial w^\alpha})=W^\alpha \frac{\partial \tau}{\partial w^\alpha}. \label{59}
\end{equation}
And
\begin{equation}
K|= K^\tau \frac{\partial}{\partial \tau} = \left( K^\alpha \frac{\partial \tau}{\partial w^\alpha} \right) \left( \frac{\partial w^\beta}{\partial \tau} \frac{\partial}{\partial w^\beta} \right), \label{60}
\end{equation}
\begin{equation}
W|= W^\tau \frac{\partial}{\partial \tau} = \left( W^\alpha \frac{\partial \tau}{\partial w^\alpha} \right) \left( \frac{\partial w^\beta}{\partial \tau} \frac{\partial}{\partial w^\beta} \right). \label{61}
\end{equation}
The 4-vector $U$ and its dual covector, $\widetilde{U}$, with respect to the metric, can be expressed in the two sets of coordinates as follows.
\begin{equation}
U=\frac{\partial}{\partial \tau}=U^\alpha \frac{\partial}{\partial w^\alpha}, \label{62}
\end{equation}
and
\begin{equation}
\widetilde{U}=-d\tau=U_\alpha dw^\alpha. \label{63}
\end{equation}
Since $\frac{\partial}{\partial \tau}=\frac{\partial w^\alpha}{\partial \tau} \frac{\partial}{\partial w^\alpha}$ and $d\tau=\frac{\partial \tau}{\partial w^\alpha} dw^\alpha$, we have
\begin{equation}
U^\alpha=\frac{\partial w^\alpha}{\partial \tau}, \label{64}
\end{equation}
and
\begin{equation}
U_\alpha=-\frac{\partial \tau}{\partial w^\alpha}. \label{65}
\end{equation}
Using \eqref{64} and \eqref{65} in \eqref{60} and \eqref{61}.
\begin{equation}
K|= -K^\alpha U_\alpha U^\beta \frac{\partial}{\partial w^\beta}. \label{65a}
\end{equation}
\begin{equation}
W|= -W^\alpha U_\alpha U^\beta \frac{\partial}{\partial w^\beta}. \label{65b}
\end{equation}
Therefore,
\begin{equation}
K|^\alpha= -K^\beta U_\beta U^\alpha, \label{65c}
\end{equation}
and
\begin{equation}
W|^\alpha= -W^\beta U_\beta U^\alpha. \label{65d}
\end{equation}
We can now express $\overline{K}$ and $\overline{W}$ as follows.
\begin{align}
\nonumber \overline{K} = K-K| &= K^\alpha \frac{\partial}{\partial w^\alpha} + K^\beta U_\beta U^\alpha \frac{\partial}{\partial w^\alpha}\\
&= \left(K^\alpha + K^\beta U_\beta U^\alpha \right) \frac{\partial}{\partial w^\alpha}, \label{66}
\end{align}
and
\begin{align}
\nonumber \overline{W} = W-W| &= W^\alpha \frac{\partial}{\partial w^\alpha} + W^\beta U_\beta U^\alpha \frac{\partial}{\partial w^\alpha}\\
&= \left(W^\alpha + W^\beta U_\beta U^\alpha \right) \frac{\partial}{\partial w^\alpha}. \label{67}
\end{align}
Therefore,
\begin{align}
\nonumber \overline{K}^\alpha &= K^\alpha + K^\beta U_\beta U^\alpha\\
\nonumber &= \delta^\alpha_\beta K^\beta + K^\beta U_\beta U^\alpha\\
&= \left( \delta^\alpha_\beta + U^\alpha U_\beta \right)K^\beta, \label{68}
\end{align}
and
\begin{align}
\nonumber \overline{W}^\alpha &= W^\alpha + W^\beta U_\beta U^\alpha\\
\nonumber &= \delta^\alpha_\beta W^\beta + W^\beta U_\beta U^\alpha\\
&= \left( \delta^\alpha_\beta + U^\alpha U_\beta \right)W^\beta. \label{69}
\end{align}
Here $\delta^\alpha_\beta$ is the usual Kronecker delta. Let
\begin{equation}
h^\alpha_\beta =\delta^\alpha_\beta + U^\alpha U_\beta, \label{70}
\end{equation}
so that
\begin{equation}
\overline{K}^\alpha = h^\alpha_\beta K^\beta, \qquad \overline{W}^\alpha = h^\alpha_\beta W^\beta. \label{71}
\end{equation}
Using the metric tensor of the spacetime, $g_{\alpha\beta}$, to lower the upper index of $h^\alpha_\beta$ gives
\begin{equation}
h_{\alpha\beta} = g_{\alpha\gamma} h^\gamma_\beta = g_{\alpha\beta}+U_\alpha U_\beta. \label{72}
\end{equation}
Let $\overline{g}_{\alpha\beta}$ be the metric of the local space of the observer, that is the metric of the subspace perpendicular to $U$ at the event of measurement. The natural requirement of $\overline{g}_{\alpha\beta}$ to be consistent with $g_{\alpha\beta}$ is to simply be a restriction of $g_{\alpha\beta}$ onto the subspace under consideration and the tangent vectors within it. That is, for the vectors $\overline{K}$ and $\overline{W}$, in the local space of the observer,
\begin{align}
\nonumber \overline{g}_{\alpha\beta} \overline{K}^\alpha \overline{W}^\beta &= g_{\alpha\beta} \overline{K}^\alpha \overline{W}^\beta\\
\nonumber &= g_{\alpha\beta} h^\alpha_\gamma K^\gamma h^\beta_\delta W^\delta\\
&= h_{\beta\gamma} K^\gamma h^\beta_\delta W^\delta. \label{73}
\end{align}
Now, $h_{\beta\gamma}=h_{\alpha \beta} h^\alpha_\gamma$, since
\begin{align}
\nonumber h_{\alpha \beta} h^\alpha_\gamma &= ( g_{\beta\alpha} + U_\beta U_\alpha ) ( \delta^\alpha_\gamma + U^\alpha U_\gamma )\\
\nonumber &= g_{\beta\gamma}+U_\beta U_\gamma+U_\beta U_\gamma-U_\beta U_\gamma\\
\nonumber &= g_{\beta\gamma}+U_\beta U_\gamma\\
&= h_{\beta\gamma}. \label{74}
\end{align}
With this, we can go back to \eqref{73} to find
\begin{equation}
\overline{g}_{\alpha\beta} \overline{K}^\alpha \overline{W}^\beta = h_{\alpha\beta} h^\alpha_\gamma K^\gamma h^\beta_\delta W^\delta = h_{\alpha\beta}\overline{K}^\alpha \overline{W}^\beta. \label{75}
\end{equation}
Since $\overline{K}$ and $\overline{W}$ are arbitrary, we have shown that
\begin{equation}
\overline{g}_{\alpha\beta} = h_{\alpha\beta}. \label{76}
\end{equation}
The considerations above should make it intuitively evident that the observer dependent tensor $h^\alpha_\beta$ is a projection tensor, which projects any 4-vector onto the local space of the observer, and the related covariant tensor $h_{\alpha\beta}$ is the metric tensor of that space. Of course, if we express $\overline{K}$ and $\overline{W}$ in the Minkowski coordinates ($x,y,z$), then the inner products in equation \eqref{50} can be taken with respect to the flat metric of the observer's space,
\begin{equation}
\mathrm{d}s^2|_\Sigma=\mathrm{d}x^2+\mathrm{d}y^2+\mathrm{d}z^2. \label{77}
\end{equation}
However, in the original coordinates of the spacetime, $w^\alpha$, at the point of measurement, the local metric of the observer's space is given by the tensor $h_{\alpha\beta}$, and the vectors $\overline{K}$ and $\overline{W}$ are given by equations \eqref{71}. With the considerations above we go back to equation \eqref{50} to find the required expression for the measurable intersection angle $\alpha$.
\begin{widetext}
\begin{align}
\nonumber \cos(\alpha) &= \frac{\overline{K} \cdot \overline{W}}{|\overline{K}||\overline{W}|} = \frac{h_{\alpha\beta}\overline{K}^\alpha \overline{W}^\beta}{\sqrt{h_{\alpha\beta}\overline{K}^\alpha \overline{K}^\beta} \sqrt{h_{\alpha\beta}\overline{W}^\alpha \overline{W}^\beta}}\\
\nonumber &= \frac{h_{\alpha\beta}K^\alpha W^\beta}{\sqrt{h_{\alpha\beta}K^\alpha K^\beta} \sqrt{h_{\alpha\beta}W^\alpha W^\beta}}\\
\nonumber &= \frac{g_{\alpha\beta}K^\alpha W^\beta+U_\alpha U_\beta K^\alpha W^\beta}{\sqrt{g_{\alpha\beta}K^\alpha K^\beta+U_\alpha U_\beta K^\alpha K^\beta} \sqrt{g_{\alpha\beta}W^\alpha W^\beta+U_\alpha U_\beta W^\alpha W^\beta}}\\
&= \frac{K \cdot W + (U \cdot K)(U \cdot W)}{\sqrt{K \cdot K + (U \cdot K)^2} \sqrt{W \cdot W + (U \cdot W)^2}}. \label{78}
\end{align}
\end{widetext}

\textbf{Theorem 1:}\\

\noindent The measurable intersection angle $\alpha$ by an observer with 4-velocity $U$, sustained by two paths with tangent 4-vectors $K$ and $W$ at the point of intersection is given by
\begin{equation*}
\cos(\alpha)=\frac{K \cdot W + (U \cdot K)(U \cdot W)}{\sqrt{K \cdot K + (U \cdot K)^2} \sqrt{W \cdot W + (U \cdot W)^2}}.
\end{equation*}\\

Equation \eqref{78} can be applied to any observer and any trajectories, whether time-like, null or space-like. However, it can be considerably simplified for the case of null trajectories, which is, fortunately, the case of interest. With $K$ and $W$ being null,
\begin{align}
\nonumber \cos(\alpha) &= \frac{K \cdot W + (U \cdot K)(U \cdot W)}{(U \cdot K)(U \cdot W)}\\
&= \frac{K \cdot W}{(U \cdot K)(U \cdot W)}+1. \label{79}
\end{align}\\

\textbf{Corollary 1:}\\

\noindent For the case of null trajectories, the above theorem reduces to
\begin{equation*}
\cos(\alpha)=\frac{K \cdot W}{(U \cdot K)(U \cdot W)}+1.
\end{equation*}\\

The above equation is the general formula we were seeking. In four dimensional spacetime, it gives the desired expression of the measurable angle, $\alpha$, in terms of the 4-velocity of the observer, $U$, and the null 4-vectors of the intersecting trajectories, $K$ and $W$. The formula is coordinate independent, which may be important in many applications.

\subsection{Applications of the general formula}
Before ending this section we consider a few applications of equation \eqref{79}. We shall demonstrate its use in the frameworks concerned with angle measurements in SdS space and the aberration of light phenomenon of special and general relativity. As always, for the sake of demonstration and simplicity, whenever there is a choice of positive or negative sign, if not stated otherwise, we shall take the positive.

\subsubsection{Static observer in SdS space revisited}
In reference to the first part of this section and the SdS metric of section \ref{sec2}, we proceed as follows. Let $U$ be the 4-velocity of a static observer, given by equation \eqref{32a} in Kottler coordinates. Again, without loss of generality we assume the trajectories are confined to the $\theta=\frac{\pi}{2}$ subspace. For the sake of comparison and simplicity let us take the 4-Vector $W$ to be exclusively in the radial direction. The inner products that we need are
\begin{equation}
K \cdot W = -f(r) K^t W^t + \frac{K^r W^r}{f(r)}, \label{82}
\end{equation}
\begin{equation}
U \cdot K = -f(r)\frac{1}{\sqrt{f(r)}}K^t = -\sqrt{f(r)} K^t, \label{83}
\end{equation}
\begin{equation}
U \cdot W = -f(r)\frac{1}{\sqrt{f(r)}}W^t = -\sqrt{f(r)} W^t. \label{83a}
\end{equation}
Using \eqref{82}, \eqref{83} and \eqref{83a} in \eqref{79}, we find
\begin{align}
\nonumber \cos(\alpha) &= \frac{-f(r) K^t W^t + \frac{K^r W^r}{f(r)} + f(r) K^t W^t}{f(r) K^t W^t}\\
&= \frac{K^r W^r}{f^2(r) K^t W^t}. \label{84}
\end{align}
And with the help of the null conditions $K \cdot K = 0$ and $W \cdot W=0$, we get
\begin{equation}
\cos(\alpha) = \frac{K^r}{\sqrt{K^{r2}+r^2 K^{\phi 2} f(r)}}. \label{84a}
\end{equation}
The above equation is identical to equation \eqref{44}, demonstrating the consistency of the general formula. As it was already shown following equation \eqref{44}, with $\alpha_E$ being the Euclidean intersection angle in the flat ($r,\phi$) plane, it is straight forward to derive equations \eqref{38}, \eqref{48}, or \eqref{38a} from equation \eqref{84a}.

\subsubsection{Relativistic aberration of light}
As discussed before, one can use the aberration equation to relate the measurable angle of the static observer to the measurable angle of an observer in relative motion to it. However, since equation \eqref{79} can be used to express the measurable angles of any two observers in any metric, one may suspect that it may be used to derive the aberration equation itself. The well known aberration equation of special relativity is the following:
\begin{equation}
\cos(\bar{\alpha})= \frac{\cos(\alpha)-v}{1-v \cos(\alpha)}. \label{85}
\end{equation}
Here, $\alpha$ and $\bar{\alpha}$ are the two different angles measured by the different observers, and $v$ is their relative speed (sometimes taken to be negative in the equation). It is commonly derived in textbooks from geometric considerations or special relativistic velocity transformations. See for example \cite{rindler}. The equation is valid under the assumption that in the frame of one of the observers the other observer is travelling in the same direction as one of the light rays. We shall derive a general aberration equation, applicable to any two observers and any two light rays in any orientation. We then demonstrate how equation \eqref{85} can be obtained, for the particular orientation assumed in the usual derivation of the aberration equation. As it was for the derivation of equation \eqref{79}, we shall assume nothing of the background metric of the spacetime, except that it is locally Minkowski and of positive signature. For simplicity and concreteness let us take the dimension to be four.

Let $U$ and $V$ be the 4-velocities of two observers at the event of intersection, with ($\tau,x,y,z$) and ($\bar{\tau},\bar{x},\bar{y},\bar{z}$) being the Minkowski coordinates of their respective local frames. Let $K$ and $W$ be the null 4-vectors, at the event of intersection, of any two trajectories of light. The derivation of the general aberration equation is immediate. With $\bar{\alpha}$ being the angle measured by the observer with 4-velocity $V$, from equation \eqref{79}
\begin{equation}
\cos(\alpha) - 1 = \frac{K \cdot W}{(U \cdot K)(U \cdot W)}, \label{86}
\end{equation}
and
\begin{equation}
\cos(\bar{\alpha}) - 1 = \frac{K \cdot W}{(V \cdot K)(V \cdot W)}. \label{87}
\end{equation}
Dividing \eqref{87} by \eqref{86} gives
\begin{equation}
\frac{\cos(\bar{\alpha}) - 1}{\cos(\alpha) - 1} = \frac{(U \cdot K)(U \cdot W)}{(V \cdot K)(V \cdot W)}. \label{89}
\end{equation}\\

\textbf{Theorem 2:}\\

\noindent The general relationship between the measurable angles $\alpha$ and $\bar{\alpha}$, related to observers with 4-velocity vectors $U$ and $V$, respectively, is given by
\begin{equation*}
\frac{\cos(\bar{\alpha}) - 1}{\cos(\alpha) - 1} = \frac{(U \cdot K)(U \cdot W)}{(V \cdot K)(V \cdot W)}
\end{equation*}\\

The above equation can be regarded as the general aberration equation. It relates the measurable angles in terms of the associated 4-vectors, it is coordinate independent and holds for any metric of general relativity. A specific aberration relationship can be obtained from \eqref{89} for any particular orientation; for the orientation assumed in the usual derivation of the aberration equation, \eqref{85}, that is, where the direction of motion of one observer coincides with a direction of a ray, it can be done as follows.

Let the direction of motion of the observer with 4-velocity $V$ in the frame of the observer with 4-velocity $U$ coincide with the direction of the light rays with 4-vector $W$. Let $v$ be the relative speed between the two observers. Solving for $\cos(\bar{\alpha})$ in \eqref{89} gives
\begin{equation}
\cos(\bar{\alpha}) = \frac{(U \cdot K)(U \cdot W)}{(V \cdot K)(V \cdot W)}(\cos(\alpha) - 1)+1. \label{90}
\end{equation}
Let us express the vectors and the inner products of equation \eqref{90} in the local Minkowski coordinates ($\tau,x,y,z$) of the observer with 4-velocity $U$. For convenience, we align the $x$ axis in the space of this observer with the direction of motion of the other observer and one of the rays. With $\bar{\tau}$ being the proper time of the observer with 4-velocity $V$, in these coordinates, we have
\begin{equation}
\mathrm{d}s^2=-\mathrm{d}\tau^2+\mathrm{d}x^2+\mathrm{d}y^2+\mathrm{d}z^2, \label{92}
\end{equation}
\begin{equation}
U^\alpha=(1,0,0,0), \label{93}
\end{equation}
\begin{equation}
V^\alpha=\left( \frac{d \tau}{d \bar{\tau}},\frac{dx}{d \bar{\tau}},0,0 \right)=(V^\tau,V^x,0,0), \label{94}
\end{equation}
\begin{equation}
K^\alpha=(K^\tau,K^x,K^y,K^z), \label{95}
\end{equation}
and
\begin{equation}
W^\alpha=(W^\tau,W^x,0,0). \label{96}
\end{equation}
The components $V^\tau$ and $V^x$ can be expressed in terms of the relative velocity, $v$, as follows. By definition
\begin{equation}
v=\frac{dx}{d \tau}=\frac{dx}{d \bar{\tau}}\frac{d \bar{\tau}}{d \tau}=\frac{V^x}{V^\tau}, \label{96a}
\end{equation}
and since $V \cdot V = -1$, we have two equations in two unknowns. Solving for $V^\tau$ and $V^x$ gives
\begin{equation}
V^\tau=\frac{1}{\sqrt{1-v^2}}, \qquad V^x=\frac{v}{\sqrt{1-v^2}}. \label{96c}
\end{equation}
The above are well known relationships of special relativity. Further, the null condition $W \cdot W = 0$ gives
\begin{equation}
W^\tau=W^x. \label{96d}
\end{equation}
The somewhat obvious expression for the angle $\alpha$ in these coordinates is obtained from equation \eqref{79} as follows.
\begin{equation}
\cos(\alpha) = \frac{-K^\tau W^\tau + K^x W^x}{K^\tau W^\tau} + 1 = \frac{K^x}{K^\tau}. \label{97}
\end{equation}
The inner products appearing in equation \eqref{90} are
\begin{equation}
U \cdot K = -K^\tau, \qquad U \cdot W = -W^\tau, \label{98}
\end{equation}
\begin{align}
\nonumber V \cdot K &= -\frac{1}{\sqrt{1-v^2}}K^\tau+\frac{v}{\sqrt{1-v^2}}K^x\\
&=-\frac{1}{\sqrt{1-v^2}}(K^\tau-vK^x), \label{99}
\end{align}
and
\begin{align}
\nonumber V \cdot W &= -\frac{1}{\sqrt{1-v^2}}W^\tau+\frac{v}{\sqrt{1-v^2}}W^x\\
&=-\frac{1}{\sqrt{1-v^2}}W^x(1-v). \label{100}
\end{align}
Using \eqref{98}, \eqref{99} and \eqref{100} in \eqref{90} gives
\begin{align}
\nonumber \cos(\bar{\alpha}) &= \frac{K^\tau W^\tau}{\left( \dfrac{1}{1-v^2} \right) (K^\tau-vK^x)W^x(1-v)}(\cos(\alpha) - 1)+1\\
\nonumber &= \frac{K^\tau (1+v)}{K^\tau-vK^x}(\cos(\alpha) - 1)+1\\
\nonumber &= \frac{(1+v)(\cos(\alpha) - 1)}{1-v\frac{K^x}{K^\tau}}+\frac{1-v \cos(\alpha)}{1-v \cos(\alpha)}\\
&= \frac{\cos(\alpha)-v}{1-v \cos(\alpha)}. \label{101}
\end{align}
Thus, we have derived the known aberration equation for the usually assumed orientation from the general equation \eqref{89}. This demonstrates the usefulness and consistency of both equations \eqref{89} and \eqref{79}. Overall, the proposed general, coordinate independent, aberration equation, \eqref{89}, may be applied to any setup and can considerably simplify the analysis in many situations.

Lastly, for completion, let us state the first order approximation in angles of equation \eqref{89}. For small angles $\alpha$ and $\bar{\alpha}$, to lowest order we find
\begin{equation}
\bar{\alpha} = \sqrt{\frac{(U \cdot K)(U \cdot W)}{(V \cdot K)(V \cdot W)}} \alpha.
\end{equation}
This simple relationship may be of use in some situations, and of course, the well known first order approximation of the usual aberration equation, \eqref{85}, can be easily derived from it.

\subsubsection{General observer in SdS space}
Going back to paths of light in SdS space, specifically in the subspace $\theta=\frac{\pi}{2}$, let us employ equation \eqref{79} to express the measurable angle by a given observer in terms of relevant parameters. We shall consider the measurable angle in reference to a ray going in the radial, increasing $r$, direction, since these rays are usually available in a realistic situation. Although the trajectories are assumed to be confined to $\theta=\frac{\pi}{2}$, the metric of the spacetime is still given by equation \eqref{tp1e1}. Let $U$ be the 4-velocity of the observer making the measurement at the point of intersection. Let $K$ and $W$ be the 4-vectors of the intersecting trajectories of light, such that $W$ corresponds to the radial trajectory. In Kottler coordinates,
\begin{equation}
U^\alpha=(U^t,U^r,U^\phi,U^\theta), \label{103}
\end{equation}
\begin{equation}
K^\alpha=(K^t,K^r,K^\phi,0), \label{104}
\end{equation}
and
\begin{equation}
W^\alpha=(W^t,W^r,0,0). \label{105}
\end{equation}
Assuming that the path corresponding to the 4-vector $K$ has a point of minimum value of $r$, $r_0$, the components of $K$ in these coordinates are subject to equation \eqref{tp2e3}. For this path
\begin{align}
\nonumber \frac{dr}{d \phi} &= r \sqrt{\left(\frac{1}{r_0^2}-\frac{2m}{r_0^3}\right)r^2+\frac{2m}{r}-1}\\
&= r^2 \sqrt{\frac{f(r_0)}{r_0^2}-\frac{f(r)}{r^2}}. \label{107}
\end{align}
Therefore,
\begin{equation}
\frac{K^r}{K^\phi}=\frac{\frac{dr}{d\lambda}}{\frac{d\phi}{d\lambda}}=\frac{dr}{d \phi}=r^2 \sqrt{\frac{f(r_0)}{r_0^2}-\frac{f(r)}{r^2}}, \label{108}
\end{equation}
where $\lambda$ is an affine parameter, parametrizing the trajectory. The null conditions $K \cdot K=W \cdot W=0$ give the following relationships
\begin{equation}
f(r)W^t=W^r, \label{109}
\end{equation}
and
\begin{align}
\nonumber f(r)K^t &= \sqrt{K^{r2}+f(r)r^2 K^{\phi 2}}\\
\nonumber &= K^\phi r^2 \sqrt{\left(\frac{K^r}{K^\phi r^2}\right)^2+\frac{f(r)}{r^2}}\\
&= K^\phi r^2 \sqrt{\frac{f(r_0)}{r_0^2}}. \label{110}
\end{align}
For convenience, we have assumed that all the components of the null vectors are positive. Let the measurable angle by the observer be $\alpha$ (we shall add the subscript 'M' when ambiguity may arise), using equation \eqref{79}, we find
\begin{widetext}
\begin{align}
\nonumber \cos(\alpha) &= \frac{K \cdot W}{(U \cdot K)(U \cdot W)}+1\\
\nonumber &= \frac{-f(r)K^t W^t+\frac{K^r W^r}{f(r)}}{(-f(r)K^t U^t+\frac{K^r U^r}{f(r)}+r^2 K^\phi U^\phi)(-f(r)W^t U^t+\frac{W^r U^r}{f(r)})}+1\\
\nonumber &= \frac{-K^\phi r^2 \sqrt{\frac{f(r_0)}{r_0^2}} W^t+K^r W^t}{(-K^\phi r^2 \sqrt{\frac{f(r_0)}{r_0^2}} U^t+\frac{K^r U^r}{f(r)}+r^2 K^\phi U^\phi)(-f(r)W^t U^t+W^t U^r)}+1\\
\nonumber &= \frac{-r^2 \sqrt{\frac{f(r_0)}{r_0^2}}+\frac{K^r}{K^\phi}}{(-r^2 \sqrt{\frac{f(r_0)}{r_0^2}} U^t+\frac{K^r U^r}{K^\phi f(r)}+r^2 U^\phi) (-f(r)U^t+U^r)}+1\\
\nonumber &= \frac{-r^2 \sqrt{\frac{f(r_0)}{r_0^2}}+r^2 \sqrt{\frac{f(r_0)}{r_0^2}-\frac{f(r)}{r^2}}}{(-r^2 \sqrt{\frac{f(r_0)}{r_0^2}} U^t+r^2 \sqrt{\frac{f(r_0)}{r_0^2}-\frac{f(r)}{r^2}}\frac{U^r}{f(r)}+r^2 U^\phi) (-f(r)U^t+U^r)}+1\\
&= \frac{-\sqrt{\frac{f(r_0)}{r_0^2}}+\sqrt{\frac{f(r_0)}{r_0^2}-\frac{f(r)}{r^2}}}{(-\sqrt{\frac{f(r_0)}{r_0^2}} U^t+ \sqrt{\frac{f(r_0)}{r_0^2}-\frac{f(r)}{r^2}}\frac{U^r}{f(r)}+U^\phi) (-f(r)U^t+U^r)}+1. \label{111}
\end{align}
\end{widetext}
In the above equation, $\Lambda$ comes in through $f(r)$ and $f(r_0)$. The measurable angle is conveniently expressed in Kottler coordinates and the relationship is applicable to any observer. Of course, due to the condition $U \cdot U=-1$ not all of the four components ($U^t,U^r,U^\phi,U^\theta$) can be independent, and at least one must depend on $\Lambda$. In different setups, any of the three space-like components, $U^r,U^\phi$ and $U^\theta$, may or may not depend on $\Lambda$, and therefore, the particular influence of $\Lambda$ depends closely on the situation being analyzed. Also, notice that the relationship between the parameters $b$ and $r_0$ can be written as follows.
\begin{equation}
\frac{1}{b^2}=\frac{f(r_0)}{r_0^2}. \label{112}
\end{equation}
This makes it slightly tempting to use the parameter $b$ to simplify equation \eqref{111}. However, considering what we know of this parameter, we see that it will partially mask the appearance of $\Lambda$, and may lead to misinterpretations when investigating the influence of $\Lambda$ on the measurable angle. Out of the three parameters $b$, $r_0$ and $B$, the parameter $r_0$ is the most appropriate and intuitive to use in the analysis at hand, and especially convenient in Kottler coordinates. To simplify the general expression given by \eqref{111}, let
\begin{widetext}
\begin{equation}
h(U)=\left(-\sqrt{\frac{f(r_0)}{r_0^2}} U^t+ \sqrt{\frac{f(r_0)}{r_0^2}-\frac{f(r)}{r^2}}\frac{U^r}{f(r)}+U^\phi \right) \left(-f(r)U^t+U^r \right). \label{113}
\end{equation}
\end{widetext}
Then
\begin{equation}
\cos(\alpha)=\frac{-\sqrt{\frac{f(r_0)}{r_0^2}}+\sqrt{\frac{f(r_0)}{r_0^2}-\frac{f(r)}{r^2}}}{h(U)}+1. \label{114}
\end{equation}
And a little algebra yields
\begin{equation}
\tan(\alpha)=\frac{\sqrt{\frac{2h(U)}{\sqrt{\frac{f(r_0)}{r_0^2}}-\sqrt{\frac{f(r_0)}{r_0^2}-\frac{f(r)}{r^2}}}-1}}{\frac{h(U)}{\sqrt{\frac{f(r_0)}{r_0^2}}-\sqrt{\frac{f(r_0)}{r_0^2}-\frac{f(r)}{r^2}}}-1}. \label{115}
\end{equation}
These expressions are particularly easy and convenient to use when $r_0$ is given as a boundary condition. Then, it is not even necessary to find a solution for the deflected trajectory, and the measurable intersection angle can found immediately. With any other boundary conditions, such as two points on the path (coordinate locations of source and observer, for example), we can use an exact solution to express $r_0$ in terms of these two points to any desired degree of accuracy. Further, although it was previously assumed that both $m$ and $\Lambda$ are relatively small for conceptual reasons, we have not yet made any mathematical approximations related to these parameters. Thus, the relationships above are exact; quantities may be calculated to any degree of accuracy and approximations can be made when convenient or necessary.

Let us apply the above results to a few specific observers. If we set the observer to be static, we get
\begin{equation}
h(U_{static})=\sqrt{\frac{f(r_0)}{r_0^2}}, \label{116}
\end{equation}
\begin{equation}
\cos(\alpha_{static})=\frac{\sqrt{\frac{f(r_0)}{r_0^2}-\frac{f(r)}{r^2}}}{\sqrt{\frac{f(r_0)}{r_0^2}}}, \label{116a}
\end{equation}
and
\begin{equation}
\tan(\alpha_{static})=\frac{\sqrt{\frac{f(r)}{r^2}}}{\sqrt{\frac{f(r_0)}{r_0^2}-\frac{f(r)}{r^2}}}. \label{117}
\end{equation}
The last equation is identical to \eqref{38a}, as expected. Using equation \eqref{116a}, equation \eqref{114} can be expressed as
\begin{equation}
\cos(\alpha)=\frac{\sqrt{\frac{f(r_0)}{r_0^2}}}{h(U)}(\cos(\alpha_{static})-1)+1. \label{117a}
\end{equation}
The above is a relationship between the intersection angle $\alpha$, measured by an observer with 4-velocity $U$, and the intersection angle $\alpha_{static}$, measured by a static observer. It may be of practical use in situations where reference to a static observer is advantageous. Notice how the relationship reminds one of the general aberration equation previously derived, from which this result could be obtained directly.

Consider now an observer on a circular trajectory, with constant coordinate $r$. That is, $U^r=0$, which gives
\begin{equation}
h(U_{circular})=\left(-\sqrt{\frac{f(r_0)}{r_0^2}} U^t+U^\phi \right) \left(-f(r)U^t \right). \label{118}
\end{equation}
In certain situations the component $U^{\phi}$ can be considered independent, since it can be determined experimentally, in others $U^\phi$ can be expressed in terms of $m$ and $\Lambda$. For example, for measurements in the solar system, $U^\phi$, can be determined from the period of rotation experimentally, or expressed in terms of the mass of the sun and $\Lambda$. In the case where the deflected ray just gazes the surface of the sun, $r_0$ can be given by other existing theories or sources, which sets a convenient boundary condition and can be used directly in the above relationships, eliminating the need for a solution. Further, if we also confine the motion of the observer to the plane of the rays, setting $U^\theta=0$, the condition $U \cdot U=-1$ gives
\begin{equation}
U^t=\frac{\sqrt{1+r^2 U^{\phi 2}}}{\sqrt{f(r)}}. \label{119}
\end{equation}
Therefore,
\begin{widetext}
\begin{align}
\nonumber h(U_{circular}) &= \left( \sqrt{\frac{f(r_0)}{r_0^2}} \frac{\sqrt{1+r^2 U^{\phi 2}}}{\sqrt{f(r)}}-U^\phi \right) \sqrt{f(r)}\sqrt{1+r^2 U^{\phi 2}}\\
&=\left( \sqrt{\frac{f(r_0)}{r_0^2}}\sqrt{1+r^2 U^{\phi 2}}-\sqrt{\frac{f(r)}{r^2}} r U^\phi \right) \sqrt{1+r^2 U^{\phi 2}}.  \label{120}
\end{align}
and
\begin{equation}
\cos(\alpha_{circular})=\frac{\sqrt{\frac{f(r_0)}{r_0^2}-\frac{f(r)}{r^2}}+\sqrt{\frac{f(r_0)}{r_0^2}} r^2 U^{\phi 2}-\sqrt{\frac{f(r)}{r^2}} r U^\phi \sqrt{1+r^2 U^{\phi 2}}}{\left( \sqrt{\frac{f(r_0)}{r_0^2}}\sqrt{1+r^2 U^{\phi 2}}-\sqrt{\frac{f(r)}{r^2}} r U^\phi \right) \sqrt{1+r^2 U^{\phi 2}}}. \label{121}
\end{equation}
\end{widetext}
The effects of $\Lambda$, $m$, and the velocity component, $r U^\phi$, on the measured angle can be studied from the above relationship, which can be considerably simplified with some standard approximations. No assumptions were taken regarding the sign of $U^\phi$. A positive sign will mean that the observer and the deflected ray move in the same angular direction, a negative sign means the opposite. If we choose to refer to a static observer at the event of measurement, then equation \eqref{117a} gives
\begin{widetext}
\begin{equation}
\cos(\alpha_{circular})=\frac{\sqrt{\frac{f(r_0)}{r_0^2}} \left(\cos(\alpha_{static})-1 \right)}{\left( \sqrt{\frac{f(r_0)}{r_0^2}}\sqrt{1+r^2 U^{\phi 2}}-\sqrt{\frac{f(r)}{r^2}} r U^\phi \right) \sqrt{1+r^2 U^{\phi 2}}}+1. \label{122}
\end{equation}
\end{widetext}
The above relationship allows an investigation into how varying the value of $U^\phi$ increases or decreases the measurable angle $\alpha_{circular}$ relative to $\alpha_{static}$. We see how the terms $\frac{f(r_0)}{r_0^2}$ and $\frac{f(r)}{r^2}$ are of some fundamental importance in this kind of analysis. Most of the relationships of interest can be expressed using combinations of these terms. Notice that in places where these terms are being subtracted from one another we have a perfect cancellation of $\Lambda$. This fact is important to keep in mind when interpreting results or making approximations involving $\Lambda$. Some approximations may prevent this sensitive cancellation, causing terms of $\Lambda$ to appear where they do not belong, and ultimately lead to misinterpretations. This observation applies to all the specific observers discussed here.

Next, consider a radially moving observer. For this observer $U^\phi=U^\theta=0$, and the condition $U \cdot U=-1$ gives
\begin{equation}
U^t=\frac{\sqrt{f(r)+U^{r2}}}{f(r)}. \label{123}
\end{equation}
Therefore,
\begin{widetext}
\begin{align}
\nonumber h(U_{radial}) &= \left(-\sqrt{\frac{f(r_0)}{r_0^2}} \frac{\sqrt{f(r)+U^{r2}}}{f(r)}+ \sqrt{\frac{f(r_0)}{r_0^2}-\frac{f(r)}{r^2}}\frac{U^r}{f(r)} \right) \left(-\sqrt{f(r)+U^{r2}}+U^r \right)\\
&= \left(\sqrt{\frac{f(r_0)}{r_0^2}}\sqrt{1+\frac{U^{r2}}{f(r)}} - \sqrt{\frac{f(r_0)}{r_0^2}-\frac{f(r)}{r^2}}\frac{U^r}{\sqrt{f(r)}} \right) \left(\sqrt{1+\frac{U^{r2}}{f(r)}} - \frac{U^r}{\sqrt{f(r)}} \right), \label{113q}
\end{align}
\begin{equation}
\cos(\alpha_{radial})= \frac{\sqrt{\frac{f(r_0)}{r_0^2}-\frac{f(r)}{r^2}}+ \left(\sqrt{\frac{f(r_0)}{r_0^2}}+\sqrt{\frac{f(r_0)}{r_0^2}-\frac{f(r)}{r^2}}\right)\left(\frac{U^{r2}}{f(r)}-\frac{U^r}{\sqrt{f(r)}} \sqrt{1+\frac{U^{r2}}{f(r)}} \right) }{\left(\sqrt{\frac{f(r_0)}{r_0^2}}\sqrt{1+\frac{U^{r2}}{f(r)}} - \sqrt{\frac{f(r_0)}{r_0^2}-\frac{f(r)}{r^2}}\frac{U^r}{\sqrt{f(r)}} \right) \left(\sqrt{1+\frac{U^{r2}}{f(r)}} - \frac{U^r}{\sqrt{f(r)}} \right)}, \label{114q}
\end{equation}
and after some algebra,
\begin{equation}
\cos(\alpha_{radial})= \frac{\cos(\alpha_{static})+ \left(1+\cos(\alpha_{static}) \right) \left( \frac{U^{r2}}{f(r)}-\frac{U^r}{\sqrt{f(r)}} \sqrt{1+\frac{U^{r2}}{f(r)}} \right) }{\left(\sqrt{1+\frac{U^{r2}}{f(r)}} - \cos(\alpha_{static}) \frac{U^r}{\sqrt{f(r)}} \right) \left(\sqrt{1+\frac{U^{r2}}{f(r)}} - \frac{U^r}{\sqrt{f(r)}} \right)}. \label{115q}
\end{equation}
\end{widetext}
The above relationships can be used to study the effects of $\Lambda$, $m$, and the velocity component, $U^r$, on the measurable intersection angle. Inspecting these equations suggests that an increasing positive $U^r$ causes the measurable angle to increase, as one would expect in this setup. This observation may lead to a method of minimizing the relative experimental uncertainty coming from the measurement of the, usually small, angle. Minimizing such uncertainties is important when trying to establish a value of $\Lambda$ experimentally. Equations \eqref{114q} and \eqref{115q} are exact relationships. Together they demonstrate the additional effect of a radial velocity on the measurable intersection angle and the way in which this aberration phenomena may be taken advantage of in an experimental attempt of measuring $\Lambda$.

Lastly, let us consider a radially moving observer, located sufficiently far from the mass where its effects are completely negligible (outside the circle $\mathcal R$ on Figure \ref{split1}), and whose motion corresponds to the Hubble flow in de Sitter space, induced by $\Lambda$. Such conditions can model a realistic astrophysical setup; for example, where the source and the deflecting mass are distant galaxies, and together with the observer the three objects are separating due to the effects of a positive cosmological constant. The main assumption here is $r \gg 2m$, such that $\frac{2m}{r} \approx 0$, and the metric at the event of measurement is approximately that of de Sitter space.
\begin{equation}
\mathrm{d}s^2 = -f_{m=0}(r) \mathrm{d}t^2 + \frac{\mathrm{d}r^2}{f_{m=0}(r)}+r^2\sin^2(\theta)\mathrm{d}\phi^2 +r^2\mathrm{d}\theta^2, \label{116q}
\end{equation}
where
\begin{equation*}
f_{m=0}(r)=1-\frac{\Lambda}{3} r^2.
\end{equation*}
From equation \eqref{app:com} in the appendix, the 4-velocity of an observer moving according to Hubble flow, also referred to as a comoving observer, far away from the mass, in Kottler coordinates is
\begin{equation}
U^\alpha_{comoving} = \left( \frac{1}{f_{m=0}(r)},\sqrt{\frac{\Lambda}{3}}r,0,0 \right). \label{117q}
\end{equation}
Notice how in this case the velocity component, $U^r$, itself depends on $\Lambda$, as to be expected, since the motion of the observer is caused by $\Lambda$. Using the above in equation \eqref{114q} produces
\begin{equation}
\cos(\alpha_{comoving}) = \frac{\sqrt{\frac{f(r_0)}{r_0^2}-\frac{f_{m=0}(r)}{r^2}}-\sqrt{\frac{f(r_0)}{r_0^2}}\sqrt{\frac{\Lambda}{3}}r}{\sqrt{\frac{f(r_0)}{r_0^2}}-\sqrt{\frac{f(r_0)}{r_0^2}-\frac{f_{m=0}(r)}{r^2}}\sqrt{\frac{\Lambda}{3}}r}. \label{118q}
\end{equation}
The above equation is exact, given \eqref{117q}, and can be considerably simplified by making approximations related to the relative magnitudes of $\Lambda$, $m$, $r$ and $r_0$. Of course, due to the chosen orientation, the above relationship, as well as \eqref{114q}, can also be obtained by means of the usual aberration equation, \eqref{85}, and the expression for $\alpha_{static}$, \eqref{116a}. The required relative speed in the aberration equation can be obtained through the same method leading to equations \eqref{96c}. Notice that the effects of $\Lambda$ in this case come from both the geometry and the velocity of the observer. Whether a positive $\Lambda$ diminishes or increases the measurable angle for such an observer can be studied from the above equation, for this particular orientation of rays. To address this question in a more general setup, equation \eqref{79} can be employed to produce similar relationships to \eqref{118q} for any orientation of interest. Also notice that in the cosmological context, where the deflecting mass may be a distant galaxy, the values of the coordinate $r$ and the parameter $r_0$ are determined indirectly, and may or may not depend on $\Lambda$ themselves as well. In the simplest case, $r_0$ can be at the edge of the deflecting galaxy, and can be found from other existing methods or tabulated data on the particular galaxy. In other cases, $r_0$ must be determined from other boundary conditions, which depending on the model and coordinates used, may themselves depend on $\Lambda$ directly or necessitate the appearance of $\Lambda$ in their relation to $r_0$. Furthermore, in the cosmological context, in a realistic case where all measurements can only be done by an observer at one point (such as on Earth in our galaxy), the determination of $r$ and $r_0$ from such measurable quantities and the dependence of these measurements on $\Lambda$ are issues that, on their own, deserve a detailed investigation. In order to avoid deviating too far off course, this investigation, which makes extensive use of our formula \eqref{79}, was reserved for a separate report, \cite{ll}. For now, however, we can learn much from the results derived in this section on the influence of $\Lambda$ and investigate the ways in which its value can be determined experimentally from some measurements of angles. The relationships obtained in this section can be used to study the influence of different parameters on measurable angles and reveal many interesting results. Various experiments concerned with the determination of $\Lambda$ from angle measurements can be analyzed, and even suggested, by means of these relationships.\\

Finally, it is clear that the results derived in this section are indispensable for a general analysis, which involves finding measurable intersection angles of light rays in SdS space. Equation \eqref{79} is a general, mathematical, result, while equation \eqref{114} specifically applies to the $\theta=\frac{\pi}{2}$ slice of SdS space and a particular orientation of light rays. Of course, by means of equation \eqref{79}, we can generalize the expressions to two arbitrary light rays in the plane of motion, without constricting one of the rays to be radial. Even further, we can generalize to arbitrary light rays confined to two different planes. However, due to the popularity of the usual conditions that lead to equation \eqref{114}, let us summarize by restating equations \eqref{tp2e3} and \eqref{114}, which constitute the complete set of tools needed to analyze paths of light and associated measurable angles in SdS space.
\begin{equation}
\left(\frac{dr}{d\phi}\right)^2=r^4\left(\frac{f(r_0)}{r_0^2}-\frac{f(r)}{r^2}\right), \tag{\ref{tp2e3}}
\end{equation}
and
\begin{equation}
\cos(\alpha)=\frac{-\sqrt{\frac{f(r_0)}{r_0^2}}+\sqrt{\frac{f(r_0)}{r_0^2}-\frac{f(r)}{r^2}}}{h(U)}+1. \tag{\ref{114}}
\end{equation}
The fact that we chose to use the parameter $r_0$ in the above expressions makes them particularly useful in applications involving symmetric trajectories with a point of closest approach, which is by far the most popular case in the literature on the topic. However, the equations above are not limited to such situations. When there is no point of closest approach, $r_0$ can be replaced by the impact parameter $B$ (or some other parameter) in both equations. Although we have argued that in SdS spacetime it may be more appropriate to choose the parameter $r_0$ over $B$ in expressions, from a mathematical perspective the parameter $B$ is more general and its use may sometimes be necessary.

To be clear, a general analysis of the kind discussed above can be carried out from basic principles by means of the Euler-Lagrange equations and equation \eqref{79}. These two tools, together with some boundary conditions, are all that is needed for a complete analysis and can be used for any setup and any coordinates. For the specific case of the $\theta=\frac{\pi}{2}$ slice of SdS space in Kottler coordinates, the differential equation governing a trajectory of light, given by Euler-Lagrange equations, reduces to \eqref{tp2e3} and the expression for a measurable angle, given by equation \eqref{79}, with reference to a radial light ray, becomes \eqref{114}. Until recently, equation \eqref{tp2e3} was generally regarded as the main tool in investigating the influence of $\Lambda$, and measurable angles were mainly found through Euclidean methods justified in certain approximations. Rindler and Ishak's work promoted attention to other sources through which $\Lambda$ can influence mathematical results. The present work, however, is the first to introduce equation \eqref{114} to this topic, which now contains the necessary and sufficient tools needed to analyze the influence of $\Lambda$ on measurements correctly for any observer. Especially when investigating the influence of $\Lambda$ on measurable angles, it is clear that equation \eqref{tp2e3} on its own is not enough. Equation \eqref{114}, in some sense, brings the concept of measurement into the analysis, and as we've seen, this is where $\Lambda$ makes an entrance. Let us re-emphasise that although $\Lambda$ does not explicitly enter the analysis through the governing differential equation, \eqref{tp2e3}, it still influences the geometry through the metric which in turn affects measurements. This influence on the geometry is accounted for in the derivation of equation \eqref{114}, through which $\Lambda$ enters the analysis explicitly. Furthermore, in situations where $r_0$ is determined from boundary conditions that may depend on $\Lambda$, $\Lambda$ can enter the analysis through $r_0$ in both equations \eqref{tp2e3} and \eqref{114}. Additionally, as we've already seen, $\Lambda$ may also enter the analysis through the components of $U$, which are not all independent due to the normality requirement and may depend on $\Lambda$ themselves through other ways. The most important lesson here is that the influence of $\Lambda$ can come from various sources, making it hard to propose general conclusions on some important issues in this topic. The influence is sensitive to a particular situation that is being analyzed, and this allows for a various possibilities of how $\Lambda$ appears in results of interest.

The applications of the general formula for the intersection angles, \eqref{79}, extend well beyond light rays in SdS space. This formula is fundamental, in a geometrical sense, and coordinate independent. It may play a central role in many types of analysis, and can simplify things considerably. It also allows the generalization and provides another perspective of special relativistic aberration of light, and can be viewed as its general relativistic counterpart. As an additional application of the general formula, we utilized it to find expressions of cosmological distances analytically and to modifying the conventional analysis of weak gravitational lensing to account for $\Lambda$. We felt that the latter deserved to be the centre of a dedicated paper on the contribution of $\Lambda$ to the lens equation, \cite{ll}. In the present paper, however, we tried to concentrate on studying the influence of $\Lambda$ on the fundamental level, which is crucial to properly understand the recent debate on $\Lambda$'s effects on bending and intersection angles, which encouraged our investigation. Some important results that are derived in \cite{ll} are included in the appendix in order to be directly referred to in the next section, where we respond to some of the recent papers on the topic.

\section{Review of recent literature} \label{sec6}
In this section we respond to some of the recent papers on the topic and compare results of significance to the ones derived in the present work. We give a brief summary of each paper we respond to, and put it in the context of the previous sections. For a detailed examination of our comparisons, we encourage the reader to refer to the papers we discuss.
\subsection{W. Rindler and M. Ishak, 2007} \label{sec6a}
In this part we summarize and respond to the paper published by W. Rindler and M. Ishak in 2007, titled ``Contribution of the cosmological constant to the relativistic bending of light revisited", \cite{ri1}. Since then, the authors have published follow-up papers on the topic, \cite{rid} \cite{ri2} \cite{ridma} \cite{ishak1}, to which the following discussion applies.

In their paper, the authors begun by noting the work of Islam, \cite{islam}, and other papers that followed, and clearly stated that they agree with the accepted conclusion that $\Lambda$ drops out of the governing differential equation for path of light. Following this claim, they presented the key idea of their new approach: Actual observations depend on the geometry (metric) in addition to the orbit equation of a light ray, and when such effects are taken into account $\Lambda$ does contribute to results of interest. They start their analysis by describing the influence of $\Lambda$ on the geometry and qualitatively describe how this influence will contribute to measurements associated with light rays. They proceed by writing an approximate solution to first order in $m$ of the orbit equation in the ($r,\phi$) plane.

(eq. (9) of \cite{ri1})
\begin{equation}
\frac{1}{r}=\frac{\sin(\phi)}{R}+\frac{3m}{2R^2}\left( 1+\frac{1}{3}\cos(2\phi) \right). \label{ri1sol}
\end{equation}
Similar to the approach in chapter 11 of \cite{rindler} by Rindler, the authors orient the path so that $r=r_0$ at $\phi=\frac{\pi}{2}$, and chose the constant of motion $R$ as the parameter in the solution. The relationship between $R$ and $r_0$ is

(eq. (10) of \cite{ri1})
\begin{equation}
\frac{1}{r_0}=\frac{1}{R}+\frac{m}{R^2}. \label{ri:rR}
\end{equation}
Their $r, \:r_0, \:\phi$ and $m$ correspond exactly to ours of the previous sections. They note that other authors used the parameter $b$ in such discussions, but argued that while $b$ is meaningful in Schwarzschild space it is not the case in SdS space, which is not asymptotically flat. Next, the authors pointed out that while their solution equally applies to both Schwarzschild space and SdS space, only in the case of Schwarzschild space the bending angle can be found by letting $r$ go to infinity in the solution; in SdS space this limit makes no sense. This way of finding the bending angle corresponds to our definition 1 of section \ref{sec4b}, which we discussed in detail and compared to other definitions. The authors then explain the need for other angles in describing the deflection of a path of light. This is an issue to which we dedicated much attention ourselves, and is the main reason for including the detailed definitions of section \ref{sec4b} in the present work.

The authors then proceed by observing that a measurable angle is found correctly through the invariant formula

(eq. (11) of \cite{ri1}, in original notation)
\begin{equation}
\cos(\psi)=\frac{g_{ij}d^i \delta^j}{\sqrt{g_{ij}d^i d^j}\sqrt{g_{ij} \delta^i \delta^j}}.
\end{equation}
Here the metric tensor components, $g_{ij}$, are those of the line element \eqref{tp1e3} in our section \ref{sec2}, $d$ and $\delta$ are the tangents of the deflected ray and a radial ray, respectively, on the ($r,\phi$) plane, and $\psi$ is the measured angle. Notice that the above equation is identical to equation \eqref{40} of our section \ref{sec5a}. This is the key step in accounting for the contribution of the geometry to the measurable angle of interest, and this is precisely where $\Lambda$ pays its role. In fact, this step is what separates Rindler and Ishak's work from all the preceding attempts to investigate the influence of $\Lambda$ on measurements associated with light rays in SdS space.

With their solution to the deflected trajectory, \eqref{ri1sol}, they find an expression for $\frac{dr}{d\phi}$ and designate it by $\frac{dr}{d\phi}=A(r,\phi)$. This allows them to write an expression for the measurable intersection angle, $\psi$, as a function of $r$ and $\phi$ as follows,

(eq. (15) of \cite{ri1})
\begin{equation}
\cos(\psi)=\frac{|A|}{\sqrt{A^2+f(r)r^2}},
\end{equation}
and

(eq. (16) of \cite{ri1})
\begin{equation}
\tan(\psi)=\frac{\sqrt{f(r)}r}{|A|}=\sqrt{f(r)}\left| \frac{rd\phi}{dr}\right|, \label{ritan}
\end{equation}
where
\begin{equation}
f(r)=1-\frac{2m}{r}-\frac{1}{3}\Lambda r^2.
\end{equation}
Notice how equation \eqref{ritan} is identical to our equations \eqref{36} and \eqref{48} of section \ref{sec5a}. At this point, the authors did not make use of their approximate solution yet, which does not carry any terms of $\Lambda$. Thus, without the need of any approximations which the authors proceeded with, the main point of their argument is established by the expressions for the measurable angle $\psi$, where $\Lambda$ explicitly appears through $f(r)$. The authors then made a definition of the \textit{one-sided bending angle} as follows,
\begin{equation}
\epsilon=\psi-\phi. \label{ri1bend}
\end{equation}
Here $\epsilon$ is the one-sided bending angle, $\psi$ is the measurable angle with the radial and $\phi$ is the angular position coordinate of the observer. The reasoning for this definition comes from their Figure 2, the important features of which can be seen in our Figures \ref{split1}, \ref{split2} and \ref{fig10a}. This definition is similar to our definitions of $\beta_M$, the \textit{measurable deflection angle by a static observer}, of section \ref{sec4b}, and its Euclidean counterpart $\beta_E$. More on this in what follows. Finally, by using their approximate solution, \eqref{ri1sol}, the authors obtained explicit results for the specific cases of $\phi=0$ and $\phi=\frac{\pi}{4}$, under the assumption that $\epsilon$ is small, see equations (17) and (19) in \cite{ri1}. The results are expressed in terms of $R$, $m$ and $\Lambda$, which allows them to discuss the influence of $\Lambda$ and compare the newly defined bending angle to the case of Schwarzschild space.

Rindler and Ishak's approach to this topic is quite original and turns out to be very significant. They brought the concept of measurement into the picture and modified the current view regarding the influence of $\Lambda$. However, let us summarize the drawbacks that we find in the following three points. First, as we already stated, the use of an approximate solution is not needed for the main argument. The influence of $\Lambda$ on an important measurable quantity is clear from equation \eqref{ritan}. Moreover, there is no need to define a new parameter $R$ and use it in final results; this task is best fulfilled by the parameter $r_0$, which has a clear and useful geometrical interpretation. Second, the authors never address the question of which observer is making the measurement. In the context of the present work the answer is obvious, it is the static observer that is implicitly taken in all the expressions for measurable angles in \cite{ri1}. However, not mentioning it explicitly, in a way, hides the fact that measurable angles are observer dependent, and the influence of $\Lambda$ through the 4-velocity of the observer may be as important to study as the influence of $\Lambda$ through the metric itself. This lack of clarity, described by the latter point, may have been a cause for some arguments by other authors who responded to \cite{ri1}, see \cite{park} \cite{kp}. Lastly, upon closer examination of equation \eqref{ri1bend}, we find that the definition of the one-sided bending angle, $\epsilon$, is somewhat peculiar, in the following sense. On the right hand side of the equation, the angle $\psi$ is directly measurable, while the angle $\phi$ is purely Euclidean. In other words, $\psi$ belongs to the local frame of the particular observer, while $\phi$ is a Euclidean angle that belongs to a diagram on the ($r,\phi$) plane. This observation was not commented on in any of the preceding papers that respond to \cite{ri1}, whether in agreement or disagreement. Although not of major consequence, this definition of a bending angle leads to some problems. Let us discuss this issue in the context of our section \ref{sec4b} and write equation \eqref{ri1bend} in our notation. To this end, we refer to Figure \ref{fig10a} of section \ref{sec4b} and consider the angles in it. $\beta_E$ is the Euclidean angle between the bending trajectory and the vector $\frac{\partial}{\partial x}$, $\beta_{E1}$ is the Euclidean angle between a radial trajectory and the vector $\frac{\partial}{\partial x}$, and $\beta_{E2}$ is the Euclidean angle between a radial trajectory and the bending trajectory. Their measurable counterparts, by a static observer, are $\beta_M$ $\beta_{M1}$ $\beta_{M2}$, respectively. The ambiguity that arises with the vector $\frac{\partial}{\partial x}$ is dealt with in the precise definition of $\beta_M$ in section \ref{sec4b}. As explained, we have chosen the reference to the vector $\frac{\partial}{\partial x}$ due to the fact that at $r=r_0$ the tangent of the trajectory is parallel to this vector, and in this sense the angle between the trajectory and the vector $\frac{\partial}{\partial x}$ is a measure of the one-sided deflection. It seems that Rindler and Ishak followed similar reasoning in their definition. Now, in terms of the angles mentioned above, a measure of the deflection we are interested in is provided by either $\beta_E$ or $\beta_M$, of which only one is physically measurable, although both can be determined analytically. A straightforward way to find these angles is through $\beta_E=\beta_{E2}-\beta_{E1}$ and $\beta_M=\beta_{M2}-\beta_{M1}$, which is where the importance of the angles $\beta_{E1}$, $\beta_{E2}$, $\beta_{M1}$ and $\beta_{M2}$ comes in and why defining these angles is necessary. In the spirit of analyzing the effect of $\Lambda$ on measurements, we used the angle $\beta_M$ rather than $\beta_E$ in defining the deflection angle at a given point, and emphasised that it is measurable. Perhaps in the same spirit, Rindler and Ishak defined their one-sided bending angle, $\epsilon$, with reference to the measurable angle $\psi$. To compare our definitions, let us relate the angles that are used in defining $\epsilon$ to the angles of Figure \ref{fig10a}. Clearly, $\psi=\beta_{M2}$ and $\phi=\beta_{E1}$. Thus, using our notation we can define an identical deflection angle to the one defined by Rindler and Ishak as $\beta_{RI}=\beta_{M2}-\beta_{E1}$. To summarize, by using the angles $\beta_{E1}$, $\beta_{E2}$ and their measurable counterparts $\beta_{M1}$ and $\beta_{M2}$, we have defined three angular quantities that serve as a measure of the deflection of a light ray at a given point. These are:
\begin{equation}
\beta_E=\beta_{E2}-\beta_{E1} \qquad \text{(Euclidean deflection)}
\end{equation}
\begin{equation}
\beta_M=\beta_{M2}-\beta_{M1} \qquad \text{(Measurable deflection)}
\end{equation}
\begin{equation}
\beta_{RI}=\beta_{M2}-\beta_{E1} \qquad \text{(Rindler and Ishak's $\epsilon$)}
\end{equation}
It is not clear as to why Rindler and Ishak chose this particular definition. Mixing measurable and Euclidean angles makes it hard to interpret results and discuss their significance. The angle $\beta_{RI}$ itself is neither measurable nor does it appear on a diagram that depicts the situation being analyzed. Hence, the geometrical significance of the Euclidean angle $\beta_E$ and the physical significance of the measurable $\beta_M$ are absent in the hybrid angle $\beta_{RI}$. Notice, however, that in the special case of $\phi=0$, which leads to equation (17) of \cite{ri1}, our bending angle, $\beta_M$, and Rindler and Ishak's bending angle $\beta_{RI}$ are equal, since in this case $\beta_{E1}=\beta_{M1}=0$. This is the case when the measurement is taken at the point of symmetry, in the language of section \ref{sec4b}, for which we defined the angle $\alpha_M$. Therefore, while equation (17) of \cite{ri1} makes perfect sense, equation (19) of \cite{ri1}, obtained for the case $\phi=\frac{\pi}{4}$, must be interpreted with extra care and its usefulness is not immediately clear. Another problem with the definition of $\beta_{RI}$ is that paths which are straight lines on the ($r,\phi$) plane may have a non-zero bending angle. A few examples can be thought of to demonstrate this fact, the simplest of which is perhaps a trajectory of light for the special case $m=0$. Overall, this non-zero bending angle occurrence can be seen from the fact that while $\beta_M=0$ for the case of a straight line, the angle $\beta_{RI}$ becomes a difference between a Euclidean angle and its measurable counterpart, which in general is non-zero when the space is curved. Thus, in light of the discussion of section \ref{sec4b}, concerning the requirements of quantities that represent deflection angles in Schwarzschild, SdS and de Sitter spaces, we see that the quantity $\beta_{RI}$, originally $\epsilon$, does not meet some of our expectations.\\

Before ending our discussion of Rindler and Ishak's work, for the sake of later argument, let us quote some important results that we obtained by means of Rindler and Ishak's methods presented in \cite{ri1}. These results are derived in detail in \cite{ll}, where we investigate the contribution of $\Lambda$ to the lens equation.

The first order solution given by \eqref{ri1sol} is all that is needed to obtain the, well know, first order single source lens equation, eq. \eqref{lens}.
\begin{equation}
y=R_S \theta_E-\frac{4mR_{SL}}{R_L \theta_E}.
\end{equation}
Referring to Figure \ref{lensing} of the appendix, the above relationship serves as a map between the distance $y$ on the lensed plane and the angular position $\theta_E$ at the point of observation, in terms of $m$ and the Euclidean parameters $R_S$, $R_L$ and $R_{SL}$. See the appendix for more details. The above relationship can be modified by utilization of equation \eqref{ritan} (arguably the equation of most significance in \cite{ri1}) to replace the Euclidean parameters on the right hand side with measurable parameters. The result is a map between $y$ and the measurable position angle $\theta_M$, in terms of angular diameter distances, all measured by a static observer.
\begin{equation}
y=D_S \theta_M-\frac{4m(D_S-D_L)}{D_L \theta_M \sqrt{1+\frac{\Lambda}{3}D_L^2}}. \label{ri:stat}
\end{equation}
See the appendix for more details. The above agrees with our equation \eqref{app:stat}. Notice the presence of $\Lambda$ in the above equation, which came about due the use of measurable parameters. Next, equation \eqref{ri:stat} can be further modified by employing the standard aberration equation to convert the quantities that are measurable by a static observer to quantities that are measurable by a comoving observer, which has a relative velocity of $v=\sqrt{\frac{\Lambda}{3}}r$ and is moving in the radial direction. The result is a map between $y$ and the measurable position angle $\theta_M$, using angular diameter distances, all measured by a comoving observer.
\begin{equation}
y=D_S \theta_M-\frac{4m(D_S-D_L)}{D_L \theta_M \left( 1-\sqrt{\frac{\Lambda}{3}}D_L \right) }. \label{ri:lcom}
\end{equation}
The above agrees with our equation \eqref{app:como}. Notice the appearance of $\Lambda$ in the above equation and how it differs from \eqref{ri:stat}. Due to the assumption of comoving motion, the above can be regarded as a cosmological gravitational lens equation for a single source, and it is noteworthy that it was derived by means of Rindler and Ishak's methods of \cite{ri1} combined with the standard aberration equation. Much of the criticism of Rindler and Ishak's conclusions is based on the fact that with a positive $\Lambda$ the source, observer and lens should be in relative, comoving, motion, which is not accounted for in \cite{ri1}. We shall use this last result when responding to some of the comments made in \cite{park} in a later section.

\subsection{M. Sereno, 2008} \label{sec6b}
In this part we summarize and respond to some aspects of the paper published by M. Sereno in 2008, titled ``Influence of the cosmological constant on gravitational lensing in small systems", \cite{sereno1}. Since then, the author has published the follow-up papers \cite{sereno2} and \cite{sereno3} on the topic, to which the following discussion also applies. Although the author supports the conclusions of Rindler and Ishak, the analysis in \cite{sereno1} provides an example of the misuse of the parameter $b$, which leads to a questionable interpretation of results.

In this paper, the author begins with a brief introduction in which he mentions Rindler and Ishak's work, \cite{ri1}. He begins his analysis with the Kottler metric, our equation \eqref{tp1e1}, and proceeds to write down the orbital equation for a light ray in ($r,\phi$) space in terms of the parameter $b$ in integral from:

(eq. (3) of \cite{sereno1}, in original notation)
\begin{equation}
\phi_S=\pm \int \frac{dr}{r^2}\left[ \frac{1}{b^2}+\frac{1}{r_\Lambda^2}-\frac{1}{r^2}+\frac{2m}{r^3} \right]^{-\frac{1}{2}}. \label{ser:equ}
\end{equation}
Here $\phi_S$ is the $\phi$ coordinate of the source, and the integral is to be taken from the $r$ coordinate of the source, $r_S$, to the $r$ coordinate of the observer, $r_O$ (in the original notation). Also, the observer is assumed to be positioned at $\phi_O=0$, without loss of generality. The parameter $r_\Lambda=\sqrt{\frac{3}{\Lambda}}$. The above equation is equivalent to our equation \eqref{tp2e1}, which we have discussed extensively, and which can also be written in terms of the parameters $B$ and $r_0$ (in our notation). Although the author defined the parameters $b_\Lambda$ and $r_{min}$, which are identical to our $B$ and $r_0$, respectively, he never used either in the expression of his solution to the orbital equation. The advantages in using either $B$ or $r_0$ instead of the parameter $b$ are discussed in detail throughout our sections \ref{sec3} and \ref{sec4}. We have shown that the parameter $b$ cannot be considered independent of $\Lambda$, and its use in results can be misleading when investigating the influence of $\Lambda$.

The author then proceeds to write an approximate solution to his equation (3) (our \eqref{ser:equ} above), expended in orders of $\epsilon_m \equiv \frac{m}{b}$ and $\epsilon_\Lambda \equiv \frac{r_O}{r_\Lambda}$, which are both represented by $\epsilon$ for simplicity.

(eq. (5) of \cite{sereno1}, in original notation)
\begin{align}
\nonumber \phi_S= &-\pi-\frac{4m}{b}+b\left( \frac{1}{r_S}+\frac{1}{r_O}\right) -\frac{15m^2 \pi}{4b^2}-\frac{128m^3}{3b^3}\\
\nonumber &+\frac{b^3}{6}\left( \frac{1}{r_S^3}+\frac{1}{r_O^3}\right)-\frac{3465m^4 \pi}{64b^4}-\frac{3584m^5}{5b^5}-\frac{2mb}{r_\Lambda^2}\\
\nonumber &-\frac{mb^3}{4}\left( \frac{1}{r_S^4}+\frac{1}{r_O^4}\right)+\frac{3b^5}{40}\left( \frac{1}{r_S^5}+\frac{1}{r_O^5}\right)\\
&-\frac{b^3}{2r_\Lambda^2}\left( \frac{1}{r_S}+\frac{1}{r_O}\right)+O(\epsilon^6). \label{ser:sol}
\end{align}
Although it appears somewhat complicated, his solution is essentially a relationship between $\phi_S$ and $r_S$ in terms of $m$, $r_O$, $b$ and $\Lambda$. This relationship is a function that represents a set of points which constitute the path of a light ray in ($r,\phi$) space. In light of our investigation of section \ref{sec3}, and given the fact that the boundary conditions the author considers are purely coordinate-like, we know that the path of light connecting the source and observer is independent of $\Lambda$. In other words, the set of points in ($r,\phi$) space that constitute the path of a light ray does not depend on $\Lambda$. The appearance of $\Lambda$ in the authors solution is entirely due to his choice of using the parameter $b$, which itself depends on $\Lambda$. The perfect cancellation of the $\Lambda$ terms in equation \eqref{ser:sol} that one would expect when transforming $b$ to either $b_\Lambda$ or $r_{min}$ is completely hidden by the approximation taken. In fact, a solution to \eqref{ser:equ} can be written without $\Lambda$, even without the use of $b_\Lambda$ or $r_{min}$, since either of which can be expressed in terms of the mass $m$, and the boundary conditions ($r_S,\phi_S$) and ($r_O,\phi_O=0$), without invoking $\Lambda$. Thus, if done correctly and with no approximations on $\Lambda$ the solution to the orbital equation should not contain any terms of $\Lambda$ at all. This is in contradiction with the conclusion made by the author following his equation (5).

Although Sereno's conclusions seem to be in agreement with those of Rindler and Ishak, we see that Rindler and Ishak took a completely different approach to this topic. They acknowledged the work of Islam and that $\Lambda$ should not contribute to the orbital equation or its solution, and they brought $\Lambda$ into the analysis through considerations of measurements. Sereno, on the other hand, without considering measurements, brought $\Lambda$ into the orbital equation by using the parameter $b$. Moreover, his approximation masked the fact that $\Lambda$ can be transformed away from the equation by using a more appropriate parameter, such as $b_\Lambda$ or $r_{min}$. In fact, if we compare Sereno's solution, \eqref{ser:sol}, to Rindler and Ishak's solution, \eqref{ri1sol}, we see while $\Lambda$ appears in one it does not appear in the other, which is quite a major conceptual disagreement. Rindler and Ishak argued against the use of the parameter $b$, which led them to define their parameter $R$. The main point here is that when investigating the appearance of $\Lambda$ in relationships of interest, the choice of the parameters used in these relationships is crucial; the advantage in using parameters that are independent of $\Lambda$ themselves is obvious.

\subsection{A. Bhadra, S. Biswas and K. Sarkar, 2010}
In this part we summarize and respond to some aspects of the paper published by A. Bhadra, S. Biswas and K. Sarkar in 2010, titled ``Gravitational deflection of light in the Schwarzschild-de Sitter space-time", \cite{bbs}. The authors of this paper seem to support Rindler and Ishak's conclusions, but there are a number of issues we find with their analysis that we shall discuss.

The authors begin by presenting the idea that $\Lambda$ does affect the orbit of a photon, as well as the resulting bending angle; unfortunately, a common idea on that side of the argument, \cite{sereno1} \cite{bp}. The authors mention Rindler and Ishak's original work, \cite{ri1}, and briefly discuss the ongoing debate regarding their conclusions. The position they seem to take is that, in addition to what was found by Rindler and Ishak, there is more to the contribution of $\Lambda$, which comes from the orbital equation. They begin their analysis by sating the Kottler metric, our equation \eqref{tp1e1}, and the orbital equation in ($r,\phi$) space in terms of the parameter $b$, our equation \eqref{tp2e1}. In defining their $b$, which is identical to our $b$, they state that it behaves as the impact parameter at large distances, which is incorrect. The quantity $\left[ \frac{1}{b^2}+\frac{\Lambda}{3} \right]^{-\frac{1}{2}}$ ($=B$) is what actually behaves as the impact parameter at large distances, see our section \ref{sec4b}. For a solution to the orbital equation, the authors used the exact same approximation as in \cite{ri1}, and even used the same parameter $R$, see equation \eqref{ri1sol}. However, they claimed that, ultimately, the parameter $R$ must be replaced with $b$ and $\Lambda$, since it is $b$ and $\Lambda$ that appear in the first order orbital equation and carry meaning. The relationship between $R$, $b$ and $\Lambda$ can be easily obtained by plugging the solution \eqref{ri1sol} into the differential equation \eqref{tp2e1}, or simply by combining equations \eqref{tp2e1a} and \eqref{ri:rR}. In either case, we find
\begin{equation}
\sqrt{\frac{1}{b^2}+\frac{\Lambda}{3}}=\frac{1}{R}+O(\left( \frac{m}{R}\right) ^2). \label{bbs:bR}
\end{equation}
The above relationship is in disagreement with the one stated by the authors:

(eq. (6) of \cite{bbs})
\begin{equation}
\frac{1}{R}-\frac{m}{R^2}=\sqrt{\frac{1}{b^2}+\frac{\Lambda}{3}}. \label{bbs:Rb}
\end{equation}
The derivation of this equation is not explicit, so the source of error is not clear. Thus, in addition to proposing the use of $b$ and $\Lambda$ instead of $R$, the authors propose an incorrect relationship to make the transformation. Furthermore, the authors claim that by virtue of equations \eqref{bbs:Rb} and \eqref{ri:rR}, the parameter $r_0$ depends on $\Lambda$ as well.

The authors proceed to investigate the bending of the orbit, and define an appropriate deflection angle for light rays in SdS space. To this end, they utilized Rindler and Ishak's method and quoted the fundamental equation of their analysis in \cite{ri1}, our \eqref{ritan}, for the measurable angle by a static observer, $\psi$. They expressed this angle in terms of $r_0$, and approximated it to first orders in $m$ and $\Lambda$.

(eq. (11) of \cite{bbs})
\begin{equation}
\tan(\psi)=\frac{r_0}{r}+\frac{m}{r}-\frac{mr_0}{r^2}-\frac{\Lambda r_0 r}{6}+\frac{\Lambda r_0^3}{6r}. \label{bbs:tan}
\end{equation}
Next, following similar reasoning to that in \cite{ri1}, the authors defined the angle $\epsilon=|\psi-\phi|$, and expressed it by using \eqref{bbs:tan} and the approximate solution \eqref{ri1sol}, given small angles $\psi$ and $\phi$.

(eq. (12) of \cite{bbs})
\begin{equation}
|\epsilon|=|\psi-\phi|=\frac{2m}{r_0}-\frac{mr_0}{r^2}-\frac{\Lambda r_0 r}{6}+\frac{\Lambda r_0^3}{6r}. \label{bbs:eps}
\end{equation}
Notice that they chose to use $r_0$ in this expression, rather than either $R$ or $b$. Also, recall that the angle $\epsilon$, defined in this way, is a mixture of measurable and coordinate-like quantities.

Up to this point, other than the different treatment and interpretations of the parameters $b$ and $R$, the results of \cite{bbs} are in perfect agreement with those of \cite{ri1}. However, following their equation (12), the authors of \cite{bbs} explain that Rindler and Ishak's decision to put the observer at $\phi=0$ in the procedure of \cite{ri1} is not justified, and ultimately conclude that the angle should be expressed in terms of the arbitrary, but far from the origin, locations of the observer and the source, ($d_{OL},\phi_O$) and ($d_{LS},\phi_S$), respectively (in original notation). Their following result, which they call the \textit{total deflection angle}, is

(eq. (13) of \cite{bbs})
\begin{align}
\nonumber |\epsilon|=&\frac{4m}{r_0}-2mr_0 \left( \frac{1}{d_{LS}^2}+\frac{1}{d_{OL}^2} \right)\\
&-\frac{\Lambda r_0}{6}(d_{OL}+d_{LS})+\frac{\Lambda r_0^3}{6} \left( \frac{1}{d_{OL}}+\frac{1}{d_{LS}} \right). \label{bbs:13}
\end{align}
The above is a sum of two angles defined by \eqref{bbs:eps}, of which one represents the deflection of the ray as it goes from the source to $r_0$, while the other represents the deflection of the ray as it goes from $r_0$ to the observer. In a sense, it is a two sided $\beta_{RI}$ angle of section \ref{sec6a}, understanding the definition of which is key to the present discussion. Notice that $r_0$ in the above expression can be found from $d_{OL}$, $\phi_O$, $d_{LS}$ and $\phi_S$, without invoking $\Lambda$, which makes it somewhat of an unnecessary parameter in this situation. Replacing $r_0$ in terms of these boundary conditions will not change the appearance of $\Lambda$ in the expression. However, the authors set forth to replace $r_0$ with the parameter $b$, by approximating the exact relationship given by \eqref{tp2e1a} to first orders in $m$ and $\Lambda$.

(eq. (14) of \cite{bbs})
\begin{equation}
\frac{1}{r_0}-\frac{m}{r_0^2}=\frac{1}{b}-\frac{\Lambda b}{6}. \label{bbs:14}
\end{equation}
By using the above relationship they rewrite \eqref{bbs:13} in terms of $b$, to which, again, they incorrectly refer as the impact parameter.

(eq. (15) of \cite{bbs})
\begin{align}
\nonumber |\epsilon|=&\frac{4m}{b}-2mb \left( \frac{1}{d_{LS}^2}+\frac{1}{d_{OL}^2} \right) +\frac{2m\Lambda b}{3}\\
&-\frac{\Lambda b}{6}(d_{OL}+d_{LS})+\frac{\Lambda b^3}{6} \left( \frac{1}{d_{OL}}+\frac{1}{d_{LS}} \right). \label{bbs:15}
\end{align}
The above is their final expression for the \textit{total deflection angle}; it is expressed to first orders in $m$ and $\Lambda$. Hence, the expression in \eqref{bbs:15} is obtained by using a two sided angle $\beta_{RI}$, and bringing the parameter $b$ (and, consequently, its dependence on $\Lambda$) to the final result. This combines the problem we find with Rindler and Ishak's analysis in \cite{ri1}, and the problem we find with Sereno's analysis in \cite{sereno1}. Similar to the case of section \ref{sec6a}, and of no surprise, the deflection angle of equations \eqref{bbs:13} and \eqref{bbs:15} is non-zero for trajectories that are straight lines. Finally, in light of our own investigation in the previous sections, it is worth saying that in the analysis of \cite{bbs} the key contribution of $\Lambda$ comes from equation \eqref{bbs:tan} and should not come from equation \eqref{bbs:14} at all. As noted by the authors in a following paragraph, the difference between their results and the ones obtained in \cite{ri1} is primarily due to the fact that they included $\Lambda$ in the orbit equation as well, by making use of the parameter $b$.

\subsection{H. Arakida and M. Kasai, 2012}
In this part we summarize and respond to the paper published by H. Arakida and M. Kasai in 2012, titled ``Effect of the cosmological constant on the bending of light and the cosmological lens equation", \cite{ak}. The authors of this paper aim to clear up the confusion in the ongoing debate on the topic, which started following Rindler and Ishak's \cite{ri1}. The authors claim that $\Lambda$ does appear in the orbital equation of light and its solution, but does not contribute to the bending angle, due to its absorption into the impact parameter $B$. These conclusions seem to be in direct contradiction with those of Rindler and Ishak, who claimed the exact opposite. Let us discuss the analysis in \cite{ak} to clarify the reasons that led the authors to their conclusions.

The authors begin by solving the orbital equation for Schwarzschild space, which is identical in form to the one in SdS space, and which they later make use of in that case. Further, turning attention to SdS spacetime and working with the Kottler metric, they defined the parameters $b$ and $B$ in the same notation as ours. They recognized that, with $\Lambda \not= 0$, $B$ is the impact parameter rather than $b$, being the distance of closest approach with $m=0$ (see our definitions in section \ref{sec4b}). Their equation (10) is their orbital equation of light in SdS space, written in terms of $b$ and $\Lambda$. It is equivalent to our equation \eqref{tp2e1}. Upon stating this equation the authors emphasised that it ``obviously" includes $\Lambda$, and stated that arguments against this fact ``would be overstated". Next, by using the results earlier obtained for the case of Schwarzschild space, the authors state an approximate solution to \eqref{tp2e1} in terms of $B$.

(eq. (12) of \cite{ak}, in original notation)
\begin{align}
\nonumber \frac{1}{r}= &\frac{1}{B}\sin(\phi)+\frac{r_g}{4B^2}(3+\cos(2\phi))\\
&+\frac{r_g^2}{64B^3}(37\sin(\phi)+30(\pi-2\phi)\cos(\phi)-3\sin(3\phi)). \label{ak:12}
\end{align}
Here, $r_g=2m$. This solution assumes the particular orientation $\phi=\frac{\pi}{2}$ at minimum $r$, and it is correct to second order in $m$. The authors note that $\Lambda$ contributes to the trajectory, \eqref{ak:12}, as well as the orbital equation by virtue of the relationship between $B$ and $b$, \eqref{tp2e2}. This argument is, unfortunately, used in a few papers on the topic, in particular \cite{bbs}, and we've already discussed the problems it carries. For instance, even if $B$ was replaced in \eqref{ak:12} with $b$ and $\Lambda$, one could solve for $b$ by plugging any known point on the path into the relationship. Putting the resulting expression for $b$ back in \eqref{ak:12} will eliminate the appearance of $\Lambda$ in the equation completely. This is all due to the specific way in which $b$ and $\Lambda$ are 'connected', which was discussed in detail in section \ref{sec4}. Also, the authors stated that some previous approximate solutions, such as Rindler and Ishak's \eqref{ri1sol}, are incorrect, since they leave residual terms of second order in $m$ when put into the governing equation. This criticism cannot be justified, since Rindler and Ishak's solution, \eqref{ri1sol}, carries only first order terms in $m$ and it is an approximation that is correct only to this order, as clearly stated.

The authors proceed by writing an expression for their deflection angle, $\alpha$, in terms of $B$:

(eq. (13) of \cite{ak}, in original notation)
\begin{equation}
\alpha=2\frac{r_g}{B}+\frac{15\pi}{16}\left( \frac{r_g}{B}\right) ^2. \label{ak:13}
\end{equation}
This expression is obtained by taking the limit $r \rightarrow \infty$ in the solution, \eqref{ak:12}, with the assumption of small $\phi$. This angle corresponds to the bending angle, $\Phi$, we defined for SdS space in section \ref{sec4b}, and it is correct to second order in $m$. As discussed in that section, this quantity is purely mathematical and has nothing to do with actual measurements of angles, it appears on the flat diagram, such as Figure \ref{split1}, and serves as a measure of the bending of the path on the plane. Based on the form of the above relationship, the authors concluded that $\Lambda$ does not contribute to the deflection angle, since it is absorbed in $B$. This raises the question as to why do the authors draw their conclusions by considering $b$ more fundamental than $B$ in the orbital equation and its solution, while they stick to $B$ in making conclusions regarding the bending angle. In other words, the authors point out the appearance of $\Lambda$ when they use $b$, and the absence of $\Lambda$ when they use $B$. The choice of their preference of which parameter to use at which occasion is unclear, and in just the same way, opposite conclusions can be made by switching the use of these parameters. The choice of $B$ over $b$ in the solution by Rindler and Ishak, for example, led to the conclusion that $\Lambda$ has no influence on the orbit, as was also concluded by Islam and many others. (To first order in $m$, Rindler and Ishak's $R$ equals our $B$, see equation \eqref{bbs:bR}.) Next, in order to compare \eqref{ak:13} to previously derived results, including equation \eqref{bbs:15} of the previous section, the authors replaced $B$ with $b$, and expended the expression to lowest orders of $m$ and $\Lambda$.

(eq. (14) of \cite{ak}, in original notation)
\begin{equation}
\alpha \simeq \frac{4GM}{c^2 b}+\frac{2GMb\Lambda}{3c^2}. \label{ak:14}
\end{equation}
Here, $\frac{2GM}{c^2}=r_g=2m$. Although they point out some agreement that they find in their comparison, it is important to make a clear distinction between the method used to derive the above and the method used to derive \eqref{bbs:15}, for example. In deriving the above, no reference to any real measurements and any possible observers was made. The influence of $\Lambda$, therefore, comes only from the use of the parameter $b$. On the other hand, in deriving \eqref{bbs:15}, a truly measurable angle was considered ($\psi$), which brought in the contribution of $\Lambda$ through its influence on the geometry, introducing factors that cannot be transformed away. The reason for any similarities between the two equations is due to the use of $b$, and the appearance of $\Lambda$ that is carried with it, in both methods. Thus, one must be careful when interpreting and comparing such relationships. Overall, up to this point, the authors did not address real measurements at all, which is what sparked the whole debate on the influence of $\Lambda$. Two important points to take from this are that the choice of parameters affects the appearance of $\Lambda$ in results of interest (once again), and that the choice of parameters must be stated explicitly in order to avoid confusion and ambiguity when making final conclusions. It is also important to note that the particular way in which the bending angle, $\alpha$, was defined in \cite{ak} is exactly what Rindler and Ishak were trying to avoid in \cite{ri1} when extending the concept to light rays in SdS space, due to the conceptual problem with the limit $r \rightarrow \infty$. While Rindler and Ishak resorted to measurable angles, through which the contribution of $\Lambda$ was found, the authors of \cite{ak} showed that $\Lambda$ appears in results of interest only when using the parameter $b$. The authors then proceed with their investigation and also found that, in regards to the cosmological lens equation, the effect of $\Lambda$ is completely absorbed in an angular diameter distance; an issue to which the discussion of the next section applies, and which we address in full detail in \cite{ll}.

\subsection{M. Park, 2008}
In this part we summarize and respond to the paper published by M. Park in 2008, titled ``Rigorous approach to gravitational lensing", \cite{park}. The author of this paper takes a different approach to the topic at hand than the ones we've seen in the papers discussed above. Rather than concerning with the contribution of $\Lambda$ to quantities such as the bending angle, the author directly derived a cosmological lens equation that accounts for $\Lambda$ and the relative comoving motion between the observer, source and the massive object. Some of the results derived in \cite{ll} are central to our response to \cite{park}, which is the main reason for having them included in the appendix.

The author used an original method to analyze the standard setup of gravitational lensing by a single source. He ultimately derived the lens equation for a comoving observer in SdS space from first principles. The lens equation applies to a comoving observer in the sense that the measurable parameters that appear in the equation are measurable by this observer. Such an equation is useful in the cosmological context, where the objects involved are distant galaxies, for example. The author started his analysis from the McVittie metric, \cite{mv}, equation (1) in \cite{park}, and specialized it to SdS spacetime by setting all the cosmological parameters except $\Lambda$ to zero, resulting in a scale factor $a(t)=e^{Ht}$, with $H=\sqrt{\frac{\Lambda}{3}}$. He then transformed to more convenient spatial coordinates, which later allow him to express angular diameter distances in an easy way. Using these coordinates he approximated the components of the metric to first order in $m$, and expressed it as follows:

(eq. (7) of \cite{park}, in original notation)
\begin{align}
\nonumber \mathrm{d}s^2 = &-\left( 1-\frac{m}{\sqrt{(x+e^{Ht}q)^2+y^2+z^2}}\right) \mathrm{d}t^2\\
&+\left( 1+\frac{m}{\sqrt{(x+e^{Ht}q)^2+y^2+z^2}}\right)(\mathrm{d}\overrightarrow{x}-H\overrightarrow{x}\mathrm{d}t)^2. \label{park:7}
\end{align}
His spatial coordinates, ($x,y,z$), are centred on a point away from the massive object. In these coordinates, the origin is a point which can describe the location of comoving observer at any time $t$. The massive object (lens) is positioned on the $x$-axis, without loss of generality, and moves away from the origin in accordance to Hubble flow. His parameter $q$ is just an arbitrary constant associated to his transformation. It can be set by knowing the relative locations of the observer (at the origin) and the lens at a given time. Note that his time-like coordinate $t$ is different to our $t$ in the Kottler metric, \eqref{tp1e1}. Far from the mass, the $t$ in \eqref{park:7} coincides with the proper time of a comoving observer, which is the FRW time coordinate in that limit. Also note that his $m$ is twice that of our $m$ in all preceding discussion; we will make it clear when using our notation or the notation of \cite{park}.

Working to first order in $m$ and confining the motion of the photon to the $x-y$ plane ($z=0$), the author formed a diagram describing the lensing setup, and found the trajectory of a light ray, satisfying the required boundary conditions. See Figure 1 in \cite{park}, which is similar to our Figure \ref{lensing} in the appendix. He proceeded to write an expression for the intersection angle $\theta$ at the origin, between the light ray coming from the source and the light ray coming from the lens, equation (26) in \cite{park}. Since this angle occurs at the origin on his diagram, by the construction of his spatial coordinates, it is equivalent to the measurable angle by an observer located at the origin, a comoving observer in the FRW sense. This allowed the author to establish the cosmological lens equation.

(eq. (29) of \cite{park}, in original notation)
\begin{align}
\nonumber \theta= &\beta+\frac{2m}{\beta d_S d_L} \{ x_S-d_L+Hd_L(x_S-d_L)\\
&+H^2d_L^2(x_S-d_L)+O(H^3)+O(\beta^2)\} +O(m^2). \label{park:29}
\end{align}
In this equation, the distance-like parameters $d_L$ and $d_S(=x_S+O(\beta^2))$ are angular diameter distances, measured by the observer at the origin. They precisely correspond to the coordinate distances used in the derivation, which explains the author's choice of transformation. Hence, the author does account for measurements by virtue of choosing his coordinates such that some Euclidean angles and coordinate distances that appear on the diagram are equivalent to some important measurable angles and distances that are needed to express final results. Note that this method of incorporating measurable quantities into the analysis can only work for a comoving observer, in a region far from the mass where its effects are completely negligible. The angle $\beta$ in the above equation is the undeflected position angle that the observer would measure in the absence of the mass.

Let us put equation \eqref{park:29} in the notation of the appendix by transforming the parameters accordingly.
\begin{equation*}
\theta \rightarrow \theta_M, \qquad \beta \rightarrow \frac{y}{D_S},
\end{equation*}
\begin{equation*}
m_{(Park)} \rightarrow 2m, \qquad H \rightarrow \sqrt{\frac{\Lambda}{3}},
\end{equation*}
\begin{equation*}
d_S \rightarrow D_S, \qquad d_L \rightarrow D_L,
\end{equation*}
\begin{equation*}
x_S=d_S+O(\theta^2) \rightarrow D_S+O(\theta_M^2).
\end{equation*}
To first order in $m$ and $\theta_M$, equation \eqref{park:29} written in our notation is:
\begin{align}
\nonumber \theta_M &=\frac{y}{D_S}\\
&+\frac{4m}{y D_L}(D_S-D_L)\left( 1+\sqrt{\frac{\Lambda}{3}}D_L+\frac{\Lambda}{3}D_L^2+O(\Lambda^\frac{3}{2})\right).  \label{park:1}
\end{align}
This equation can be solved for $y$ and compared to the relationships stated in the appendix. Again, to first order in $m$ and $\theta_M$, we find
\begin{align}
\nonumber y &=D_S \theta_M\\
&-\frac{4m(D_S-D_L)}{\theta_M D_L}\left( 1+\sqrt{\frac{\Lambda}{3}}D_L+\frac{\Lambda}{3}D_L^2+O(\Lambda^\frac{3}{2})\right). \label{park:2}
\end{align}
The above is the cosmological gravitational lens equation, expressed entirely in terms of directly measurable parameters; it assumes the measurements are taken by a comoving observer. This equation is in perfect agreement with our equation \eqref{app:comoa}, which is an approximation of equation \eqref{app:como}, obtained by series expansion in $\Lambda$. This leads us to conclude that Park's result is correct to the highest order of his approximation. It is worth noting that our approach in deriving \eqref{app:como} in \cite{ll} is significantly different than the method used by Park to derive \eqref{park:29}. It is reassuring to see completely diverse procedures lead to identical final result.

However, following the establishment of equation \eqref{park:29}, the author set to replace some appearances of the distances $x_S$ and $d_L$ in the equation with the distance $d_{SL}$ (in his notation). $d_{SL}$ is the angular diameter distance from the source to the lens, it corresponds exactly to our $R_{SL}$ in the appendix; in principle it could be measured directly by an observer at the source or at the location of the lens. Hence, $d_{SL}$ is a measurable quantity, but the observer that can measure it must be located away from the assumed point of observation. Thus, if all observations are assumed to be taken at a single point, as in the cosmological context, then the angular diameter distance $d_{SL}$ must be determined indirectly, from other measurements.

In order to include $d_{SL}$ in \eqref{park:29}, the author used the relationship given by his equation (30) in \cite{park}, which is equivalent to our equation \eqref{app:dls} in the appendix. His final result is

(eq. (31) of \cite{park}, in original notation)
\begin{equation}
\theta= \beta+\frac{2md_{SL}}{\beta d_S d_L}(1+O(H^3)+O(\beta^2)) +O(m^2), \label{park:31}
\end{equation}
which in our notation, to first order in $m$ and $\theta_M$, is
\begin{equation}
\theta_M=\frac{y}{D_S}+\frac{4mR_{SL}}{y D_L}\left( 1+O(\Lambda^\frac{3}{2})\right). \label{park:5}
\end{equation}
Solving the above for $y$, we find, to first order in $m$ and $\theta_M$,
\begin{equation}
y=D_S \theta_M-\frac{4mR_{SL}}{\theta_M D_L}\left( 1+O(\Lambda^\frac{3}{2})\right). \label{park:6}
\end{equation}
Again, the above equation is in perfect agreement with our results, which can be seen by using equation (A.13) to include $R_{SL}$ in our cosmological lens equation \eqref{app:como}. In fact, since our results are exact in $\Lambda$, we see that if Park were to work with any higher order terms of $\Lambda$, he would have found that all these terms would be zero in his approximation as well. Notice how $\Lambda$ gets thoroughly absorbed into the angular diameter distance $R_{SL}$. Thus, only when expressing the lens equation entirely in terms of the angular diameter distances $D_S$ and $D_L$ does $\Lambda$ make an appearance; an appearance that can be completely transformed away by using the angular diameter distance $R_{SL}$. Clearly, given the relationship between $R_{SL}$, $D_S$ and $D_L$ (equation (A.13)), using only two of the three parameters is enough to express any result of interest. This raises the following question: which parameters should be used in expressing the cosmological lens equation? Or more specifically: should the parameter $R_{SL}$ be used at all? Of the three parameters $D_S$, $D_L$ and $R_{SL}$, only $D_S$ and $D_L$ are directly measurable at the assumed point of observation. And although $R_{SL}$ can be found indirectly from other measurements that can be made at the point of observation, the value of $R_{SL}$ can be established only with knowledge of $\Lambda$ (as in equation (A.13), for example). With this in mind, we can address the above question by considering two possible cases in which the lens equation may be used.

First, in a case where all the parameters of interest, such as the three $D_S$, $D_L$ and $R_{SL}$, are available from some tabulated data or another source, one can use the lens equation in either form, with or without $R_{SL}$. In this case using $R_{SL}$ in the cosmological lens equation is preferable, since it simplifies the expression. This will allow the predictions of images and masses by means of the lens equation, but will not allow studying the effects of $\Lambda$ on measurable quantities directly, which are completely absorbed in $R_{SL}$. Then, although $\Lambda$ will not appear in the lens equation, if it is to be accounted for, its value must still be used at some point to establish the tabulated data, specifically the value of $R_{SL}$. Thus, we see that the lack of appearance of $\Lambda$ in a relationship does not necessary imply its lack of influence on the phenomenon being studied.

Second, in a case where no pre-recorded parameters are available, it is clearly advantageous to use parameters that are measurable directly in the cosmological lens equation. Therefore, in this case, equation \eqref{app:como} (or \eqref{park:2}) is preferable, in which $\Lambda$ appears explicitly and its influence on measurable quantities can be studied directly. In short, we see that $\Lambda$ has an effect on the cosmological lens equation in any case, and needs to be accounted for directly or indirectly. This should be kept in mind when choosing parameters in the expression of the cosmological lens equation and making any conclusions. Then, using or not using the angular diameter distance $R_{SL}$ in the final expression is really a matter of preference in a given situation.

Further, following his equation (31) in \cite{park}, the author states that ``[his] result is in contradiction to the recent claims by \cite{ri1} which assert that there should be a $O(\Lambda)$ correction to the conventional lensing analysis". This statement is somewhat inequitable, since in \cite{ri1} Rindler and Ishak never concern with the gravitational lens equation directly, and consider a setup that is quite different, for which they produce results applicable only to a static observer. Later in his discussion, the author explains that the disagreement between his and Rindler and Ishak's results may be due to the following two problems:
\begin{enumerate}
\item The setup in \cite{ri1} is not realistic, since they consider a static observer and neglect the relative comoving motion between the observer, source and lens.
\item The relationships in \cite{ri1} are not expressed in terms of angular diameter distances, which is necessary for comparison with conventional results.
\end{enumerate}
He then explained that in their follow-up paper \cite{ridma}, they failed to address these problems properly, and suggested that it is possible to modify their existing results for an appropriate comparison. He pointed out that using relativistic aberration to modify their results can help resolve the first problem, but converting parameters to angular diameter distances could be tricky, which, as he explains, makes his approach favourable. In a paper published by Ishak et. al. in 2010, \cite{rid}, the authors argued that the apparent disagreement between the conclusions of \cite{ri1} and those of Park can be due to the fact that Park dropped terms of order $\beta^2$ from his final result, equation \eqref{park:31}, which carry terms of $\Lambda$. However, to properly compare the results of \cite{ri1} and \cite{park} we have used the method in \cite{ri1} to derive a lens equation subject to the same conditions as in \cite{park}, and found perfect agreement. More on it bellow, recall the end of section \ref{sec6a}.

Much of the analysis of \cite{ll} involves finding relationships between measurable and coordinate-like distances. We found that the methods presented in \cite{ri1} allow for converting a coordinate distance to the angular diameter distance measured by a static observer. This finding allowed for the derivation outlined at the end of section \ref{sec6a}. Equation \eqref{ri:lcom} is a cosmological lens equation, which we derived through Rindler and Ishak's methods and the standard aberration equation. It accounts for the effects of $\Lambda$ on the geometry and the relative comoving motion, induced by $\Lambda$, between the observer, source and lens. The distance-like and angular quantities on the right side of equation \eqref{ri:lcom}, as well as on the right side of equation \eqref{park:2}, are all measurable by a comoving observer. Since equation \eqref{ri:lcom} agrees with our \eqref{app:como}, which agrees with equation \eqref{park:2}, we find perfect agreement between Park's result and the one we've obtained through Rindler and Ishak's methods. Although the two methods are quite different, when done correctly they produce identical results. Finally, let us re-emphasize that it should not be concluded from Park's results that the influence of $\Lambda$ on the cosmological lens equation is of $O(\Lambda^\frac{3}{2})$ or higher. In fact, what Park found, as we did as well, is that there is a term of $\Lambda^\frac{1}{2}$ in the lens equation, when considering a comoving observer. This is an important fact when comparing it to the lens equation for a static observer, for which we found through our methods, as well as Rindler and Ishak's methods, that the lowest order $\Lambda$ term that appears is $\Lambda^1$. Hence, given the investigations of our previous sections we were able to make a clear comparison between the results and conclusions in \cite{ri1} and \cite{park}.

\subsection{I. B. Khriplovich and A. A. Pomeransky, 2008}
In this part we summarize and respond to the paper published by I. B. Khriplovich and A. A. Pomeransky in 2008, titled ``Does the cosmological term influence gravitational lensing?", \cite{kp}. The results of this paper are often referred to in arguments against the conclusions of \cite{ri1}. The authors of this paper used both the Kottler metric and the FRW metric, equation (8) in \cite{kp}, to investigate the appearance of $\Lambda$ in a given expression of interest. Far away from the mass, the Kottler metric is well approximated by the de Sitter metric, which is equivalent to the FRW metric with a scale factor $a(t)=e^{Ht}$ ($H=\sqrt{\frac{\Lambda}{3}}$). By arriving at specific relationships through both the use of de Sitter coordinates and FRW coordinates separately, the authors compared the contribution of $\Lambda$ in the two different cases, and made conclusions based on this comparison.

The authors begun their analysis by considering the invariant $g^{\mu \nu}k_{1\mu}k_{2\nu}$, where $k_1$ and $k_2$ are tangents of two intersecting null geodesics. They designate the positive root of this invariant by $I$. In a local frame of some observer, it is trivial to show that for a small intersection angle between the light rays the invariant $I$ can be expressed (up to a factor of $2$) as

(eq. (1) of \cite{kp}, in original notation)
\begin{equation}
I=\omega \theta. \label{kp:1}
\end{equation}
Here, $\omega$ and $\theta$ are the energy of the photons and the intersection angle between them, respectively, that the observer measures. This equation can be easily obtained from the first order in angle approximation of our equation \eqref{79}, keeping in mind equation \eqref{energy}, and its true for any observer as long as $\theta$ is small. It is assumed here that the two intersecting photons are of the same energy. Note that the quantities appearing on the right hand side of the above equation are directly measurable, and their values are observer dependent, while the quantity on the left hand side of the equation is a constant for the particular intersecting trajectories. For different observers, the measurements of $\theta$ and $\omega$ shift accordingly, so that their product always remains the same.

The authors first considered the standard setup of gravitational lensing in Kottler coordinates. See Figure 1 in \cite{kp}, which is similar to our Figure \ref{lensing} in the appendix. After approximating the solution to the orbital equation of light, far away from the mass, the authors express the measurable intersection angle, $\theta$, between the bending trajectory and a purely radial trajectory, in these coordinates:

(eq. (6) of \cite{kp}, in original notation)
\begin{equation}
\theta=\frac{d\phi \sqrt{|g_{\theta \theta}|}}{dR \sqrt{|g_{RR}|}}=\frac{\rho}{R\sqrt{|g_{RR}|}}=\theta_0\sqrt{1-\lambda^2 R^2}. \label{kp:6}
\end{equation}
Here $\theta_0$ is the Euclidean intersection angle appearing on their diagram, their $\lambda$, $R$ and $\rho$ are equal to our $\sqrt{\frac{\Lambda}{3}}$, $r$ and $r_0$ of the previous sections, respectively. Note that the subscripts of the metric component $g_{\theta \theta}$ in the above equation as well as in Figure 1 of \cite{kp} are most likely a mistype, this component should be $g_{\phi \phi}$. We immediately recognize that the above relationship refers to a static observer in Kottler (or de Sitter) coordinates. This relationship is in perfect agreement with Rindler and Ishak's main result of \cite{ri1}, equation \eqref{ritan}, and of course with our results of section \ref{sec5}. This equation is the most basic example of a relationship between a measurable angle and a Euclidean angle that appears on a flat plane, on which a diagram of the setup is drawn. Notice that the solution to the orbital equation is not necessary to form this particular relationship. It is also important to note that the main reason for this agreement between the results is due to the fact that the same static observer is involved in both approaches, which is unfortunately not specifically stated in neither \cite{ri1} nor \cite{kp}. With the above expression for $\theta$, the authors proceeded to express the invariant $I$ as follows:

(eq. (7) of \cite{kp}, in original notation)
\begin{equation}
I=\omega_{dS}\theta_0\sqrt{1-\lambda^2 R^2}. \label{kp:7}
\end{equation}
Here, the subscript of $\omega_{dS}$ refers to the fact that the analysis is carried out with de Sitter coordinates. It should be clear that given the fact that it is the static observer that is involved in the angle measurement, $\omega_{dS}$ is the energy that is measured by a static observer as well. Evidently, Kottler (or de Sitter) coordinates were employed in this paper merely in order to form relationships for a static observer in SdS space; to form the same relationships for a different observer the authors employed other coordinates, as we discuss below. It is not perfectly clear as to why the authors chose to use the Euclidean angle $\theta_0$ in the expression for $I$ above, and what purpose this expression serves. Since the energy $\omega$ has no obvious non-measurable counterpart, it is only $\theta$ that can be switched around with its Euclidean counterpart, $\theta_0$, in the expression for $I$. As should be abundantly clear by now, a relationship between such measurable and Euclidean angles should always involve $\Lambda$ when working in Kottler (or de Sitter) coordinates. Thus, when a given expression involves one of the angles $\theta$ or $\theta_0$, but does not involve $\Lambda$, by replacing the angle involved with its counterpart $\Lambda$ is forced into the expression. The reason for choosing one angle over the other as a parameter in a given expression should always be clarified before drawing any conclusion from the expression. It is often advantageous to express some relationship with purely measurable parameters or, conversely, with purely Euclidean (or coordinate-like) parameters. The expression for $I$ above mixes measurable and Euclidean parameters with no satisfactory reason.

Next, the authors proceeded their investigation by employing FRW coordinates to produce an expression for the invariant $I$ with reference to a comoving observer. Far away from the mass, the Kottler metric is well approximated by the de Sitter metric, which is equivalent to the FRW metric with the scale factor $a(t)=e^{\lambda t}$ (in the notation of \cite{kp}). See equations (8) and (9) in \cite{kp}. In that region of space, a comoving observer is simply an observer with constant FRW spatial coordinates, and it is in this way, as recognized by the authors of \cite{kp}, it is easy to produce results for this observer by using the FRW metric. Through the use of this metric the authors find:

(eq. (16) of \cite{kp}, in original notation)
\begin{equation}
I=\omega_{FRW}\frac{\rho}{r_0}. \label{kp:16}
\end{equation}
Here, the subscript of $\omega_{FRW}$ refers to the fact that the analysis is carried out with FRW coordinates. The authors argue that these coordinates are the most appropriate for the description of observations, but given the tools of our section \ref{sec5}, we recognize that these coordinates are simply convenient to use when dealing with comoving observers. Identical results can be obtained with any equivalent metric as long as the observer is the same, and its 4-velocity is transformed appropriately and accounted for in the derivation. The parameter $r_0$ in the above equation is not the same $r_0$ that was used in the previous sections. This $r_0$ is the constant FRW coordinate distance between the comoving observer and the lens, while the distance of closest approach to the lens, with reference to areal radius coordinate, is represented by $\rho$. (In the FRW sense, $\rho$ is the 'distance', which is the coordinate separation multiplied by the scale factor, at the time of closest approach of the photon to the lens.) The proper interpretation of $\rho$ in this context deserves further attention, but we shall not digress into it here. Since $\omega_{FRW}$ is the measurable energy by a comoving observer, the quantity $\frac{\rho}{r_0}$ in \eqref{kp:16} equals the intersection angle that is measurable by this observer as well. As before, the choice of parameters in the above expression for $I$ as well as its purpose are not perfectly clear, and we see a mix between measurable and non-measurable quantities. Note that to arrive at the above equation one simply needs to express the measurable intersection angle, $\theta$, appearing in \eqref{kp:1} as $\frac{\rho}{r_0}$, which can be easily done by drawing the diagram of the lensing setup with reference to FRW coordinates. In fact, the solution to the orbital equation is not needed to find the required expression. And finally, although $\Lambda$ does not explicitly appear in the above expression for $I$, it does not tell us anything about its influence on measurements of angles or about its possible appearance in other relationships of interest. This absence of $\Lambda$ in the above expression, in contrast to its appearance in equation \eqref{kp:7}, seems to be wrongfully interpreted throughout the literature.

It is clear that with our general formula for the measurable angle, equation \eqref{79}, we can easily produce results by using any coordinates for any observer. It saves the trouble of transforming to a specific coordinate system merely to consider the measurement of a specific observer, as was done by the authors of \cite{kp} and \cite{park}, for example. Although the authors of \cite{kp} did consider measurements by both static and comoving observers, neither the bending angle in SdS space nor the lens equation were specifically addressed. And while they also touched up on the actual trajectory of light, see equations (3-5) and (18) in \cite{kp}, which led them to define the parameter $\rho$, they did not really need these relationships to establish their ultimate results, equations \eqref{kp:7} and \eqref{kp:16}. The quantities $\theta$ and $\omega$ in \eqref{kp:1} are directly measurable and local, and as long as the intersection angle at the point of observation is small, the rest of the trajectories does not matter. Clearly, it also does not matter what metric one chooses to work with if the metrics are equivalent. The authors of \cite{kp} decided to use the FRW metric, far away from the mass, merely to consider measurements made by a comoving observer. In this sense, in the cosmological context, these coordinates are the ones that are more appropriate to describe measurements, as they claim. However, let us re-emphasize that from the results of \cite{kp} it cannot be concluded that $\Lambda$ has no effect on gravitational lensing; more specifically, it is incorrect to reason that the results of \cite{kp} imply the non-contribution of $\Lambda$ to the cosmological lens equation.

\section{Discussion}
In a universe with a cosmological constant, the space outside a spherically symmetric non-rotating mass is well described by the Kottler metric. With the recent increasing interest in the cosmological constant, SdS spacetime became a popular background for investigating the various effects of gravity. A natural way to study the effects of $\Lambda$ is to revisit the classical tests of general relativity. One of the most popular predicted phenomena associated with such tests is the deflection of light by a massive object. It was a long time ago that the question of whether or not $\Lambda$ plays a roll in this phenomenon has been asked, but unfortunately until this day this topic seems to be suffering from misconceptions and disagreements. We see that the answer to the above question is not simply in the positive or negative, but is very sensitive to the particular situation that is being considered. In the course of the ongoing investigation it became clear that in order to properly address the above question one must consider both the geometry of the underlying space and the act of observation by a given observer, on top of the orbital equation for a light ray and its solution. It is mainly due to the work of Islam, \cite{islam}, that it was generally agreed upon that $\Lambda$ has no affect on the orbit of a light ray, as acknowledged by Rindler and Ishak in \cite{ri1}. And it is due to the findings of Rindler and Ishak in \cite{ri1} that it was realized by many that real measurements must be considered as well in investigating the contribution of $\Lambda$. Given that the previous investigations and conclusions by Islam in \cite{islam} and Rindler and Ishak in \cite{ri1} are correct, what we have done in the present work is address the following question: in what way do results and expressions of interest depend on which observer is making the measurement? In other words, since according to Islam the path of light is not affected by $\Lambda$, and according to Rindler and Ishak the measurement of an angle is affected by $\Lambda$, the circumstances naturally leads to the question above. Investigating this question in detail led us to the results of section \ref{sec5}, the most important of which are not found in the literature and are fundamental to the topic at hand.

We have begun our investigation from fundamental considerations, and revisited the original issue of whether or not $\Lambda$ affects the path of a light ray itself in section \ref{sec3}. It was found that the dependence of a path on $\Lambda$ was entirely involved in the boundary conditions that are being used in a given situation. Evidently, whether or not $\Lambda$ enters the orbital differential equation does not matter, due to the particular way in which its appearance can be entirely absorbed into a new parameter. Specifically, even if the orbital equation is written in terms of a parameter (such as $b$) that brings in a term of $\Lambda$ with it, this term of $\Lambda$ will vanish from the solution completely when certain boundary conditions are enforced. Such boundary conditions are purely Euclidean, or rather coordinate related, which are the most popular in the literature and most appropriate in common situations; for such boundary conditions varying the value of $\Lambda$ would not affect the set of points through which the light passes. For this reason, we recognized that it is acceptable to conclude, but with caution, that $\Lambda$ does not affect the path of light and best not be used in the orbital equation. However, it is also important to understand that when considering directly measurable quantities as boundary conditions, $\Lambda$ usually enters the equation describing the path. In addition, of course, in situations where the boundary conditions themselves depend on $\Lambda$ directly, $\Lambda$ will also appear in the equation describing the path. An important lesson here is that the contribution of $\Lambda$ to results of interest depends closely on the situation being analyzed, and any general conclusions should be drawn carefully. Our investigations illuminate many possible sources of confusion and misinterpretation regarding this issue, which unfortunately seem to have had a great affect on recent literature.

Let us re-emphasise that perhaps the most important result of this work is equation \eqref{79}. It opens up a way to a more general analysis and is essential to properly investigate the effects of $\Lambda$ on measurable angles. In addition, it allows for an elegant approach to many situations when analyzing gravitational lensing, and yields an invariant general relativistic aberration equation. It is interesting to note that in some recent papers, such as \cite{park} and \cite{kp}, the authors used a transformation of coordinates in order to be able to find a measurable angle by a given observer. It seems that trying to express a measurable angle by an arbitrary observer in an analytic, and coordinate independent, way is generally avoided in the literature. Often, the coordinate transformations that make it easy to express a given measurable angle abandon the use of spherical symmetry and complicate the overall analysis considerably, see \cite{park}. This undesirable consequence and other complications that a coordinate transformation may bring can be easily avoided by working with the general formula \eqref{79}; it can be put to use in any coordinate system and produce results related to any observer of interest. More on this in \cite{ll}, where we demonstrate the latter point in the context of weak gravitational lensing, and compare results obtained by means of equation \eqref{79} to results obtained by means of a coordinate transformation (as was done in \cite{park}, for example).

In addition to the papers discussed in section \ref{sec6}, there are other papers on the topic that are worth looking at, including \cite{schucker1}, \cite{kcd}, \cite{simpson}, \cite{bp}, \cite{mirag}, \cite{lake2} and some references therein. Although our responses to some of these papers are not included in the present report, the material we presented here is useful in understanding and interpreting their results, and it is of fundamental importance for making proper comparison of the different conclusions the authors arrive to. It is also worth mentioning that when studying the effects of $\Lambda$, approximations on $\Lambda$ should be avoided or made with care. Due to the sensitive way in which $\Lambda$ vanishes from exact results, within a given approximation $\Lambda$ may end up appearing in relationships where it does not belong. And although such an approximation might be justified, due to the smallness of $\Lambda$ or some other parameter, and may be numerically accurate, this appearance of $\Lambda$ in resulting relationships may be theoretically misleading; see \cite{sereno1} and our section \ref{sec6b}.

Finally, we hope that the material presented in this work will provide a proper perspective when addressing questions regarding the influence of $\Lambda$, and that it will aide in gaining a clear understanding of, and ultimately settling, the recent debate on the topic.

\appendix
\section{Additional results}
The relationships that are stated below are derived in \cite{ll}, where we turn attention to the role of $\Lambda$ in cosmological distance measurements and the gravitational lens equation. The following figure is referred to in the definitions and the relationships below.
\begin{figure}[!ht]
\includegraphics[width=85mm]{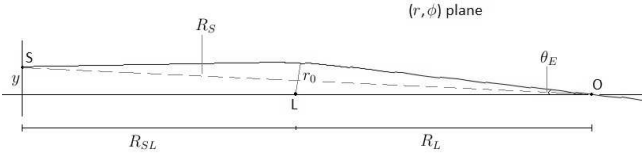}
\caption{Standard setup of gravitational lensing, containing source (S), observer (O) and lensing object (L). The background metric is SdS and the coordinates used in the diagram are Kottler's $r$ and $\phi$ with the lensing object at the origin. The deflected trajectory of light is the solid curve, the parameters appearing on the diagram are defined below.} \label{lensing}
\end{figure}
\subsection*{List of parameters}
\begin{longtable}{l p{77mm}}
$m$ &- Mass of lensing object.\\
$\Lambda$ &- Cosmological constant.\\
$U$ &- 4-velocity vector of a given observer.\\
$r_0$ &- Appearing on the diagram, coordinate (Euclidean) distance of the point of closest approach.\\
$y$ &- Appearing on the diagram, the position coordinate of the source on the lensed plane.\\
$R_S$ &- Appearing on the diagram, the Euclidean distance between point of emission and point of observation on the plane of the diagram (the ($r,\phi$) plane).\\
$R_L$ &- Appearing on the diagram, the Euclidean distance between the point of observation and the lensing object on the plane of the diagram (the ($r,\phi$) plane).\\
$R_{SL}$ &- Appearing on the diagram, the Euclidean distance between the point of emission and the lensing object on the plane of the diagram (the ($r,\phi$) plane). This also happens to be the measurable angular diameter distance, for the specific case of comoving relative motion between the source and the lens. In that case, the distance can be directly measured by an observer moving with the source or the lens, or determined through other measurements.\\
$D_S$ &- The measurable angular diameter distance to the source by a given observer.\\
$D_L$ &- The measurable angular diameter distance to the lensing object by a given observer.\\
$\theta_E$ &- Appearing on the diagram, the Euclidean angle at which the deflected ray arrives at the point of observation.\\
$\theta_M$ &- The measurable position angle of the source in the local frame of a given observer, measured relative to the location of the lensing object (the measurable counterpart of $\theta_E$).\\
\end{longtable}
\subsection*{Important relationships}
The relationships listed below refer to the setup of Figure \ref{lensing}. All of the parameters used are defined above.\\

4-velocity vectors of a static, and a far from the origin comoving observer, in Kottler coordinates:
\begin{equation}
U_{static}=\left( \frac{1}{\sqrt{1-\frac{2m}{r}-\frac{\Lambda}{3}r^2}},0,0,0 \right)
\end{equation}
\begin{equation}
U_{comoving}=\left( \frac{1}{1-\frac{\Lambda}{3}r^2}+O\left( \frac{m}{r}\right) ,\sqrt{\frac{\Lambda}{3}}r+O\left( \frac{m}{r}\right),0,0 \right) \label{app:com}
\end{equation}\\

Gravitational lens equation in terms of Euclidean parameters, to first order in $m$ and $\theta_E$:
\begin{equation}
y=R_S \theta_E-\frac{4mR_{SL}}{R_L \theta_E}. \label{lens}
\end{equation}
Gravitational lens equation in terms of measurable parameters by an observer with 4-velocity $U$, to first order in $m$ and $\theta_M$:
\begin{equation}
y=D_S \theta_M-\frac{4m(D_S-D_L)}{D_L \theta_M}h(\Lambda,U), \label{lensE}
\end{equation}
where
\begin{equation}
h(\Lambda,U)=\left| \frac{f(r)}{f(r)U^t-U^r}\right|_{m=0},\quad  f(r)=1-\frac{2m}{r}-\frac{\Lambda}{3}r^2.
\end{equation}

Specific case of \eqref{lensE} for a static observer, first order in $m$ and $\theta_M$, exact in $\Lambda$:
\begin{equation}
y=D_S \theta_M-\frac{4m(D_S-D_L)}{D_L \theta_M \sqrt{1+\frac{\Lambda}{3}D_L^2}}. \label{app:stat}
\end{equation}
Approximation of the above, first order in $m$, $\theta_M$ and $\Lambda$:
\begin{equation}
y=D_S \theta_M-\frac{4m(D_S-D_L)}{D_L \theta_M}\left( 1-\frac{D_L^2}{6}\Lambda \right).
\end{equation}

Specific case of \eqref{lensE} for a comoving observer, first order in $m$ and $\theta_M$, exact in $\Lambda$:
\begin{equation}
y=D_S \theta_M-\frac{4m(D_S-D_L)}{D_L \theta_M \left( 1-\sqrt{\frac{\Lambda}{3}}D_L \right) }. \label{app:como}
\end{equation}
Approximation of the above, first order in $m$ and $\theta_M$, lowest powers of $\Lambda$:
\begin{equation}
y=D_S \theta_M-\frac{4m(D_S-D_L)}{D_L \theta_M}\left( 1+\frac{D_L}{\sqrt{3}}\sqrt{\Lambda} +\frac{D_L^2}{3}\Lambda \right). \label{app:comoa}
\end{equation}\\

Useful relationship for angular diameter distances:
\begin{equation}
R_{SL}=h(\Lambda,U)(D_S-D_L)+O(\theta_E^2). \label{rsl}
\end{equation}
Specific case of \eqref{rsl} for a static observer:
\begin{align}
R_{SL}&=\frac{D_S-D_L}{\sqrt{1+\frac{\Lambda}{3}D_L^2}}+O(\theta_E^2)\\
&= (D_S-D_L) \left( 1-\frac{D_L^2}{6}\Lambda \right) +O(\Lambda^2)+O(\theta_E^2).
\end{align}
Specific case of \eqref{rsl} for a comoving observer:
\begin{equation}
R_{SL}=\frac{D_S-D_L}{(1-\sqrt{\frac{\Lambda}{3}}D_L)}+O(\theta_E^2)
\end{equation}
\begin{equation}
= (D_S-D_L) \left(1+\frac{D_L}{\sqrt{3}}\sqrt{\Lambda} +\frac{D_L^2}{3}\Lambda \right) +O(\Lambda^{\frac{3}{2}})+O(\theta_E^2). \label{app:dls}
\end{equation}

\end{document}